\documentclass[a4paper,11pt]{article}
\pdfoutput=1
\usepackage[utf8]{inputenc}
\usepackage[T1]{fontenc}
\usepackage[english]{babel}
\usepackage[colorlinks=true
,urlcolor=blue
,anchorcolor=blue
,citecolor=blue
,filecolor=blue
,linkcolor=blue
,menucolor=blue
,linktocpage=true
,pdfproducer=medialab
,pdfa=true
]{hyperref}
\usepackage{amsmath,amsthm,amsbsy,amssymb,amsfonts,graphicx,mathrsfs,bm,cite,color,accents}
\usepackage{empheq}
\usepackage[customcolors]{hf-tikz}
\hfsetfillcolor{blue!5}
\hfsetbordercolor{white}
\hyphenchar\font=\string"7F
\usepackage{tikz}
\usetikzlibrary{calc}
\DeclareMathAlphabet{\mathdsl}{U}{bbm}{m}{sl}

\usepackage[utf8]{inputenc}
\renewcommand{\thefootnote}{\arabic{footnote}}

\makeatletter
\catcode`\@=11
\@addtoreset{equation}{section}

\makeatother

\newcommand{\dd}{\mathrm{d}}

\newcommand{\rmd}{\mathrm{d}}

\newcommand{\cD}{\mathcal D}

\newcommand{\cI}{\mathcal I}\newcommand{\cJ}{\mathcal J}
\newcommand{\cK}{\mathcal K}\newcommand{\cL}{\mathcal L}

\newcommand{\nn}{\nonumber}

\newcommand{\GL}{\text{GL}}





\newcommand{\Lie}{\cL}
\newcommand{\gLie}{\hat{\cL}}
\newcommand{\tr}{\text{tr}}

\newcommand{\sfa}{\mathsf{a}}
\newcommand{\sfb}{\mathsf{b}}
\newcommand{\sfc}{\mathsf{c}}
\newcommand{\sfd}{\mathsf{d}}
\newcommand{\sfe}{\mathsf{e}}
\newcommand{\sff}{\mathsf{f}}
\newcommand{\sfg}{\mathsf{g}}

\newcommand{\sfm}{\mathsf{m}}
\newcommand{\sfn}{\mathsf{n}}
\newcommand{\sfp}{\mathsf{p}}

\allowdisplaybreaks[3]

\begin{document}

\hypersetup{pageanchor=false}
\begin{titlepage}
\renewcommand{\thefootnote}{\fnsymbol{footnote}}

\vspace*{1.0cm}

\centerline{\LARGE\textbf{All maximal gauged supergravities with uplift}}

\vspace{1.0cm}

\centerline{
{Falk Hassler}%
\footnote{E-mail address: falk.hassler@uwr.edu.pl} 
and
{Yuho Sakatani}%
\footnote{E-mail address: yuho@koto.kpu-m.ac.jp}%
}

\begin{center}
${}^\ast${\it University of Wroc\l{}aw, Faculty of Physics and Astronomy,}\\
{\it Maksa Borna 9, 50-204 Wroclaw, Poland}

${}^\dagger${\it Department of Physics, Kyoto Prefectural University of Medicine,}\\
{\it 1-5 Shimogamohangi-cho, Sakyo-ku, Kyoto 606-0823, Japan\\}
\end{center}

\begin{abstract}
Generalised parallelisable spaces permit to uplift many maximal gauged supergravities to ten or eleven dimensions. While some of the former are explicitly known, the literature is still lacking an systematic construction and a complete classification. There are existence proves but they are not sufficient for applications like generalised U-dualities. We address this issue and present a fully explicit construction, and with it a classification, of generalised parallelisable spaces for maximal gauged supergravities in four or more dimensions. All embedding tensors that can be realised without breaking the section condition of exceptional field theory are identified in a distinguished E$_n(n)$ frame and the corresponding generalised frame fields are constructed. This gives new insight into the old question: ``Which maximal gauged supergravities have uplifts to 10/11d?'' and provides the basis to explore solution generating techniques based on generalised dualities. 
\end{abstract}

\thispagestyle{empty}
\end{titlepage}
\hypersetup{pageanchor=true}

\setcounter{footnote}{0}

\newpage
\hrule
\tableofcontents\vspace{2em}
\hrule

\section{Introduction}
Gauged supergravities provide important models in low dimensions and have been actively studied (see \cite{Samtleben:2008pe,Gallerati:2016oyo} for lecture notes). Their application stretches from holography to flux compactifications. A reason for their success is that, unlike in the case of the ungauged supergravities, non-trivial scalar potentials arise. This helps in solving the moduli problem or in producing non-vanishing cosmological constants in string theory. Despite their nice features, the status of gauged supergravities as a low-energy limit of string theory is not completely clear. At low energies, string theory or M-theory are governed by ten- or eleven-dimensional supergravity. However, there is a long-standing question which can be stated as follows:
\begin{center}
\emph{Which gauged supergravities admit an uplift to ten- or eleven-dimensional supergravity?}
\end{center}
We address this question for the case of maximal gauged supergravities in four or more dimensions. If there exists an uplift, we provide the corresponding (generalised) Scherk-Schwarz ansatz by explicitly constructing the twist matrix $E_A{}^I$, also known as the generalised frame field of the underlying generalised parallelisable space. There are several key elements required to approach the question and we will explain them in the following.

First is the embedding tensor formalism of gauged supergravity \cite{Nicolai:2000sc,Nicolai:2001sv,deWit:2002vt,deWit:2004nw,deWit:2005ub,deWit:2007kvg}. It is based on the idea that only a subgroup $G$ of the global U-duality symmetry $E_{n(n)}\times \mathbb{R}^+$ is promoted to a local gauge symmetry. The constants that specify how the former is embedded into the latter are called the embedding tensor. It is realised as a rank-three tensor $X_{AB}{}^C$ that can be interpreted as a set of matrices $(X_A)_B{}^C := X_{AB}{}^C$. Closure of the gauge symmetry requires that these matrices satisfy the quadratic constraint $[X_A,\,X_B]=-X_{AB}{}^C\,X_C$\,. But because the constants $X_{AB}{}^C$ are in general not antisymmetric in their first two indices, they cannot be interpreted as the structure constants of the Lie algebra of $G$. Instead, they describe a Leibniz algebra. Still not any Leibniz algebra is admissible. Supersymmetry imposes additional linear constraints on the embedding tensor, projecting out certain irreducible representations of the duality group. The embedding tensor captures all maximal gauged supergravities in nine dimensions or less; All their properties are encoded in it. A full classification of all these theories would require to identify all solutions to the quadratic constraint. This problem is comparable with classifying all Lie algebras and known to be notoriously difficult. We will therefore not address it here.

The second element is the notion of generalised parallelisable spaces $M$ \cite{Grana:2008yw,Lee:2014mla}. In analogy with group manifolds in differential geometry, a manifold $M$ that admits a globally-defined generalised frame $E_A$ satisfying
\begin{align}\label{eq:framealgebra}
    \gLie_{E_A} E_B = -X_{AB}{}^C\,E_C
\end{align}
is called generalised parallelisable. Here, $\gLie_{E_A}$ denotes the generalised Lie derivative in exceptional generalised geometry \cite{Koepsell:2000xg,Hull:2006tp,Hull:2007zu,PiresPacheco:2008qik,Coimbra:2011ky} or exceptional field theory \cite{Hohm:2013pua,Hohm:2013vpa,Hohm:2013uia,Hohm:2014fxa} (see also \cite{West:2003fc,Kleinschmidt:2003jf,West:2004kb,Riccioni:2007ni,Hillmann:2009ci,Berman:2010is,Berman:2011cg,Berman:2011jh,Aldazabal:2013mya,Cederwall:2013naa} for earlier works on extended geometries). These spaces are so valuable because we can use their respective frame $E_A$ to obtain a generalised Scherk-Schwarz reduction ansatz that results in the gauged supergravity specified by $X_{AB}{}^C$ (see \cite{Hohm:2014qga} for the details of the reduction ansatz). Therefore, we are left with three remaining questions:
\begin{enumerate}
    \item Which embedding tensors $X_{AB}{}^C$ describe generalised parallelisable spaces?
    \item How to construct the corresponding generalised frame fields $E_A$?
    \item Is $E_A$ unique or are there different choices for the same embedding tensor?
\end{enumerate}
Especially for the first two, much progress has been made in \cite{Godazgar:2013oba,Lee:2014mla,Baron:2014yua,Hohm:2014qga,Baron:2014bya,Baguet:2015sma,Lee:2015xga,duBosque:2017dfc,Inverso:2017lrz,Cassani:2019vcl} over the years, and many generalised frames have been constructed explicitly. One of the major results is that for a class of gaugings, called non-geometric gaugings, the generalised frame $E_A$ necessarily depends on the extended coordinates in exceptional field theory. Then $E_A$ is defined on the extended space and violates the section condition. Consequently, the connection to supergravity, generalised geometry and the low-energy limit of string theory is lost. All other gaugings, which we call geometric, correspond to generalised parallelisable spaces. However, besides many explicit examples\footnote{There are too many examples to give a complete list here. To get an impression of the current state of the art refer to \cite{Inverso:2016eet,Malek:2017cle,DallAgata:2019klf}. We thank Gianguido Dall'Agata for pointing out these references to us.}, a complete classification of geometric gaugings and the systematic construction of their frames $E_A$ is still lacking. There are existence proves, most notably \cite{Inverso:2017lrz}. But they do not give complete expressions for $E_A$'s components which are needed for many applications. To fill this gap, is the main objective of this paper.

The third question about the uniqueness has been ignored in the beginning. Most likely, because the construction of a frame on group manifolds is unique up to choosing between right- or left-action. Its relevance became clear in \cite{Hassler:2017yza,duBosque:2017dfc,Sakatani:2019jgu}, where it was shown that in generalised geometry different frames may exist on different manifolds that still result in the same structure constants. They are related by Poisson-Lie T-duality \cite{Klimcik:1995ux} and reveal an intriguing connection between (generalised) dualities in string theory, gauged supergravities and consistent truncations. This idea was generalised to exceptional field theory and results in the notion of generalised U-duality \cite{Sakatani:2019zrs,Malek:2019xrf,Malek:2020hpo,Sakatani:2020wah}. It is concerned with a class of Leibniz algebras, called the Exceptional Drinfel'd Algebras (EDAs). For them the explicit construction of $E_A$, satisfying the algebra \eqref{eq:framealgebra}, is known. Each EDA gives rise to a generalised parallelisable space, and with it a generalised Scherk-Schwarz reduction. However, EDAs only cover a restricted class of embedding tensors and we still do not have a general construction of $E_A$ (see for example section 8 of \cite{Bugden:2021nwl} for the current status).

Finally, generalised parallelisable spaces beyond Drinfel'd algebras\footnote{Drinfel'd algebras and their decomposition into two maximally isotropic subgroups, called Manin triple, describe the local structure of a Poisson-Lie group. That is also the origin of the name Possion-Lie T-duality.} has been explored \cite{duBosque:2017dfc,Demulder:2018lmj,Hassler:2019wvn,Borsato:2021vfy} in double field theory \cite{Siegel:1993xq,Siegel:1993th,Siegel:1993bj,Hull:2009mi,Hull:2009zb,Hohm:2010pp} and generalised geometry. They represent the most general class of embedding tensors that can be realised in generalised geometry and double field theory without violating the section condition. Our main results here, are based on all the insights from above and can be summarised as follows.

First, we identify the most general embedding tensor $X_{AB}{}^C$ that the frame algebra \eqref{eq:framealgebra} can realise without $E_A$ violating the section condition. We call this class \underline{geometric gaugings}. The embedding tensor is captured by certain representations of the global symmetry $E_{n(n)}\times\mathbb{R}^+$. But in this form, it contains various locally non-geometric fluxes. Therefore, we identify linear constraints that remove all locally non-geometric fluxes. Thereby, we find the most general form of the geometric gaugings. Due to the absence of the locally non-geometric fluxes, $E_A$ does not depend on the extended coordinates and the corresponding gauged supergravity can be uplifted to supergravity in ten or eleven dimensions.\footnote{To be more precise, it is uplifted to supergravity only when the trombone gauging is absent, $X_{AB}{}^B=0$\,.} We will see that geometric gaugings are a natural extension of the embedding tensor studied in \cite{Demulder:2018lmj,Hassler:2019wvn,Borsato:2021vfy}. 

Second, we clarify the local structure of the manifold $M$. As is discussed in \cite{Grana:2008yw,Lee:2014mla}, generalised parallelisable spaces $M$ should be coset spaces $M=G/H$ for some Lie groups $G$ and $H$. We elucidate how to construct $G$ for a given $X_{AB}{}^C$, and obtain a distinguished representation of its generators that we call the Leibniz representation (see \cite{Strobl:2016aph,Strobl:2019hha} for a closely related discussion). This clearly shows that $G$ is a subgroup of $E_{n(n)}\times\mathbb{R}^+$. For geometric gaugings, there additionally exists a subgroup $H$ of $G$. Its choice may be not unique and the resulting ambiguity provides the mechanism underlying generalised U-duality. Here we choose a basis in which $H$ is manifest and thereby provide the same starting point as was used in the recent discussion of solution generating techniques in \cite{Borsato:2021vfy}. In gauged supergravity, $G$ plays the role of the gauge group and the coset space $G/H$ is the internal manifold in the reduction required to get it from higher-dimensional supergravity. We also reveal an interesting mathematical structure associated with the Lie group $G$. Using our matrix representation of $G$, we construct a one-form $v^A$ that is an extension of the Maurer-Cartan form of $G$. It requires an additional two-form $w^A$ that satisfies together with $v^A$ a modified Maurer-Cartan equation. This structure is related to enhanced Leibniz algebras as explained in appendix~\ref{app:ELA}. In particular, the one-form $v^A$ plays an important role in our construction of $E_A$. 

Finally, we provide a systematic construction of the generalised frame $E_A$ for all geometric gaugings. Thereby, all gauged supergravities with a higher-dimensional origin, can be explicitly uplifted by using $E_A{}^I$ as the twist matrix in the generalised Scherk-Schwarz ansatz. We restrict the discussion in the main text to maximal supergravities in four or higher dimensions, but the extension to three dimensions is straightforward and we discuss it in appendix~\ref{app:algebra}. Conceptually, it follows the same pattern as $n\le7$, it is only algebraically more involved. For EDAs, the three-dimensional case is studied in \cite{Sakatani:2020wah}. 

This article is organised as follows: In section \ref{sec:ODDmotiv}, we begin with the motivating example of the generalised geometry/double field theory and review the results of \cite{Demulder:2018lmj,Borsato:2021vfy}. All steps taken there will be repeated and if required extended in the later sections. Section \ref{sec:geogaugings} establishes the most general form of geometric gaugings. In exceptional field theory, there exist two inequivalent possibilities to define the physical coordinates (solutions to the section condition). One is suitable for describing M-theory, while the other is for type IIB. Accordingly, we construct two types of geometric gaugings. For each of them, we count how many non-geometric fluxes are removed by the respective linear constraints. We elucidate the construction of the Lie algebra Lie($G$) and its subalgebra Lie($H$) in section \ref{sec:Lie}. Understanding them is crucial for the construction of the generalised frame because it is defined on the coset space $M=G/H$. At this point, we also identify the mathematical structure that underlies the construction of the generalised frame in section~\ref{sec:genframes}. Here, we give explicit expressions for $E_A{}^I$ in M-theory and type IIB. In particular, several field strengths, such as $F^\mu_3$ and $F_5$, are fixed on $M$. They have to satisfy the corresponding Bianchi identities, such as $\dd F^\mu_3=0$ and $\dd F_5 = \tfrac{1}{2}\,\epsilon_{\mu\nu}\,F^\mu_3\wedge F^\nu_3$\,. We show that the Bianchi identities are ensured by the Leibniz identity for $X_{AB}{}^C$ (also known as the quadratic constraint in gauged supergravity). Therefore, we can construct the generalised frame $E_A$ satisfying the algebra \eqref{eq:framealgebra} for arbitrary geometric gaugings $X_{AB}{}^C$. Section \ref{sec:conclusions} is devoted to a brief summary and an outlook on interesting problems that our results make accessible.

\section{Generalised geometry as motivating example}\label{sec:ODDmotiv}
The construction of generalised frames in generalised geometry that satisfy the algebra \eqref{eq:framealgebra} is well-understood. Here the generalised Lie derivative has the explicit form
\begin{equation}\label{eq:genLieODD}
  \gLie_U V^I = U^J \partial_J V^I - \left( \partial_J U^I - \partial^I U_J \right) V^J
\end{equation}
and the frame field $E_A{}^I$ is an element of the Lie group O($D$,$D$). Its construction is a guiding principle for exceptional generalised geometry. Therefore, we review it in this section. The starting points for the discussion are two main ingredients:
\begin{enumerate}
  \item \label{item:LieG} A Lie algebra Lie($G$), defined by
    \begin{equation}
        [T_A, T_B] = F_{AB}{}^C\, T_C\,,
    \end{equation}
    whose adjoint action leaves the $\eta$-metric invariant. The latter arises from the natural paring between vectors and one-forms on the generalised tangent space $T M \oplus T^* M$, which gives rise to
    \begin{equation}\label{eq:genLie_GG}
      U_I\, V^I = \eta_{IJ}\, U^I\, V^J = u_i\, v^i + u^i\, v_i \,.
    \end{equation}
    Assuming $U^I = \begin{pmatrix} u^i & u_i \end{pmatrix}$ and $V^I = \begin{pmatrix} v^i & v_i \end{pmatrix}$, we find
    \begin{equation}
      \eta_{IJ} = \begin{pmatrix}
        0 & \delta_i^j \\
        \delta_j^i & 0
      \end{pmatrix} \quad \text{and the flat version} \quad
      \eta_{AB} = E_A{}^I\, E_B{}^J\, \eta_{IJ}\,.
    \end{equation}
    Both are used together with their inverses to lower/raise curved and flat indices respectively. As a direct consequence of $\eta_{AB}$ being invariant under the action of Lie($G$), namely $F_{A(B}{}^D \eta_{C)D}= 0$, $F_{ABC}$ is totally antisymmetric once its last index is lowered.
  \item\label{item:geogauging} An explicit decomposition of the O($D$,$D$) indices $F_{ABC}$ carries into GL($D$) indices, resulting in the four independent components
    \begin{equation}\label{eq:nongeofluxes}
      F_{abc} = H_{abc}\,, \qquad
      F_{ab}{}^c = f_{ab}{}^c \,, \qquad
      F_a{}^{bc} = Q_a{}^{bc} \,, \qquad \text{and} \qquad
      F^{abc} = R^{abc}\,.
    \end{equation}
    For historic reasons they are called $H$-, $f$-, $Q$- and $R$-fluxes. We will see shortly, that it is only possible to construct a generalised frame which satisfies the section condition in double field theory with $\partial_I = \begin{pmatrix} \partial_i & 0 \end{pmatrix}$ if there is a decomposition to GL($D$) indices that satisfies
    \begin{equation}\label{eq:defeta_AB}
      \eta_{AB} = \begin{pmatrix}
        0 & \delta_a^b \\
        \delta_b^a & 0
      \end{pmatrix} \qquad \text{and} \qquad
      R^{abc} = 0\,.
    \end{equation}
    We call all Lie algebras that admit this form \underline{geometric}, because the R-flux is known as a non-geometric flux. Note that this decomposition is in general not unique and one can sometimes find different choices for the $H$-, $f$- and $Q$-fluxes on the same Lie algebra Lie($G$) with $R^{abc}=0$. This effect is the basis of the Poisson-Lie T-duality. Moreover, there might be non-geometric generalised frames that violate the section condition. But in this work we are not interested in them. In particular, we only use double and exceptional field theory such that they admit a trivial solution of the section condition and therefore are equivalent to generalised geometry.
\end{enumerate}
Based on these ingredients, we will use the ansatz \cite{duBosque:2017dfc}
\begin{equation}\label{eq:ansatzoddframe}
  E_A{}^I = M_A{}^B \begin{pmatrix}
    v_b{}^j & 0 \\
    0 & v^b{}_j
  \end{pmatrix}
  \begin{pmatrix}
    \delta_j^i & -B_{ji} \\
    0 & \delta_i^j
  \end{pmatrix}
\end{equation}
for the generalised frame on the coset $G/H$. Let us explain its constituents in more detail. First, we have the maximally isotropic subgroup $H$, which is generated by $T^a$. Because $R^{abc} = 0$, these generators indeed form a subgroup. Moreover, $M_A{}^B$ implements the adjoint action
\begin{equation}
  M_A{}^B\, T_B = m^{-1}\, T_A\, m
\end{equation}
of a coset representative $m \in G/H$ and $v^a{}_i$ captures one half of the corresponding left-invariant Maurer-Cartan form
\begin{equation}
  T_A\,v^A = m^{-1}\,\dd m = T_a\, v^a{}_i\, \dd x^i + T^a\, A_{i a}\, \dd x^i
\end{equation}
(we call the second part $A_{ia}$ or simply $A_{a}=A_{ia}\,\dd x^i$). Its dual vector fields are denoted by $v_a{}^i$ and satisfy $v_a{}^i\, v^b{}_i = \delta_a^b$. Finally, there is the two-form $B$-field,
\begin{equation}
  B = \frac1{2!}\, B_{ij}\, \dd x^i \wedge \dd x^j \,,
\end{equation}
which still has to be fixed.

To relate this ansatz to the frame algebra in the introduction, we now show that it is possible to fix $B$ such that \eqref{eq:framealgebra} is satisfied for $X_{AB}{}^C = F_{AB}{}^C$. Before we look at the details, let us explain why structure constants for a Lie algebra Lie($G$) are indeed the most general candidate for $X_{AB}{}^C$. The symmetric part $X_{(AB)}{}^C$ can be written in generalised geometry as
\begin{equation}
  \gLie_{E_A} E_B{}^I + \gLie_{E_B} E_A{}^I = -2\, X_{(AB)}{}^C\, E_C{}^I = \partial_I \left( E_{AJ}\, E_B{}^J \right) = \partial_I \eta_{AB} = 0\,.
\end{equation}
Thus, we learn that $X_{AB}{}^C$ has to be antisymmetric with respect to its first two indices. Additionally, the Leibniz identity for the generalised Lie derivative imposes the Jacobi identity $3\, X_{[AB}{}^D\, X_{C]D}{}^E = 0$ on the constants $X_{AB}{}^C$. This shows why $X_{AB}{}^C = F_{AB}{}^C$ holds without any loss of generality. One of the challenges in dealing with the extension to exceptional generalised geometry is that $X_{(AB)}{}^C = 0$ does not hold anymore. Finally, we can check that $X_{ABC}$ is totally antisymmetric and, after taking into account the definition of the generalised Lie derivative \eqref{eq:genLieODD}, is given by
\begin{equation}\label{eq:Xabc_GG}
  X_{ABC} = 3\,\Omega_{[ABC]} 
\end{equation}
with
\begin{equation}
  \Omega_{AB}{}^C = E_A{}^I\,E_B{}^J\,\partial_I E^C{}_J\,.
\end{equation}
To further evaluate this expression, it is convenient to decompose the generalised frame into two parts \cite{duBosque:2017dfc},
\begin{equation}
  E_A{}^I = M_A{}^B\, V_B{}^I\,,
\end{equation}
which gives rise to
\begin{equation}
  \Omega_{AB}{}^C = M_A{}^D\, M_B{}^E\, (M^{-1})_F{}^C \left( \hat\Omega_{DE}{}^F - V_D{}^I\,(M^{-1}\,\partial_I M)_E{}^F \right)\,,
\end{equation}
where
\begin{equation}
  \hat\Omega_{AB}{}^C := V_A{}^I\,V_B{}^J\,\partial_I V^C{}_J\,.
\end{equation}
The inverse adjoint action $(M^{-1})_A{}^B$ that appears here for the first time can be either computed directly by $(M^{-1})_A{}^B\,T_B = m\,T_A\,m^{-1}$ or by imposing $(M^{-1})_A{}^C\, M_C{}^B = (M^{-1}\,M)_A{}^B = \delta_A^B$. Based on the definition of the adjoint action, we compute the second term in the parenthesis as
\begin{equation}
  - V_A{}^I\,( M^{-1}\,\partial_I M )_B{}^C = \begin{cases}
    A_a{}^D\,F_{DB}{}^C \\
    0\,,
  \end{cases}
\end{equation}
with
\begin{equation}
  A_a{}^B = \begin{pmatrix} \delta_a^b \\ v_a{}^i A_{i b} \end{pmatrix}.
\end{equation}

Now it is time to plug the result back into \eqref{eq:Xabc_GG}. Doing so, we notice that the left-hand side is invariant under the adjoint action. Hence,
we only have to verify the simpler conditions
\begin{equation}\label{eq:XaBCcomponents}
  X_{aB}{}^C = \begin{cases}
    Q_a{}^{bc} \\
    \hat X_{ab}{}^c + 2\,A_{i d}\,v_{[a}{}^i\, Q_{b]}{}^{cd} + 2\, f_{ab}{}^c\\
    \hat X_{abc} + 3\,A_{i d}\, v_{[a}{}^i\, f_{bc]}{}^d + 3\, H_{abc}\,,
  \end{cases}
\end{equation}
where $\hat X_{ABC}$ are the generalised fluxes for the generalised frame $V_A{}^I$, namely
\begin{equation}
  \hat X_{ABC} = 3\,\hat\Omega_{[ABC]} \,.
\end{equation}
At this point, it becomes obvious why $X^{abc} = R^{abc} = 0$ has to be imposed. There is no contribution on the right-hand side with all indices up, because all non-vanishing terms there have at least one of their three indices down. Thus, $H$-, $f$- and $Q$-fluxes are permitted but we cannot have $R$-flux. To check the remaining components, we have to compute
\begin{equation}
    \begin{aligned}
        \hat X_{ab}{}^c &= 2\,v_a{}^i\, v_b{}^j\, \partial_{[i} v^c_{j]} = -\iota_{v_a} \iota_{v_b} \dd v^c \\
        \hat X_{abc} &= 3\,v_a{}^i\,v_b{}^j\, v_c{}^k\, \partial_{[i} B_{jk]}\,.
    \end{aligned}
\end{equation}
With the help of the Maurer-Cartan equation,
\begin{equation}\label{eq:dvA}
  \dd v^A = -\frac12\, F_{BC}{}^A\, v^B \wedge v^C\,,
\end{equation}
the first equation for $\hat{X}_{ab}{}^c$ can be further evaluated to
\begin{equation}
  \hat{X}_{ab}{}^c = - f_{ab}{}^c - 2\, A_{i d}\, v_{[a}{}^i\, Q_{b]}{}^{cd} \,.
\end{equation}
This is great because it is exactly what is needed to obtain the desired result $X_{ab}{}^c = f_{ab}{}^c$ from \eqref{eq:XaBCcomponents}. We see that all components of $X_{AB}{}^C$, except for the last one $X_{abc}$\,, already work out fine only by assuming $R^{abc}=0$. To complete the argument, we have to fix the two-form $B$ such that
\begin{equation}\label{eq:H-ODD}
    \begin{aligned}
        \dd B = H &= \frac{1}{6} \left( v^a\, H_{abc} - 3\, v^A\, F_{Abc} \right) \wedge v^b \wedge v^c \\
        &= -\frac13\, H_{abc}\, v^a \wedge v^b \wedge v^c - \frac12\, f_{ab}{}^c\, v^a \wedge v^b \wedge A_c
    \end{aligned}
\end{equation}
holds.

We can find at least locally a $B$ with this property if, and only if, $H$ is closed,
\begin{equation}\label{eq:Hclosed}
  \dd H = 0\,.
\end{equation}
To see that this is the case, we decompose the equation
\begin{equation}
  \frac34\, F_{[AB}{}^E\, F_{CD]E} = J_{ABCD} = 0\,,
\end{equation}
which follows from the Jacobi identity of the Lie algebra Lie($G$), into the four non-trivial contributions
\begin{align}
  J_{abcd} = \frac32\, f_{[ab}{}^e\, H_{cd]e} &= 0\,, \nn \\
  J_{abc}{}^d = \frac34 \left( H_{e[ab}\, Q_{c]}{}^{de} - f_{[ab}{}^e\, f_{c]e}{}^d \right) &= 0\,, \label{eq:brokenjacobif}\\
  J_{ab}{}^{cd} = \frac14\left( f_{ab}{}^e\, Q_e{}^{cd} - 4 f_{e[a}{}^{[c}\, Q_{b]}{}^{d]e} \right) &= 0\,, \nn \\
  J_a{}^{bcd} = \frac34\, Q_a{}^{e[b}\, Q_e{}^{cd]} &= 0 \,. \label{eq:jacobiLieH}
\end{align}
They permit us to better understand the role the different fluxes play. Thus, it is worth having a short digression. The last relation, \eqref{eq:jacobiLieH}, describes the Jacobi identity for the structure constants $Q_a{}^{bc}$ of the maximally isotropic subalgebra Lie($H$) generated by $T^a$. From \eqref{eq:brokenjacobif}, we learn that the remaining generators $T_a$ generally do not span a second Lie algebra. They only do when $H_{abc}$ vanishes and we interpret it as an obstruction to obtain a second maximally isotropic subgroup. But back to \eqref{eq:Hclosed}. By applying the Maurer-Cartan equation \eqref{eq:dvA}, we find the following contributions: 
\begin{equation}\label{eq:dH=0}
  \dd H = - \frac12\, J_{abcd}\, v^a \wedge v^b \wedge v^c \wedge v^d - \frac43\, J_{abc}{}^d\, v^a \wedge v^b \wedge v^c \wedge A_d - J_{ab}{}^{cd}\, v^a \wedge v^b \wedge A_c \wedge A_d = 0\,.
\end{equation}
We conclude that the three-form $H$ is closed due to the Jacobi identity of the Lie algebra Lie($G$). Therefore, it is possible to construct at least patch-wise the $B$-field that enters our ansatz \eqref{eq:ansatzoddframe} for the generalised frame.

\section{Geometric gaugings in maximal supergravities}\label{sec:geogaugings}
We now explain how the construction in generalised geometry can be extended to exceptional generalised geometry. In doing so, we have to overcome two conceptual challenges. They can be directly related to the two protagonists of section~\ref{sec:ODDmotiv} which are listed on page~\pageref{item:LieG}:
\begin{enumerate}
  \item As we discussed above, for generalised geometry the constants $X_{AB}{}^C$ in \eqref{eq:framealgebra} are antisymmetric with respect to their first two indices. This allowed us to interpret them as structure constants of a Lie algebra, Lie($G$), whose Lie group $G$ was one of the major ingredients in the construction of the generalised frame. But in exceptional generalised geometry, the structure of the generalised Lie derivative is more complicated and therefore $X_{AB}{}^C$ is in general not antisymmetric in $A$ and $B$.
  \item Moreover, we need to relate the extended symmetry group to the structure group of the manifold $M$ on which the generalised frame $E_A{}^I$ is defined. For the T-duality group O($D$,$D$) this requires a decomposition of duality covariant, flat indices of $X_{AB}{}^C$ into GL($D$) indices carried by tensors on the manifold $M$. Only after this branching, we were able to exclude non-geometric frames. Those cannot be constructed without violating the section condition, and if it is violated, it is only possible to uplift the corresponding gauge supergravity in extended field theories. Then any direct relation to supergravity as low energy effective theory of string and M-theory is lost. Therefore, we limit our discussion to conventional geometric frames and thus require that certain non-geometric flux components of $X_{AB}{}^C$ vanish. More specifically for O($D$,$D$)$\,\rightarrow\,$GL($D$), $X_{AB}{}^C$ decomposes four independent tensors, $H_{abc}$, $f_{ab}{}^c$, $Q_a{}^{bc}$ and $R^{abc}$.  Non-geometric frames are exclusively characterised by $R^{abc}$ which we require to vanish. Extending this approach to the U-duality groups E$_{n(n)}$, we have to overcome two problems:
  \begin{enumerate}
    \item There are two different ways to branch E$_{n(n)}$, namely
      \begin{align}
        \mathrm{E}_{n(n)} &\rightarrow \mathrm{GL}(n) \qquad&& \text{M-theory section} \label{eq:branchM}\\
        \mathrm{E}_{n(n)} &\rightarrow \mathrm{GL}(n-1) \times \mathrm{SL}(2) && \text{IIB section\,.}\label{eq:branchB}
      \end{align}
      We will analyse them separately and in particular
    \item identity the non-geometric fluxes which have to vanish in order to obtain a geometric target space. We call the corresponding $X_{AB}{}^C$ \underline{geometric gaugings} in the M-theory/IIB section.
  \end{enumerate}
\end{enumerate}
First, the second challenge will be addressed in the remainder of this section. Its resolution is crucial to solve the first problem in the next section. At this point, we will identify a Lie subalgebra, Lie($G$), of the full Leibniz algebra encoded in the constants $X_{AB}{}^C$. Its associated Lie group $G$, together with a properly chosen subgroup $H$, will form the manifold $M=G/H$ on which the generalised frame is defined. Compared to the last section, the additional layer of complexity is to extract these two subalgebras. Finally, we complete the construction of the corresponding generalised frame fields in section~\ref{sec:genframes}.

Before, looking at the details of geometric gauging, it is mandatory to define the generalised Lie derivative that dominates the frame algebra \eqref{eq:framealgebra}, namely \cite{Berman:2012vc}
\begin{equation}
  \gLie_U V^I = U^J \partial_J V^I - V^J \partial_J U^I + Y^{IJ}_{KL}\,\partial_J U^K\,V^L\,.
  \label{eq:gLie}
\end{equation}
Comparing this definition with \eqref{eq:genLieODD}, we see that for generalised geometry/double field theory
\begin{equation}
  Y^{IJ}_{KL} = \eta^{IJ}\, \eta_{KL}
\end{equation}
holds. Its generalisation to the U-duality groups E$_{n(n)}$ is given by \cite{Berman:2012vc}
\begin{equation}\label{eq:defY}
  Y^{IJ}_{KL} = (t^{\bm\alpha})_L{}^I\, (t_{\bm\alpha})_K{}^J + \beta\, \delta_L^I\, \delta_K^J + \delta_K^I\, \delta_L^J\,.
\end{equation}
Here $(t_{\bm\alpha})_A{}^B$ denote the generators of the U-duality group and we use their Cartan-Killing metric $\kappa_{\bm\alpha\bm\beta}$ and its inverse $\kappa^{\bm\alpha\bm\beta}$ to lower and raise adjoint valued indices like $\bm\alpha$, $\bm\beta$, \dots\,. For \eqref{eq:defY} to be well-defined, we also have to fix the normalisation of $\kappa_{\bm\alpha\bm\beta}$. A convenient choice that avoids a prefactor in front of the first term of \eqref{eq:defY}'s right-hand side is \cite{Sakatani:2020wah}
\begin{equation}
  \kappa_{\bm\alpha\bm\beta} = - \frac{1}{\alpha}\, \tr_{R_1} \left( t_{\bm\alpha}\, t_{\bm\beta} \right)\,.
\end{equation}
The trace in this expression is evaluated in the same representation, $R_1$, in which also all capital indices transform. Table~\ref{tab:dualitygroups} summarises the relevant representations and constants for all duality groups we deal with in this article.
\begin{table} 
  \centering\begin{tabular}{c||ccccc}
    & O($D$,$D$) & E$_{4(4)}$ & E$_{5(5)}$ & E$_{6(6)}$ & E$_{7(7)}$ \\
    \hline
    $\alpha$ & 2 & 3 & 4 & 6 & 12 \\
    $\beta$  & 0 & 1/5 & 1/4 & 1/3 & 1/2 \\
    $R_1$    & $\bm{2D} $ & $\overline{\bm{10}}$ & $\bm{16}$ & $\bm{27}$ & $\bm{56}$ \\
    $R_2$    & $\bm{1}$   & $\bm{5}$ & $\bm{10}$ & $\overline{\bm{27}}$ & $\bm{133}$
  \end{tabular}
  \caption{Relevant representations and constants for all discussed duality groups.}\label{tab:dualitygroups}
\end{table}

Geometric gaugings are defined such that, the generalised frame fields $E_A$ in \eqref{eq:framealgebra} can be constructed without breaking the section condition
\begin{equation}\label{eq:SC}
  Y^{IJ}_{KL}\, \partial_I \,\cdot\, \partial_J \,\cdot\, = 0
\end{equation}
of exceptional field theory. Any violation of this would spoil the relation to exceptional generalised geometry and make a supergravity interpretation impossible. Therefore, we restrict our discussion to geometric gaugings although they put strong restrictions on the derivative $\partial_I$ of $E_A$ and thereby limit the possible form of $X_{AB}{}^C$. In the following, we present the most general form of the latter that permits geometric generalised frames. To this end, we first note that $X_{AB}{}^C$ is not an arbitrary rank-3 tensor in the tensor product $R_1\otimes R_1 \otimes\overline{R_1}$. Because the generalised Lie derivative has to level the $Y$-tensor invariant, we have
\begin{equation}
    X_{AB}{}^C = W_{AB}{}^C + (t^{\bm{\alpha}})_D{}^E\, (t_{\bm{\alpha}})_B{}^C\,W_{EA}{}^D - \beta\,W_{DA}{}^D\,(t_0)_B{}^C \,,
\label{eq:X-general}
\end{equation}
where $W_{AB}{}^C$ are certain constants that can be expressed as
\begin{equation}
    W_{AB}{}^C := W_A{}^{\bm{\alpha}}\,(t_{\bm{\alpha}})_B{}^C + W_A{}^0\,(t_0)_B{}^C\,.
\end{equation}
In particular, we have here complemented the generators $t_{\bm\alpha}$ of the duality group E$_{n(n)}$ by the scaling symmetry generator $(t_0)_A{}^B = -\delta_A^B$. Using this convention, the section condition \eqref{eq:SC} for the generalised frame fields $E_A$ is equivalent to
\begin{equation}\label{eq:SC-W}
    Y^{AB}_{CD}\,W_A{}^{\dot{\bm{\alpha}}}\,W_B{}^{\dot{\bm{\beta}}} = 0\,,
\end{equation}
where $\dot{\bm{\alpha}}=\begin{pmatrix}\bm{\alpha} & 0\end{pmatrix}$\,. To see how this pivotal constraint arises, note that $W_{AB}{}^C$ is actually the value of the generalised Weitzb{\"o}ck connection
\begin{equation}
  \Omega_{AB}{}^C := E_A{}^I\,E_B{}^J\,\partial_I E_J{}^C \qquad \text{where} \qquad
  E_I{}^A:= (E^{-1})_I{}^A\,,
\end{equation}
at the distinguished point where $E_A{}^I$ is the identity element of E$_{n(n)}$. Without loss of generality, we assign the coordinates $x^i = 0$ to this point and thereby define
\begin{equation}
  W_{AB}{}^C = \left. \Omega_{AB}{}^C \right|_{x^i=0} \,.
\end{equation}
Now one can easily verify that \eqref{eq:SC-W} is equivalent to the section condition imposed at $x^i=0$ for the generalised frames,
\begin{equation}
  \left. Y^{IJ}_{KL}\, \partial_I E_A{}^M\, \partial_J E_B{}^N \right|_{x^i=0} = 0\,.
\end{equation}
This condition is clearly necessary, and we will see later that it is also sufficient.

As proposed in \eqref{eq:branchM} and \eqref{eq:branchB}, there are two inequivalent ways to branch the $R_1$ representation to GL($n$) and GL($n-1$)$\times$SL(2), respectively. We consider these branchings, because the general linear group in $n$ dimensions captures diffeomorphisms on an $n$-dimensional manifold $M$. On this manifold, we will construct the generalised frame $E_A$. There are two different branchings, because of the well-known fact that the section condition in exceptional field theory admits two inequivalent solutions:
\begin{itemize}
    \item M-theory section: As an example, let us take the $R_1$ of E$_{7(7)}$ and the branching
    \begin{equation}\label{eq:R1-M}
      \bm{56} \rightarrow \bm{7} \oplus \overline{\bm{21}} \oplus \bm{21} \oplus \overline{\bm{7}}
    \end{equation}
    from E$_{7(7)}$ to GL(7). On the seven-dimensional manifold $M$\footnote{In the context of exceptional field theory, this is also denoted as the internal space.} these irreps correspond to the exceptional generalised tangent bundle $TM \oplus \Lambda^2 T^* M \oplus \Lambda^5 T^*M \oplus (\Lambda^7 T^* M\otimes T^*M)$ \cite{Hull:2007zu,PiresPacheco:2008qik,Coimbra:2011ky} whose sections are fixed by a vector, a two-form, a five-form and a mixed-symmetry tensor on $M$ (see also \cite{West:2003fc,Cook:2008bi} for the decomposition of the $R_1$ representation based on the $E_{11}$ programme \cite{West:2001as}). Accordingly, the derivative $\partial_I$ on the extended space decomposes into the four contributions $\partial_I = \begin{pmatrix} \partial_i & \tilde\partial^{i_1 i_2} & \tilde\partial^{i_1 \dots i_5} & \tilde\partial^{i_1 \dots i_7,i'} \end{pmatrix}$. Only the first corresponds to momentum modes dual to the position on $M$. All other contributions describe various brane charges. Therefore, up to an $E_{7(7)}$ transformation, any solution of the section condition satisfies $\tilde\partial^{i_1 i_2} \, \cdot \, = 0$, $\tilde\partial^{i_1 \dots i_5} \, \cdot \, = 0$ and $\tilde\partial^{i_1\dots i_7,i} \, \cdot \, = 0$. In particular, this tells us that $Y^{ij}_{KL} = 0$. Going from curved to flat indices by an arbitrary E$_{7(7)}$ transformation, like it is mediated by the generalised frame $E_A{}^I$, we find exactly the same situation, $Y^{ab}_{CD} = 0$, because the Y-tensor is an $E_{7(7)}$ invariant. We therefore conclude from \eqref{eq:SC-W} that $W_A{}^{\dot{\bm{\alpha}}}$ is constraint by
    \begin{equation}\label{eq:WA-M}
        W_A{}^{\dot{\bm{\alpha}}} = \delta_A^b\,W_b{}^{\dot{\bm{\alpha}}} \qquad (a,b=1,\dotsc,n)\,.
    \end{equation}
    Here we have dropped the assumption $n=7$\,, because the same discussion goes through for $n\le 7$. For example, for $n=6$\,, the last component of the $R_1$ representation $\tilde\partial^{i_1 \dots i_7,i'}$, which contains seven totally antisymmetric indices, disappears and the branching
    \begin{equation}
        \bm{27} \rightarrow \bm{6} \oplus \overline{\bm{15}} \oplus \overline{\bm{6}}
    \end{equation}
    of the $R_1$ of E$_{6(6)}$ to GL(6) is recovered. In this way, we obtain the same conclusion \eqref{eq:WA-M} for any $n\leq 7$\,. For completeness, we emphasise that the lowercase indices $a$, $b$, \dots{} and their curved counterparts $i$, $j$, \dots{} enumerate all basis vectors of the fundamental representation of GL($n$).
  
    \item Type IIB section: Again, we take $R_1$ of E$_{7(7)}$ as an illustrative example and consider the branching
    \begin{equation}\label{eq:R1-IIB}
        \bm{56} \rightarrow (\bm{6}, \bm{1}) \oplus (\overline{\bm{6}}, \bm{2}) \oplus (\bm{20},\bm{1}) \oplus (\bm{6}, \bm{2}) \oplus (\overline{\bm{6}}, \bm{1})
    \end{equation}
    from E$_{7(7)}$ to GL(6)$\times$SL(2). On the resulting six-dimensional manifold $M$, these irreps correspond to the type IIB generalised tangent bundle $T M \oplus \left( T^*M \right)^2 \oplus \Lambda^3 T^*M \oplus \left(\Lambda^5 T^* M \right)^2 \oplus (\Lambda^6 T^* M\otimes T^*M)$ \cite{Hull:2007zu,Grana:2009im,Aldazabal:2010ef,Grana:2011nb} (see also \cite{Cook:2008bi,Tumanov:2014pfa}). Its sections are fixed by a vector, a pair of one-forms, a three-form, a pair of five-forms, and a mixed-symmetry tensor. Under the S-duality subgroup SL(2), the respective pairs transform as doublets. Due to this structure, the derivative on the extended space decomposes into the five contributions $\partial_I = \begin{pmatrix} \partial_{\sfm} & \tilde\partial^{\sfm}_\mu & \tilde\partial^{\sfm_1 \dots \sfm_3} & \tilde\partial^{\sfm_1\dots \sfm_5}_\mu & \tilde\partial^{\sfm_1\dots \sfm_6,\sfm'} \end{pmatrix}$, where we introduced the index $\mu=1,\,2$ for the fundamental representation of SL(2). The section condition again requires that after a suitable $E_{7(7)}$ transformation all derivatives with a tilde applied to any field vanish. Consequently, we obtain $Y^{\sfm\sfn}_{KL} = 0$ and, following the same argument as for the M-theory section above, $Y^{\sfa\sfb}_{CD} = 0$. Applying this result to \eqref{eq:SC-W} gives rise to the linear constraint
    \begin{equation}\label{eq:WA-B}
        W_A{}^{\dot{\bm{\alpha}}} = \delta_A^{\sfb}\,W_{\sfb}{}^{\dot{\bm{\alpha}}}\qquad (\sfa,\sfb=1,\dotsc,n-1)\,.
    \end{equation}
    As before, the result can be generalised for $n\le7$. Now indices $\sfa$, $\sfb$, \dots{} and $\sfm$, $\sfn$, \dots{} label the basis of GL($n-1$)'s fundamental representation.
\end{itemize}
Both cases give different linear constraints on geometric gaugings that are completely encoded by $W_A{}^{\dot{\bm\alpha}}$. Any geometric gauging has to solve at least one of these constraints. Of course, it can also solve both simultaneously. Due to their different algebraic structure, we further discuss \eqref{eq:WA-M} and \eqref{eq:WA-B} separately in the following two subsections.

\subsection{Geometric gaugings in the M-theory section}\label{sec:geoMth}
We already discussed how to split the $R_1$ irrep that governs capital indices in \eqref{eq:R1-M}. In the same vein, we also branch the adjoint representation, $\bm{133}$, of E$_{7(7)}$ to deal with the $\dot{\bm\alpha}$ index on the right-hand side of \eqref{eq:WA-M}. It gives rise to
\begin{equation}
  \bm{133} \rightarrow \bm{7} \oplus \overline{\bm{35}} \oplus (\bm{48}+\bm{1}) \oplus \bm{35} \oplus \overline{\bm{7}}
\end{equation}
and we decompose the generators of E$_{7(7)}$ accordingly into
\begin{equation}\label{eq:Enn-gen-M}
    t_{\bm{\alpha}} = \begin{pmatrix}
        R^{a_1\dots a_6} & R^{a_1\dots a_3} & K^a{}_b & R_{a_1\dots a_3} & R_{a_1\dots a_6}
    \end{pmatrix}.
\end{equation}
Combining this result with the four different contributions to the index $A$ discussed below \eqref{eq:R1-M}, we find that $W_a{}^{\dot{\bm{\alpha}}}=\begin{pmatrix} W_a{}^{\bm{\alpha}} & W_a{}^0\end{pmatrix}$ consists of six independent GL(7) tensors, namely
\begin{equation}
    f_{a,b_1\dots b_6},\,f_{a,b_1\dots b_3},\,f_{a_1 a_2}{}^b,\,f_a{}^{b_1\dots b_3},\,f_a{}^{b_1\dots b_6}\,, \quad \text{and}\quad Z_a\,.
\end{equation}
By substituting $W_a{}^{\dot{\bm{\alpha}}}$ back into \eqref{eq:X-general}, we compute all the components of the structure constants $X_{AB}{}^C$ that are permitted for geometric gaugings. It turns out that only the totally antisymmetric part of $f_{a,a_1\dots a_6}$, $f_{a,a_1\dots a_3}$, and $f_{a,b}{}^c$ contributes to $X_{AB}{}^C$, and thus $X_{AB}{}^C$ consists of the structure constants
\begin{equation}
 f_{a_1\dots a_7},\,f_{a_1\dots a_4},\,f_{a_1 a_2}{}^b,\,f_a{}^{b_1\dots b_3},\,f_a{}^{b_1\dots b_6}\,, \quad \text{and} \quad Z_a\,.
\label{eq:M-structureC}
\end{equation}
We then introduce the generators
\begin{equation}\label{eq:gen-M}
    T_A = \begin{pmatrix}
        T_a & T^{a_1a_2} & T^{a_1\dots a_5} & T^{a_1\dots a_7,a'}
    \end{pmatrix},
\end{equation}
in one-to-one correspondence with the generalised tangent bundle that we obtained from the branching \eqref{eq:R1-M} above. Again, we can treat all U-duality groups with $n\le7$ in a unified fashion. For example, to find all geometric gaugings of the M-theory section of the U-duality group SL(5)=$E_{4(4)}$, we remove the generators $T^{a_1\dots a_5}$, $T^{a_1\dots a_7,a'}$ and the corresponding structure constants $f_{a_1 \dots a_7}$, $f_a{}^{b_1\dots b_6}$.

With $X_{AB}{}^C$ completely fixed, we define the algebra $T_A\circ T_B = X_{AB}{}^C\,T_C$ of the generators that we eventually want to represent in terms of the generalised frame fields $E_A$. For the sake of brevity, we present here only its contributions for $n\leq 6$\,, 
\begin{equation}\label{eq:geoalg-M}
    \begin{aligned}\tikzmarkin{geoalg-M}(7.4,-0.3)(-1,0.4)
        T_a \circ T_b &= f_{ab}{}^c\,T_c + f_{abc_2}\,T^{c_2} \,,\\
        T_a \circ T^{b_2} &= -f_a{}^{b_2c}\,T_c + \delta^{b_2}_{de}\,f_{ac}{}^{d}\,T^{ec} + 3\,Z_a\,T^{b_2} - f_{a c_3}\,T^{b_2 c_3} \,,\\
        T_a \circ T^{b_5} &= f_a{}^{b_5c}\,T_c + \delta^{b_5}_{c_3d_2}\,f_{a}{}^{c_3}\,T^{d_2} - \delta^{b_5}_{de_4}\,f_{ac}{}^{d}\,T^{e_4c}  +6\,Z_a\,T^{b_5} \,,\\
        T^{a_2} \circ T_b &= f_b{}^{a_2c}\,T_c + \delta^{a_2}_{de}\,\delta_{c_2b}^{f_2e}\,f_{f_2}{}^{d}\,T^{c_2} - 3\,Z_c\,\delta^{ca_2}_{be_2}\,T^{e_2} + \delta^{a_2c_4}_{bd_5}\,f_{c_4}\,T^{d_5} \,,\\
        T^{a_2} \circ T^{b_2} &= \delta^{b_2}_{de}\, f_c{}^{a_2d}\, T^{ec} - \delta^{a_2}_{de}\,f_{c_2}{}^{d}\,T^{eb_2c_2} + 3\,Z_c\,T^{a_2b_2c} \,,\\
        T^{a_2} \circ T^{b_5} &= -\delta^{b_5}_{de_4}\,f_c{}^{a_2 d}\, T^{e_4c} \,,\\
        T^{a_5} \circ T_b &= - f_b{}^{a_5c}\,T_c - \delta^{a_5}_{c_3d_2}\,f_b{}^{c_3}\,T^{d_2} - \delta^{a_5}_{d_3bf}\,f_c{}^{d_3}\,T^{fc}\\
        &\quad + \delta^{a_5}_{de_4}\,f_{bc}{}^{d}\,T^{e_4c} + \delta^{a_5}_{dbf_3}\,f_{c_2}{}^{d}\, T^{f_3 c_2} - 6\,\delta^{ca_5}_{bd_5}\,Z_c\, T^{d_5}\,,\\
        T^{a_5} \circ T^{b_2} &= - \delta^{b_2}_{de}\,f_c{}^{a_5 d}\,T^{ec}  + \delta^{a_5}_{d_3e_2}\,f_c{}^{d_3}\, T^{e_2 b_2c} \,,\\
        T^{a_5} \circ T^{b_5} &= \delta^{b_5}_{de_4}\,f_c{}^{a_5 d}\, T^{e_4c} \,,\tikzmarkend{geoalg-M}
    \end{aligned}
\end{equation}
while the full algebra for $n\leq 7$ is given in appendix~\ref{app:algebra}. To keep the form of this algebra as clear as possible, we introduced here a shorthand notation that compresses the $n$ indices of $T^{a_1\dots a_n}$ to $T^{a_n}$. Contractions of the arising multi-indices contain additionally, the proper combinatorial factor, such as $V_{a_n}\,W^{a_n}=\frac{1}{n!}\,V_{a_1\dots a_n}\,W^{a_1\dots a_n}$\,. 
Furthermore, all Kronecker deltas with multi-indices are defined by $\delta_{a_n}^{b_n}=n!\,\delta_{a_1\dots a_n}^{b_1 \dots b_n}$\,.

If the generators $T_a$ form a Lie algebra, the structure constants $f_{a_4}$ and $f_{a_7}$ disappear and we reproduce the class of exceptional Drinfel'd algebras (EDA) \cite{Sakatani:2019zrs,Malek:2019xrf,Malek:2020hpo,Sakatani:2020wah}. They generalise Lie bi-algebras that arise as the infinitesimal structure around the identity of a Poisson-Lie group. If one instead removes the (globally) non-geometric fluxes $f_a{}^{b_1\dots b_3}$ and $f_a{}^{b_1\dots b_6}$ (which generalise $Q_a{}^{bc}$ in \eqref{eq:nongeofluxes}), and also $Z_a$\,, the situation studied in \cite{Godazgar:2013oba} arises. Another interesting outcome of \eqref{eq:M-structureC} is an upper bound on the dimension of geometric gaugings. It is only a bound, because the Leibniz identity for the structure constants $X_{AB}{}^C$ imposes additional constraints. However, we can compare this bound with the number of the embedding tensor components that survive the linear constraints. They are listed in table~\ref{tab:ratio-geo}. To reproduce for example the result for $n=7$ given there, we have to remember that the embedding tensor (with the trombone gauging) furnishes the representations $\bm{912}+\bm{56}$. Because they contain both geometric and non-geometric gaugings, we need to impose additional linear constraints. They only permit the structure constants given in \eqref{eq:M-structureC}, whose representations under $\GL(n)$ are
\begin{equation}
    \bm{1} \oplus \overline{\bm{35}} \oplus (\bm{140} \oplus \bm{7}) \oplus (\overline{\bm{224}} \oplus \overline{\bm{21}}) \oplus (\bm{28} \oplus \bm{21}) \oplus \bm{7}\,.
\end{equation}
In total this amounts to $484$ components, and the other $484$ components correspond to (locally) non-geometric fluxes. 
\begin{table}
    \centering
    \begin{tabular}{c||c|cc|cc}
        $n$ & embedding tensor & \multicolumn{2}{c|}{M-theory} & \multicolumn{2}{c}{type IIB} \\\hline
        2 & $(\bm{3}\oplus\bm{2})\oplus (\bm{2}\oplus\bm{1})$ & 4 & (50\%) & 4 & (50\%) \\
        3 & $((\bm{6},\bm{2})\oplus(\bm{3},\bm{2}))\oplus (\bm{3},\bm{2})$ & 15 & (63\%) & 14 & (58\%) \\
        4 & $(\bm{\overline{40}}\oplus\bm{\overline{15}})\oplus \bm{\overline{10}}$ & 45 & (69\%) & 41 & (63\%) \\
        5 & $\bm{144}\oplus\bm{16}$ & 110 & (69\%) & 96 & (60\%) \\
        6 & $\bm{351}\oplus \bm{27}$ & 237 & (63\%) & 191 & (51\%) \\
        7 & $\bm{912} \oplus \bm{56}$ & 484 & (50\%) & 340 & (35\%)
    \end{tabular}
    \caption{The representation of the embedding tensor (with trombone) \cite{Nicolai:2001sv,deWit:2002vt,deWit:2003hr,LeDiffon:2008sh} and the number of geometric gaugings in each dimension.\label{tab:ratio-geo}}
\end{table}

\subsection{Geometric gaugings in the type IIB section}\label{sec:geoIIB}
For type IIB sections, we proceed as in the last subsection and branch the adjoint representation of E$_{7(7)}$ to the group of diffeomorphisms GL(6) and S-duality transformations SL(2),
\begin{equation}
    \bm{133} \rightarrow 2\,(\bm{1}, \bm{2}) \oplus 2\,(\bm{15}, \bm{1}) \oplus 2\,(\overline{\bm{15}}, \bm{2}) \oplus (\bm{35}+\bm{1}, \bm{1}) \oplus (\bm{1}, \bm{3}) \,.
\end{equation}
According to this decomposition, the generators $t_{\bm\alpha}$ split into the eight contributions
\begin{equation}\label{eq:Enn-gen-IIB}
    t_{\bm{\alpha}} = \begin{pmatrix}
        R^{\sfa_1\dots\sfa_6}_\alpha & R^{\sfa_1\dots\sfa_4} & R_\alpha^{\sfa_1\sfa_2} & K^{\sfa}{}_{\sfb} & R^{\alpha}{}_{\beta} & R^\alpha_{\sfa_1\sfa_2} & R_{\sfa_1\dots \sfa_4} & R_{\sfa_1\dots\sfa_6}^\alpha
    \end{pmatrix},
\end{equation}
where $\alpha=1,2$\,, and correspondingly $W_{\sfa}{}^{\dot{\bm{\alpha}}}=\begin{pmatrix} W_{\sfa}{}^{\bm{\alpha}} & W_{\sfa}{}^0\end{pmatrix}$ is formed by
\begin{equation}
    f^\beta_{\sfa,\sfb_1\dots\sfb_6},\,f_{\sfa,\sfb_1\dots\sfb_4},\,f^\beta_{\sfa,\sfb_1\sfb_2},\,f_{\sfa,\sfb}{}^{\sfc},\,f_{\sfa,\beta}{}^{\gamma},\,f_{\sfa}{}_\beta^{\sfb_1\sfb_2},\,f_{\sfa}{}^{\sfb_1\dots\sfb_4},\,f_{\sfa}{}_\beta^{\sfb_1\dots\sfb_6}\,, \quad \text{and}\quad Z_{\sfa}\,.
\end{equation}
Plugging this result back into \eqref{eq:X-general}, we again find that only those parts of the various $f$ whose lowered indices are totally antisymmetrised contribute. Hence, we obtain the structure constants
\begin{equation}\label{eq:IIB-structureC}
    f_{\sfa_1\dots \sfa_5},\,f^\alpha_{\sfa_1\dots\sfa_3},\,f_{\sfa_1\sfa_2}{}^{\sfb},\,f_{\sfa\beta}{}^{\gamma},\,f_{\sfa}{}_\beta^{\sfb_1\sfb_2},\,f_{\sfa}{}^{\sfb_1\dots\sfb_4},\,f_{\sfa}{}_{\beta}^{\sfb_1\dots \sfb_6}\,,\quad \text{and} \quad Z_{\sfa}
\end{equation}
for geometric gaugings in the type IIB section. Guided by \eqref{eq:R1-IIB}, we also build the generators $T_A$ that mimic the algebra \eqref{eq:framealgebra} of the generalised frame $E_A$ from the five components 
\begin{equation}\label{eq:gen-IIB}
    T_A = \begin{pmatrix}
        T_{\sfa} & T_\alpha^{\sfa} & T^{\sfa_1\dots\sfa_3} & T_\alpha^{\sfa_1\dots\sfa_5} & T^{\sfa_1\dots\sfa_6,\sfa'}
    \end{pmatrix}.
\end{equation}
With all structure constants and generators fixed, we can finally present the algebra $T_A\circ T_B = X_{AB}{}^C\,T_C$ of geometric gaugings in the IIB section. For $n\leq 6$ it reads
\begin{equation}
\begin{aligned}\tikzmarkin{geomalg-IIB}(9.4,-0.4)(-0.7,0.4)
    T_{\sfa}\circ T_{\sfb} &=f_{\sfa\sfb}{}^{\sfc}\,T_{\sfc} + f^\gamma_{\sfa\sfb\sfc}\,T_\gamma^{\sfc} + f_{\sfa\sfb\sfc_3}\,T^{\sfc_3} \,,\\
    T_{\sfa}\circ T^{\sfb}_\beta &= f_{\sfa}{}_{\beta}^{\sfb\sfc}\,T_{\sfc} - f_{\sfa\beta}{}^{\gamma}\,T_\gamma^{\sfb} - f_{\sfa\sfc}{}^{\sfb}\,T_\beta^{\sfc} + 2\,Z_{\sfa}\,T^{\sfb}_\beta - \epsilon_{\beta\gamma}\,f^{\gamma}_{\sfa\sfc_2}\, T^{\sfb\sfc_2} + f_{\sfa\sfc_4}\,T_\beta^{\sfb\sfc_4}\,,\\
    T_{\sfa}\circ T^{\sfb_3} &= f_{\sfa}{}^{\sfb_3\sfc}\, T_{\sfc} - \delta^{\sfb_3}_{\sfc_2\sfd}\,\epsilon^{\gamma\delta}\,f_{\sfa}{}_{\gamma}^{\sfc_2}\, T_{\delta}^{\sfd} - \delta^{\sfb_3}_{\sfd\sfe_2}\,f_{\sfa\sfc}{}^{\sfd}\, T^{\sfe_2 \sfc} + 4\,Z_{\sfa}\,T^{\sfb_3} - f^\gamma_{\sfa\sfc_2}\,T_\gamma^{\sfb_3\sfc_2} \,,\\
    T_{\sfa}\circ T_\beta^{\sfb_5} &= f_{\sfa}{}_{\beta}^{\sfb_5\sfc}\,T_{\sfc} - \delta^{\sfb_5}_{\sfc_4\sfd}\, f_{\sfa}{}^{\sfc_4}\,T_\beta^{\sfd} + \delta^{\sfb_5}_{\sfc_2\sfd_3}\, f_{\sfa}{}_{\beta}^{\sfc_2}\,T^{\sfd_3}\\
    &\quad  - f_{\sfa\beta}{}^{\gamma}\, T_\gamma^{\sfb_5} - \delta^{\sfb_5}_{\sfd\sfe_4}\,f_{\sfa\sfc}{}^{\sfd}\,T_\beta^{\sfe_4\sfc} + 6\,Z_{\sfa}\,T^{\sfb_5}_\beta \,,\\
    T^{\sfa}_\alpha\circ T_{\sfb} &= - f_{\sfb}{}_{\alpha}^{\sfa\sfc} \, T_{\sfc} - \delta^{\sfa\sfd}_{\sfb\sfc}\,f_{\sfd\alpha}{}^{\gamma}\, T_\gamma^{\sfc} + f_{\sfb\sfc}{}^{\sfa}\,T_\alpha^{\sfc} + 2\,\delta^{\sfa\sfc}_{\sfb\sfd}\,Z_{\sfc}\, T^{\sfd}_\alpha - \delta^{\sfa\sfd_3}_{\sfb\sfc_3}\,\epsilon_{\alpha\gamma}\,f^{\gamma}_{\sfd_3}\,T^{\sfc_3} \,,\\
    T^{\sfa}_\alpha\circ T^{\sfb}_\beta &= f_{\sfc}{}_{\alpha}^{\sfa\sfb}\,T_\beta^{\sfc} + f_{\sfc\alpha}{}^\gamma\,\epsilon_{\gamma\beta}\,T^{\sfc \sfa\sfb} + \epsilon_{\alpha\beta}\,f_{\sfc_2}{}^{\sfa}\,T^{\sfc_2\sfb} - 2\,\epsilon_{\alpha\beta}\,Z_{\sfc}\,T^{\sfa\sfb\sfc} + \epsilon_{\alpha\gamma}\,f^{\gamma}_{\sfc_3}\,T_\beta^{\sfa\sfb\sfc_3} \,,\\
    T^{\sfa}_\alpha\circ T^{\sfb_3} &= \delta^{\sfb_3}_{\sfd\sfe_2}\,f_{\sfc}{}_{\alpha}^{\sfa \sfd}\,T^{\sfe_2 \sfc} + f_{\sfc\alpha}{}^\gamma\,T_\gamma^{\sfa\sfc \sfb_3} - f_{\sfc_2}{}^{\sfa}\,T_\alpha^{\sfc_2 \sfb_3} + 2\,Z_{\sfc}\,T^{\sfa \sfb_3\sfc}_\alpha\,,\\
    T^{\sfa}_\alpha\circ T_\beta^{\sfb_5} &= \delta^{\sfb_5}_{\sfd\sfe_4}\, f_{\sfc}{}_{\alpha}^{\sfa \sfd}\,T_\beta^{\sfe_4\sfc} \,,\\
    T^{\sfa_3} \circ T_{\sfb} &= - f_{\sfb}{}^{\sfa_3\sfc}\,T_{\sfc} - \delta^{\sfa_3}_{\sfd_2\sfe}\,\delta_{\sfb\sfc}^{\sfe\sff}\,\epsilon^{\gamma\delta}\,f_{\sff}{}_{\gamma}^{\sfd_2}\,T_\delta^{\sfc} + \delta^{\sfa_3}_{\sfd\sfe_2}\,f_{\sfb\sfc}{}^{\sfd}\, T^{\sfe_2\sfc}\\
    &\quad - \delta^{\sfa_3}_{\sfb\sfd\sfe}\, f_{\sfc_2}{}^{\sfd}\, T^{\sfe\sfc_2} + 4\,\delta^{\sfa_3\sfc}_{\sfb\sfd_3}\,Z_{\sfc}\, T^{\sfd_3}\,,\\
    T^{\sfa_3} \circ T_\beta^{\sfb} &= f_{\sfc}{}^{\sfa_3 \sfb}\,T_\beta^{\sfc} - \delta^{\sfa_3}_{\sfd_2\sfe}\, f_{\sfc}{}_{\beta}^{\sfd_2}\,T^{\sfe \sfb\sfc} + \delta^{\sfa_3}_{\sfd\sfe_2}\, f_{\sfc_2}{}^{\sfd}\,T_\beta^{\sfe_2 \sfb \sfc_2} - 4\,Z_{\sfc}\,T_\beta^{\sfa_3\sfb\sfc}\,,\\
    T^{\sfa_3} \circ T^{\sfb_3} &= \delta^{\sfb_3}_{\sfd\sfe_2}\, f_{\sfc}{}^{\sfa_3 \sfd}\, T^{\sfe_2 \sfc} - \delta^{\sfa_3}_{\sfd_2\sfe}\, \epsilon^{\gamma\delta}\,f_{\sfc}{}_{\gamma}^{\sfd_2}\,T_\delta^{\sfe\sfb_3\sfc} \,,\\
    T^{\sfa_3} \circ T_\beta^{\sfb_5} &= \delta^{\sfb_5}_{\sfd\sfe_4}\,f_{\sfc}{}^{\sfa_3 \sfd}\,T_\beta^{\sfe_4 \sfc}\,,\\
    T_\alpha^{\sfa_5} \circ T_{\sfb} &= - f_{\sfb}{}_{\alpha}^{\sfa_5 \sfc}\,T_{\sfc} - \delta^{\sfa_5}_{\sfd_4\sfe}\,\delta^{\sfe\sff}_{\sfb\sfc}\,f_{\sff}{}^{\sfd_4}\, T_\alpha^{\sfc} + \delta^{\sfa_5}_{\sfb\sfd_2\sfe_2}\,f_{\sfc}{}_{\alpha}^{\sfd_2}\, T^{\sfe_2\sfc} - \delta^{\sfa_5}_{\sfc_2\sfd_3}\,f_{\sfb}{}_{\alpha}^{\sfc_2}\,T^{\sfd_3}\\
    &\quad - \delta^{\sfa_5}_{\sfb\sfd_4}\,f_{\sfc\alpha}{}^{\gamma}\, T_\gamma^{\sfd_4\sfc} + f_{\sfb\alpha}{}^{\gamma}\,T_\gamma^{\sfa_5} + \delta^{\sfa_5}_{\sfd\sfe_4}\,f_{\sfb\sfc}{}^{\sfd}\, T_\alpha^{\sfe_4\sfc} + \delta^{\sfa_5}_{\sfd\sfb\sfe_3}\,f_{\sfc_2}{}^{\sfd}\, T_\alpha^{\sfe_3 \sfc_2} \,,\\
    T_\alpha^{\sfa_5} \circ T_\beta^{\sfb} &= f_{\sfc}{}_{\alpha}^{\sfa_5\sfb}\,T_\beta^{\sfc} + \delta^{\sfa_5}_{\sfd_4\sfe}\,\epsilon_{\alpha\beta}\,f_{\sfc}{}^{\sfd_4}\,T^{\sfe \sfb\sfc} - \delta^{\sfa_5}_{\sfd_2\sfe_3}\,f_{\sfc}{}_{\alpha}^{\sfd_2}\,T_\beta^{\sfe_3 \sfb\sfc}\,,\\
    T_\alpha^{\sfa_5} \circ T^{\sfb_3} &= \delta^{\sfb_3}_{\sfd\sfe_2}\,f_{\sfc}{}_{\alpha}^{\sfa_5 \sfd}\, T^{\sfe_2\sfc} - \delta^{\sfa_5}_{\sfd_4\sfe}\,f_{\sfc}{}^{\sfd_4}\, T_\alpha^{\sfe \sfb_3\sfc} \,,\\
    T_\alpha^{\sfa_5} \circ T_\beta^{\sfb_5} &= \delta^{\sfb_5}_{\sfd\sfe_4}\,f_{\sfc}{}_{\alpha}^{\sfa_5 \sfd}\, T_\beta^{\sfe_4\sfc} \,. \tikzmarkend{geomalg-IIB}\label{eq:geomalg-IIB}
\end{aligned}
\end{equation}
Its extension to $n\leq 7$ is straightforward but rather bulky. Therefore, we moved it to appendix~\ref{app:algebra}. Furthermore, note that we use here the same shorthand notation as in \eqref{eq:geoalg-M}. To go to any $n\leq 6$, one just removes all structure constants and generators that vanish due to their antisymmetric indices\footnote{Clearly, the number $m$ of totally antisymmetric indices $\sfa_1 \dots \sfa_m$ has to be $m\le n-1$.}. Type IIB EDAs are recovered when the structure constants $f_{\sfa_3}^\alpha$ and $f_{\sfa_5}$ vanish. From the first line of the algebra \eqref{eq:geomalg-IIB}, we see that in this case the generators $T_a$ generate a Lie algebra.

Counting the resulting geometric gaugings, we complete the second part of table~\ref{tab:ratio-geo}. It is not hard to verify the results for $n=7$ given there: All admissible structure constants for geometric type IIB gaugings are given in \eqref{eq:IIB-structureC}. Their representations under GL(6)$\times$SL(2) are
\begin{align}
 (\overline{\bm{6}},\bm{1}) \oplus (\bm{20},\bm{2}) \oplus (\bm{84}+\bm{6},\bm{1}) \oplus (\bm{6},\bm{3}) \oplus (\overline{\bm{84}} \oplus \overline{\bm{6}},\bm{2}) \oplus (\bm{70} \oplus \bm{20},\bm{1}) \oplus (\bm{6},\bm{2}) \oplus (\bm{6},\bm{1}) \,.
\end{align}
In total there are $442$ components, while the other $526$ components correspond to (locally) non-geometric fluxes in type IIB.

\subsection{Generalised geometry revisited}
The notion of geometric gaugings that we have developed in this section is a natural extension of the second ingredient on page \pageref{item:geogauging} in the construction of generalised frames for generalised geometry. As a consistency check, we will quickly show that it gives indeed the expected result $R^{abc}=0$.

Our starting point for the derivation is the branching of the $R_1$ representation, the vector representation of O($D$,$D$), to the geometric subgroup GL($D$),
\begin{equation}\label{eq:ODD->GLD}
    \bm{2D} \rightarrow \bm{D} \oplus \overline{\bm{D}} \,.
\end{equation}
As expected, it gives rise to the generalised tangent bundle $T M \oplus T^* M$, whose sections are given in terms of a vector and a one-form. By following the same logic as for the M- and type IIB section at the beginning of this section, we split the partial derivative into $\partial_I = \begin{pmatrix} \partial_i & \tilde\partial^i \end{pmatrix}$ and find that the section condition can always be chosen as $\tilde\partial^i \cdot = 0$. Consequently, geometry gaugings in generalised geometry are defined by
\begin{equation}
    W_A{}^{\dot{\bm\alpha}} = \delta_A^b\,W_b{}^{\dot{\bm\alpha}} \qquad (a,b=1,\dots,D)\,.
\end{equation}
Remarkably, this is exactly the same expression as we obtained for M-theory in \eqref{eq:WA-M}. We then branch the adjoint representation as
\begin{equation}
    \bm{D(2D-1)} \rightarrow \overline{\bm{\frac{D(D-1)}2}} \oplus (\overline{\bm{D}}\otimes\bm{D}) \oplus \bm{\frac{D(D-1)}2}
\end{equation}
and split the generators $t_{\bm{\alpha}}$ accordingly,
\begin{equation}\label{eq:ODD-gen}
    t_{\bm\alpha} = \begin{pmatrix}
        R^{a_1 a_2} & K^a{}_b & R_{a_1 a_2}
    \end{pmatrix}\,.
\end{equation}
This is nothing else than the well-known decomposition of any O($D$,$D$) element into an invertible $D\times D$-matrix, complemented by a $B$- and a $\beta$-shift. Respectively, these elements are generated by $K^a{}_b$, $R^{a_1 a_2}$ and $R_{a_1 a_2}$. After taking into account the appropriate normalisation, they are represented by the $2D\times 2D$-matrices
\begin{equation}\label{eq:gen-ODD}
    \begin{aligned}
        (K^c{}_d)_A{}^B = \begin{pmatrix}
            \delta^c_a\,\delta_d^b & 0 \\
            0 & -\delta^a_d\,\delta_b^c
        \end{pmatrix}\,, \qquad &&
        (R^{c_1c_2})_A{}^B &= \begin{pmatrix}
            0 & -2\,\delta^{c_1c_2}_{ab} \\
            0 & 0 \end{pmatrix}\,,\\ &&
        (R_{c_1c_2})_A{}^B &= \begin{pmatrix}
            0 & 0 \\
            -2\,\delta_{c_1c_2}^{ab} & 0
        \end{pmatrix}
    \end{aligned}
\end{equation}
with the commutators
\begin{equation}
    \begin{aligned}
        [K^a{}_b\,,R^{c_1c_2}] &= 2\,\delta^{c_1c_2}_{bd}\,R^{ad}\,,\qquad &&
        [K^a{}_b\,,R_{c_1c_2}] &= -2\,\delta_{c_1c_2}^{ad}\,R_{bd}\,, \\
        [R_{a_1a_2}\,,R^{b_1b_2}] &= 4\,\delta_{[a_1}^{[b_1}\,K^{b_2]}{}_{a_2]}\,.
    \end{aligned}
\end{equation}
This is all we need to identify the contributions to the embedding tensor $W_a{}^{\dot{\bm\alpha}}$ as
\begin{equation}
    f_{a,b_1 b_2}\,, f_{a,b}{}^c\,, f_a{}^{b_1 b_2}\,, \quad \text{and} \quad Z_a\,.
\end{equation}
More precisely, we fix $W_A = W_A{}^{\bm\alpha}\,t_{\bm\alpha}$ to
\begin{equation}\label{eq:Wa-ODD}
    W_a = -\tfrac{1}{2!}\,f_{abc}\,R^{bc} + \tfrac{1}{2}\,f_{ab}{}^c \, K^b{}_c - \tfrac{1}{3!}\,f_a{}^{d_1d_2}\, R_{d_1d_2} \,,\qquad 
 W^a =0\,.
\end{equation}
Here we drop the trombone gauging $Z_a$ because it was not included in section~\ref{sec:ODDmotiv}. An extension with non-vanishing $Z_a$ has been studied in \cite{Fernandez-Melgarejo:2021zyj} where the Poisson-Lie T-duality is extended to the Jacobi-Lie T-duality. We then find that $X_{AB}{}^C$ is fixed completely in terms of the totally antisymmetric parts
\begin{equation}\label{eq:geogaugings-ODD}
    f_{a_1 \dots a_3}\,, f_{a_1 a_2}{}^b\,, \quad \text{and} \quad f_a{}^{b_1 b_2}
\end{equation}
as
\begin{equation}
    \begin{aligned}
        X_a &= -\tfrac{1}{2}\,f_{ab_1 b_2}\,R^{b_1 b_2} + f_{ab}{}^c \, K^b{}_c - \tfrac{1}{2}\,f_a{}^{b_1 b_2}\, R_{b_1 b_2} \,,\\
        X^a &= - f_b{}^{ac}\, K^b{}_c - \tfrac{1}{2}\,f_{b_1 b_2}{}^a\, R^{b_1 b_2}\,,
    \end{aligned}
\end{equation}
with $(X_A)_B{}^C = X_{AB}{}^C$. Finally, we use the matrices in \eqref{eq:gen-ODD} to compute $X_{AB}{}^C$ explicitly. After lowering its last index with $\eta_{AB}$ given in \eqref{eq:defeta_AB}, we obtain the expected three independent contributions
\begin{equation}
    X_{ABC} = \begin{cases}
        f_{abc} = H_{abc} \\
        f_{ab}{}^c \\
        f_a{}^{bc} = Q_a{}^{bc}\,.
    \end{cases}
\end{equation}
But this is nothing else than imposing $R^{abc} = 0$, as we did in section~\ref{sec:ODDmotiv}.

We should also not miss to get the ratio of geometric gaugings and compare it to the ones listed in table~\ref{tab:ratio-geo}. Counting all components of the admissible gaugings in \eqref{eq:geogaugings-ODD}, we find
\begin{equation}
    N_{\mathrm{geo}} = \frac{D}{6}\,(D - 1) (7 D - 2)\,.
\end{equation}
On the other hand, the total number of $X_{AB}{}^C$'s independent components is (remember that $X_{ABC}$ is totally antisymmetric)
\begin{equation}
   N = \frac{2 D}{3}\, (D - 1 ) (2 D - 1)  \,,
\end{equation}
and their ratio is
\begin{equation}
    \frac{N_{\mathrm{geo}}}N = \frac78 + \frac{3}{16 D - 8}\,. 
\end{equation}
For $D\ge2$, it is monotonically decreasing from 100\% to the lower bound of 87.5\%. 

\section{Lie subalgebras of the Leibniz algebra}\label{sec:Lie}
As mentioned in point one at the beginning of section~\ref{sec:geogaugings}, a major novelty in the construction of generalised frames in exceptional generalised geometry compared to generalised geometry is that the structure constants $X_{AB}{}^C$ are not antisymmetric with respect to their first two indices. Hence, their Leibniz identity 
\begin{equation}\label{eq:LeibnizID}
    X_{AC}{}^E\, X_{BE}{}^D - X_{BC}{}^E\, X_{AE}{}^D = - X_{AB}{}^E\, X_{EC}{}^D
\end{equation}
(or shorter $(T_A \circ T_B) \circ T_C + T_B \circ (T_A \circ T_C) = T_A \circ ( T_B \circ T_C )$) does not automatically turn in a Jacobi identity for a Lie algebra as it was the case in generalised geometry.

\subsection{Lie(\texorpdfstring{$G$}{G}) and Lie(\texorpdfstring{$H$}{H})}
Still, we can show that the Leibniz algebra defined by the structure constants $X_{AB}{}^C$ contains at least one Lie subalgebra. A naive proposal would be to use the antisymmetric part $X_{[AB]}{}^C$ as the latter's structure constants. However, they do not satisfy the Jacobi identity
\begin{equation}\label{eq:JacX[AB]C}
    X_{[AB]}{}^E\, X_{[EC]}{}^D + X_{[BC]}{}^E\, X_{[EA]}{}^D + X_{[CA]}{}^E\, X_{[EB]}{}^D = X_{[AB}{}^E\, Z_{C]E}{}^D
\end{equation}
due to the term on the right-hand side. For convenience, we have defined here the symmetric part of the structure constants
\begin{equation}
    Z_{AB}{}^C := X_{(AB)}{}^C \,,
\end{equation}
that captures the failure of the Jacobi identity. We still find that the Lie bracket
\begin{equation}
    [ T_A,\, T_B ]_- := \frac12 \left( T_A \circ T_B - T_B \circ T_A \right) \qquad \text{with} \qquad
    [T_A,\, T_B]_- = X_{[AB]}{}^C\, T_C
\end{equation}
is well-defined. At the first glance, this seems like a contradiction, because its structure constants do not satisfy the Jacobi identity \eqref{eq:JacX[AB]C}. The solution to this puzzle \cite{Samtleben:2008pe} is that the contraction
\begin{equation}\label{eq:ZX=0}
    Z_{AB}{}^E\, X_{EC}{}^D = 0
\end{equation}
vanishes\footnote{This relation follows immediately from the Leibniz identity \eqref{eq:LeibnizID} by symmetrising in the indices $A$ and $B$ on the left- and right-hand side.}. Thus, to obtain a Lie subalgebra of the Leibniz algebra generated by $T_A$, we have to remove all generators $T_{\acute\alpha}$ for which $Z_{BC}{}^{\acute\alpha} \ne 0$. Clearly, these generators span a vector space which we denote by $\cI$. The Lie algebra Lie($G$), we are looking for, is spanned by the generators in the complement, denoted by $T_{\grave a}$. It always includes the generators $T_a$ or $T_{\sfa}$ because, as one checks with \eqref{eq:geoalg-M} or \eqref{eq:geomalg-IIB}, the symmetrised product
\begin{equation}
	[ T_A,\, T_B ]_+ = \frac12 \left( T_A \circ T_B + T_B \circ T_A \right) = Z_{AB}{}^C\, T_C
\end{equation}
never contains $T_a$ or $T_{\sfa}$ for any geometric gaugings. Hence, $T_A$ can be decomposed into
\begin{equation}\label{eq:decomp-T}
    T_A = \begin{pmatrix} T_{\grave a} & T_{\acute\alpha} \end{pmatrix}
        = \begin{pmatrix} T_a & T_{\grave\alpha} & T_{\acute\alpha} \end{pmatrix}
        = \begin{pmatrix} T_a & T_{\check\alpha} \end{pmatrix}.
\end{equation}
Note that we here give the decomposition for M-theory sections. For type IIB sections the index $a$ has to be replaced by $\sfa$. In total, three fundamental indices appear in \eqref{eq:decomp-T}, namely
\begin{equation}
    \begin{aligned}
        a&=1,\dots,n & (\sfa&=1,\dots,n-1), \\
        \grave\alpha&=1,\dots,\dim R_1 - \dim \cI - n \quad & (\grave\alpha&=1,\dots,\dim R_1 - \dim \cI - n + 1) \qquad \text{and}\\
        \acute\alpha&=1,\dots,\dim\cI
    \end{aligned}
\end{equation}
for M-theory (type IIB) sections\footnote{We will use this notation repeatedly in the following; first the expression for M-theory sections and after it, in brackets, the corresponding one for the type IIB sections.}. Furthermore, we define the two composite indices
\begin{equation}
    T_{\grave a} = \begin{pmatrix}
        T_a & T_{\grave\alpha}
    \end{pmatrix} \qquad \text{and} \qquad
    T_{\check\alpha} = \begin{pmatrix}
        T_{\grave\alpha} & T_{\acute\alpha}
    \end{pmatrix} .
\end{equation}
With them, we extract
\begin{equation}\label{eq:restrictionsZandX}
    Z_{AB}{}^{\grave c} = 0 \qquad \text{and} \qquad
    X_{\acute\alpha B}{}^C = 0
\end{equation}
from \eqref{eq:ZX=0}. Therefore, if one restricts the Leibniz algebra (spanned by the generators $T_A$) to the generators $T_{\grave a}$, the structure constants $X_{\grave a\grave b}{}^{\grave c}$ are antisymmetric in their first two indices and solve the Jacobi identity
\begin{equation}
    3\, X_{[\grave a\grave b}{}^{\grave e}\, X_{\grave c]\grave e}{}^{\grave d} = 0
\end{equation}
which arises from restricting \eqref{eq:LeibnizID} to $A=\grave a$, $B=\grave b$, $C=\grave c$ and $D=\grave d$. Hence, as anticipated, the generators $T_{\grave a}$ span a Lie algebra Lie($G$).
 
This argument agrees with recent work on exceptional algebroids \cite{Bugden:2021wxg,Bugden:2021nwl}, also called elgebroids. They use a more formal, index-free approach compared to our setting. Both are related in appendix~\ref{app:elgebroids}. Elgebroids denote the Leibniz algebra by $E$ and have thus $T_A \in E$. They define an ideal $\cI$ as a linear subspace $\cI \subset E$ that is spanned by all linear independent generators $Z_{AB}{}^C\, T_C$. But these are exactly the ones we call $T_{\acute\alpha} \in \cI$. From \eqref{eq:restrictionsZandX}, one sees that the right- and left-action of any element of $\cI$ on an arbitrary element of $E$ still is in $\cI$, namely
\begin{equation}
    T_{\acute\alpha} \circ T_B = X_{\acute\alpha B}{}^C\, T_C = 0 \subset \cI
        \qquad \text{and} \qquad
    T_B \circ T_{\acute\alpha} = 2\, Z_{\acute\alpha B}{}^C\, T_C = 2\, Z_{\acute\alpha B}{}^{\acute\gamma}\, T_{\acute\gamma} \subset \cI\,.
\end{equation}
Therefore, the quotient (ring) of $E$ by $\cI$, $E/\cI$, is well-defined and gives rise to the Lie algebra Lie($G$). In our convention, it is spanned by $T_{\grave a}\in$\,Lie($G$).
 
However, to proceed in the same way as in section~\ref{sec:ODDmotiv} for generalised geometry, we also need a subgroup $H\subset G$ to describe the space $M$, on which the generalised frame field is defined, as the coset $M=G/H$. To identify $H$, we first note that for any geometric gauging
\begin{equation}\label{eq:subalg-geom}
    X_{\check\alpha\check\beta}{}^c = 0 \qquad (X_{\check\alpha\check\beta}{}^{\sfc} = 0)
\end{equation}
holds. To prove these relations, first note that $T_{\check\alpha}$ is formed from all components in \eqref{eq:gen-M} (\eqref{eq:gen-IIB}) after excluding $T_a$ ($T_{\sfa}$),
\begin{equation}
    \begin{aligned}
        T_{\check\alpha} &= \begin{pmatrix}
            T^{a_1 a_2} & T^{a_1 \dots a_5} & T^{a_1\dots a_7,b}
        \end{pmatrix}\\
        \bigl(T_{\check\alpha} &= \begin{pmatrix}
            T^{\sfa}_\alpha & T^{\sfa_1\dots\sfa_3} & T^{\sfa_1\dots\sfa_5}_\alpha & T^{\sfa_1\dots\sfa_6,\sfb}
        \end{pmatrix}\bigr)\,.
    \end{aligned}
\end{equation}
Therefore, one has to check that all products $T^{a_n} \circ T^{b_m}$ ($T^{\sfa_n}_\bullet \circ T^{\sfb_m}_\bullet$\footnote{$\bullet$ is a place holder for $\alpha$/$\beta$ that is required depending on the value of $n$/$m$.}) do not have any $T_c$ contribution. Indeed this is the case and we conclude that for any geometric gauging the generators $T_{\check\alpha}$ form a Leibniz subalgebra. In the context of elgebroids, this subalgebra is denoted by $V$ and called a co-Lagrangian subalgebra \cite{Bugden:2021wxg,Bugden:2021nwl}.\footnote{There is a condition $\text{Tr}_E\, \text{ad}_v = \tfrac{\lambda}{\lambda-1}\, \text{Tr}_V\, \text{ad}_v$ ($\lambda := -\beta\,\dim R_1$) on $V$ for the consistency of elgebroids \cite{Bugden:2021nwl}. In our notation, this reads $X_{\check{\alpha}B}{}^B = \tfrac{\lambda}{\lambda-1}\, X_{\check{\alpha}\check{\beta}}{}^{\check{\beta}}$ and is indeed satisfied for an arbitrary $n\leq 7$ both in M-theory and type IIB.} Because $\cI$ does not contain the generators $T_a$ ($T_{\sfa}$), it is not just a two-sided ideal of $E$ but also that of $V$. Thus, the quotient $V/\cI$ is identified with the second Lie algebra, Lie($H$). With the explicit decomposition of $T_A$ introduced in \eqref{eq:decomp-T}, we see that $H$ is generated by $T_{\grave\alpha}$, a subset of $G$'s generators $T_{\grave a}$. This renders $H$ a subgroup of $G$ and the coset $M=G/H$ is well-defined.

\subsection{Leibniz representation}
To make further progress, we need to find a representation for these two Lie algebras and their corresponding Lie groups. In this article, we are only interested in local results. Thus, we use the exponential map
\begin{equation}
    G \ni g = \exp\left( x^{\grave a}\,T_{\grave a} \right) = \sum_{n=0}^\infty \frac{(x^{\grave a}\,T_{\grave a})^n}{n!}
\end{equation}
to relate Lie algebras, like Lie($G$) here, with the corresponding Lie group. To evaluate the matrix exponential explicitly, we have to express the generators $T_{\grave a}$ in terms of matrices that represent the Lie algebra under consideration. A natural representation is given in terms of the structure constants of the Leibniz algebra as
\begin{equation}
    (T_{\grave a})_A{}^B = - X_{\grave a B}{}^C = \begin{pmatrix}
        - X_{\grave a\grave b}{}^{\grave c} & - X_{\grave a\grave b}{}^{\acute\gamma} \\
        0 & - X_{\grave a\acute\beta}{}^{\acute\gamma}
    \end{pmatrix}.
\end{equation}
Note the 0 in the lower-left corner. It is due to the property $X_{\grave a\acute\beta}{}^{\grave c} = 2\, Z_{\grave a \acute\beta}{}^{\grave c} = 0$. We will call this representation the \underline{Leibniz representation} $R$. For the required
\begin{equation}
    [ T_{\grave a}, T_{\grave b} ] = X_{\grave a\grave b}{}^{\grave c}\,T_{\grave c} 
\end{equation}
to hold, the three conditions
\begin{align}
    2 X_{[\grave a|\grave c}{}^{\grave e} X_{|\grave b]\grave e}{}^{\grave d} &= - X_{\grave a\grave b}{}^{\grave e} X_{\grave e\grave c}{}^{\grave d} \label{eq:Radj}\,, \\
    2 X_{[\grave a|\acute\gamma}{}^{\acute\epsilon} X_{|\grave b]\acute\epsilon}{}^{\acute\delta} &= - X_{\grave a\grave b}{}^{\grave e} X_{\grave e\acute\gamma}{}^{\acute\delta} \label{eq:R2}\,,\quad \text{and}\\
    2 X_{[\grave a|\grave c}{}^{\grave e} X_{|\grave b]\grave e}{}^{\acute\delta} + 2 X_{[\grave a|\grave c}{}^{\acute\epsilon} X_{|\grave b]\acute\epsilon}{}^{\acute\delta} &= - X_{\grave a\grave b}{}^{\grave e} X_{\grave e\grave c}{}^{\acute\delta}\label{eq:intertwine}
\end{align}
are needed. The first of them is just \eqref{eq:subalg-geom}. We already explained that it originates from the Leibniz identity after properly restricting the indices. $X_{\grave a\grave b}{}^{\grave c}$, as the structure constants of the Lie algebra Lie($G$), form a representation on their own, the adjoint representation. In the same way, \eqref{eq:R2} tells us that $X_{\grave a\acute\beta}{}^{\acute\gamma}$ has to be second representation of Lie($G$), $R_{\cI}$, with $\dim R_{\cI} = \dim R_1 - \dim G$. Again, we find that this constraint holds due to the Leibniz identity \eqref{eq:LeibnizID} after restricting it to $A=\grave a$, $B=\grave b$, $C=\acute\gamma$, $D=\acute\delta$. The Leibniz representation $R$ is however not just a direct sum of the adjoint representation and $R_{\cI}$. Both are intertwined using the third tenor $X_{\grave a\grave b}{}^{\acute\gamma}$. For $R$ to be a representation of Lie($G$), it has to satisfy \eqref{eq:intertwine} which it indeed does due to the Leibniz identity \eqref{eq:LeibnizID}. In Loday algebra cohomology, it gives rise to a cocycle as is explained in appendix~\ref{app:ELA}.

Remarkably, $(T_{\grave a})_A{}^B$ already contains all non-vanishing components of the Leibniz algebra's structure constants $X_{AB}{}^C$ because $X_{\acute\alpha B}{}^C = 0$ due to \eqref{eq:restrictionsZandX}. Therefore, we have a one-to-one correspondence
\begin{equation}
    \begin{tikzpicture}
        \node[draw,rectangle] (Leibniz) {\parbox{10em}{%
            \centering Leibniz algebra\\
            $T_A \circ T_B = X_{AB}{}^C T_C$
        }};
        \node[at=(Leibniz.east),anchor=west,xshift=7em,draw,rectangle] (Lie) {\parbox{12em}{%
            \centering Lie algebra Lie($G$) with\\Leibniz representation $R$\,.
        }};
        \draw[<->,thick] ($(Leibniz.east)+(0.25,0)$) -- ($(Lie.west)-(0.25,0)$);
    \end{tikzpicture}
\label{eq:one-to-one}
\end{equation}
For the construction of generalised frame fields, the right-hand side is what we need. In geometric gaugings, we identified the second Lie algebra Lie($H$) as a subalgebra of Lie($G$). As such, the representation $R$ carries directly over to it. Another property of $G$ is that it is a subgroup of E$_{n(n)}\times\mathbb{R}^+$ and $X_{\grave{a}B}{}^C= X_{\grave{a}}{}^{\dot{\bm\alpha}}\, (t_{\dot{\bm\alpha}})_B{}^C$ describes its embedding. Therefore what we call structure constants is also known as the embedding tensor. 

\subsection{Group element and differential identities}\label{sec:group+diff}
In the construction of generalised frame fields that satisfy \eqref{eq:framealgebra}, we rely on differential and algebraic identities. We discuss the former here and the latter in the next section. For frames in generalised geometry, we know from section~\ref{sec:ODDmotiv} that we need the two identities
\begin{equation}\label{eq:adj-MC}
    (M^{-1})_A{}^C\,\dd M_C{}^B = - v^C\, X_{CA}{}^B = v^C\, (T_C)_A{}^B
\end{equation}
and
\begin{equation}\label{eq:vATA}
    \dd v^A\, T_A = -\frac12\, X_{BC}{}^A\, v^B \wedge v^C\, T_A\,.
\end{equation}
The first of them defines the right-invariant Maurer-Cartan form $v^A\, T_A$, while the second is just the consequence of $\dd^2=0$ and $(M^{-1})_A{}^C\, M_C{}^B = \delta_A^B$\,. In generalised geometry, the matrices $T_A$ are in the adjoint representation of Lie($G$). If it is faithful, all $A=1,\dots,2 D$ one-forms $v^A$ are fixed. Otherwise, one has to use a faithful representation, which according to Ado's theorem is finite-dimensional. In the end, $v^A$ satisfies the Maurer-Cartan equation
\begin{equation}
    \dd v^A = -\frac12\, X_{BC}{}^A\, v^B\wedge v^C \,. 
\end{equation}

To write down analogue identities for the frame $E_A$ in exceptional generalised geometry, we keep in mind that $M_A{}^B$ is nothing else as a coset element in the adjoint representation. Thus, we can define\footnote{Here $T_i = \delta_i^a T_a$ is the same as $T_a$. We just change the index to relate it to the coordinates which come with curved indices.}
\begin{equation}\label{eqn:cosetelement}
    M_A{}^B := \exp\left( x^i\,T_i \right)_A{}^B
\end{equation}
where $G$'s generators $(T_c)_A{}^B$ are taken in the Leibniz representation $R$. Of course, any other choice of the coset representative is also fine because it can be related by a diffeomorphism to \eqref{eqn:cosetelement}. Without loss of generality, we can parameterise the matrix $M_A{}^B$ and its inverse $(M_A{}^B)^{-1}$ as
\begin{equation}\tikzmarkin{MaB}(0.1,-0.7)(-0.1,0.9)
    M_A{}^B = \begin{pmatrix}
        M_{\grave a}{}^{\grave b} & I_{\grave a}{}^{\acute\gamma}\, M_{\acute\gamma}{}^{\acute\beta} \\
        0 & M_{\acute\alpha}{}^{\acute\beta}
    \end{pmatrix}\,,\qquad
    (M^{-1})_A{}^B = \begin{pmatrix}
        (M^{-1})_{\grave a}{}^{\grave b} & - (M^{-1})_{\grave a}{}^{\grave c}\, I_{\grave c}{}^{\acute\beta} \\
        0 & (M^{-1})_{\acute\alpha}{}^{\acute\beta}
    \end{pmatrix}\,.\tikzmarkend{MaB}
\end{equation}
With $M_A{}^B$ fixed, we proceed by computing $v^A\, T_A$. In perfect agreement with \eqref{eq:adj-MC}, we find
\begin{equation}\label{eq:dMM-1}
    (M^{-1})_A{}^C\,\dd M_C{}^B = \begin{pmatrix} 
        \dd M_{\grave c}{}^{\grave b} (M^{-1})_{\grave a}{}^{\grave c} & (M^{-1})_{\grave a}{}^{\grave c} \dd I_{\grave c}{}^{\acute\delta} M_{\acute\delta}{}^{\acute\beta} \\
        0 & \dd M_{\acute\gamma}{}^{\acute\beta} (M^{-1})_{\acute\alpha}{}^{\acute\gamma}
    \end{pmatrix} 
    = - v^C\,X_{CA}{}^B = - v^{\grave{c}}\,X_{\grave{c}A}{}^B\,.
\end{equation}
Because, $X_{\acute\alpha B}{}^C = 0$ this relation only fixes the part $v^{\grave a}$ of $v^A$. For definiteness, we set $v^{\acute\alpha} = 0$ but any other choice\footnote{A choice which is useful for actual application is $v_i^A = \delta_i^A +\sum_{n=1}^\infty \frac{1}{(n+1)!}\,\bigl[(x^k\, T_{k})^n\bigr]_i{}^A$, which generally gives $v^{\acute\alpha}\neq 0$. The benefit of this is that we can automatically compute $v_i^A$ without explicitly identifying the Lie group $G$.} would lead eventually to the same results ($w^{\grave a} = 0$ and \eqref{eq:dw} but an expression for $w^{\acute\alpha}$ that differs from \eqref{eq:wA}). $v^{\grave a}$ is already fixed by 
\begin{align}
    (M^{-1})_{\grave a}{}^{\grave c}\,\dd M_{\grave c}{}^{\grave b} = - v^{\grave c}\,X_{\grave c\grave a}{}^{\grave b}\,,
\end{align}
arising from $A=\grave{a}$ and $B=\grave{b}$ in \eqref{eq:dMM-1}. Hence, it is nothing else that the left-invariant Maurer-Cartan form on the coset $M=G/H$.

Instead of the Maurer-Cartan equation
\begin{equation}\label{eq:MC-vgravea}
    \dd v^{\grave a} = -\frac12\, X_{\grave b\grave c}{}^{\grave a}\, v^{\grave b} \wedge v^{\grave c}
\end{equation}
the identity 
\begin{equation}\label{eq:modified-MC1}
    \dd v^A + \frac12\, X_{BC}{}^A \, v^B\wedge v^C + w^A = 0\,,
\end{equation}
is better suited to construct the generalised frame in the next section. Comparing these two equations, we find\footnote{This shows that $w^A$ can be spanned by the basis of $\cI$ as $w^A=w^{BC}\,Z_{BC}{}^A$. Then \eqref{eq:modified-MC1} takes the same form as the flatness condition for the modified field strength in gauged supergravity \cite{deWit:2004nw,deWit:2005hv}.}
\begin{equation}\label{eq:wA}
    w^{\grave a} = 0 \qquad \text{and} \qquad w^{\acute\alpha} = - \frac12\, X_{\grave b\grave c}{}^{\acute\alpha}\, v^{\grave b} \wedge v^{\grave c}\,.
\end{equation}
The reason why \eqref{eq:modified-MC1} is preferable over \eqref{eq:MC-vgravea} is that it contains the Leibniz algebra's full structure constants and thus makes the action of the duality group E$_{n(n)}$ manifest. On the other hand, there is a price to pay, a new two-form $w^A$. Its exterior derivative is
\begin{equation}\label{eq:dw}
    \begin{aligned}
	    \dd w^A &= X_{[BC]}{}^A\, w^B \wedge v^C + \frac12\, X_{[BC]}{}^E\, X_{[ED]}{}^A\, v^B\wedge v^C\wedge v^D\\
        &= -Z_{BC}{}^A\, w^B \wedge v^C + \frac{1}{3!}\, X_{[BC}{}^E\, Z_{D]E}{}^A\, v^B\wedge v^C\wedge v^D\,.
    \end{aligned}
\end{equation}
Recalling \eqref{eq:JacX[AB]C}, we can see that the second term measures the violation of the Jacobi identity of $X_{[AB]}{}^C$. 
As we note in Appendix \ref{app:ELA}, the pair of one-form $v^A$ and two-form $w^A$ plays the role of a generalised Maurer-Cartan form associated with the Leibniz algebra. The appearance of an additional two-form is mathematically interesting, but due to the constraint
\begin{equation}
	w^A\, X_{AB}{}^C = 0\,,
\end{equation}
we will see that it is not relevant for the construction of the generalised frame fields discussed in the next section. Thus, the two-form part does not encode new physics but is rather a necessary burden to have a fully manifest, linearly acting duality group.

\section{Construction of the generalised frame fields}\label{sec:genframes}
We are finally ready to construct the generalised frame field $E_A$ on the coset $M=G/H$ for all geometric gaugings. Our starting point is a coset representative $M_A{}^B \in M$ in the Leibniz representation. Moreover, we will extensively use the differential constraints \eqref{eq:adj-MC} and \eqref{eq:modified-MC1}. For the reader's convenience we repeat them here,
\begin{align}\tikzmarkin{modified-MC}(0.2,-0.4)(-0.2,0.5)
	(M^{-1})_A{}^C\,\dd M_C{}^B &= -v^C\,X_{CA}{}^B\,, &&\text{and} \label{eq:Mi-dM}\\
	\dd v^A &= - \frac{1}{2}\,X_{[BC]}{}^A\,v^B\wedge v^C - w^A 
	\quad &&\text{with} \qquad w^A\, X_{AB}{}^C = 0\,.\tikzmarkend{modified-MC}
\label{eq:modified-MC}
\end{align}
Because $M_A{}^B$ is governed by the same identities as the one introduced in section~\ref{sec:ODDmotiv}, it is suggestive to use the same ansatz for the generalised frame,
\begin{equation}\label{eq:ansatzframe}
	E_A{}^I=M_A{}^B\,V_B{}^I\,,
\end{equation}
as in \eqref{eq:ansatzoddframe}. $V_A{}^I$ contains the components $v^a_i$ of $v^A = v^A_i\,\dd x^i$, its dual vector fields $v_a^i$ ($v_a^i\,v_i^b = \delta_a^b$)\footnote{On the coset $M=G/H$, we are guaranteed that $v^a_i$ is invertable.} and various form fields. The latter crucially depends on the section. Hence, we first treat it here as a completely general element of E$_{n(n)}\times \mathbb{R}^+$ and later specify it further for M-theory sections in section~\ref{sec:frame-M-th} and for type IIB in section~\ref{sec:frame-IIB}. For this generalised frame, we compute the associated generalised flux
\begin{align}\label{eq:bmXabC}
 \bm{X}_{AB}{}^C = \Omega_{AB}{}^C + (t^{\bm{\alpha}})_D{}^E\, (t_{\bm{\alpha}})_B{}^C\,\Omega_{EA}{}^D - \beta\, \Omega_{DA}{}^D\,(t_0)_B{}^C \,,
\end{align}
where
\begin{equation}
	\Omega_{AB}{}^C := E_A{}^I\,E_B{}^J\,\partial_I E_J{}^C\,.
\end{equation}
In the end, our goal is to fix $V_A{}^I$ such that $\bm{X}_{AB}{}^C$ matches $X_{AB}{}^C$ for any geometric gauging. But for now, it is good to still distinguish between the two. Taking into account the ansatz \eqref{eq:ansatzframe}, $\Omega_{AB}{}^C$ can be further evaluated to
\begin{equation}\label{eq:Omega-general}
    \begin{aligned}
        \Omega_{AB}{}^C &= M_A{}^D\,M_B{}^E\,(M^{-1})_F{}^C\,\bigl[\hat{\Omega}_{DE}{}^C - V_D^I\, (M^{-1}\,\partial_I M)_E{}^F \bigr]\\
        &= M_A{}^D\,M_B{}^E\,(M^{-1})_F{}^C\,\delta_D^d\,\bigl(\hat{\Omega}_{dE}{}^C + A_d{}^G\,X_{GE}{}^C \bigr)\,.
    \end{aligned}
\end{equation}
At this point, we use the first differential identity \eqref{eq:Mi-dM} to go from the first line to the second line. Moreover, it is convenient to introduce the two new quantities
\begin{equation}\label{eq:Omega-hat}
	\hat{\Omega}_{aB}{}^C := v_a^i\,V_B{}^J\,\partial_i V_J{}^C \qquad \text{and} \qquad
 A_a{}^B := v_a^i\,v_i^B\,.
\end{equation}
To evaluate them further, we shall choose an M-theory section or a type IIB section. 

\subsection{M-theory section}\label{sec:frame-M-th}
We now take into account the branching from E$_{n(n)}${}$\rightarrow\,$GL($n$) that governs M-theory sections. In section~\ref{sec:geoMth}, we explained how it affects the generators $T_A$ in \eqref{eq:gen-M}. Because, the Maurer-Cartan one-form $v = T_A\, v^A$ takes values in them, we split\footnote{We use the convention $v = T_A\,v^A= T_a\,v^a + \frac{1}{2!}\,T^{a_1 a_2}\,A_{a_1a_2}+\frac{1}{5!}\,T^{a_1\dots a_5} A_{a_1\dots a_5}+\frac{1}{7!}\,T^{a_1\dots a_7,a'}\, A_{a_1\dots a_7,a'}$ that includes the factorial for each antisymmetric indices.}
\begin{equation}
	v_i^A = \begin{pmatrix}
	   v_i^a & A_{i a_1 a_2} & A_{i a_1 \dots a_5} & A_{i a_1 \dots a_7,a'}
	\end{pmatrix},
\end{equation}
accordingly. We use $v^a_i$ and $v_a^i$ to relate flat and curved GL($n$)-indices. In particular, this prescription results in
\begin{equation}
	A_a{}^B = v_a^i\,v_i^B = \begin{pmatrix}
	    \delta_a^b & A_{ab_1b_2} & A_{ab_1\dots b_5} & A_{ab_1\dots b_7,b'}
	\end{pmatrix}.
\end{equation}
We then introduce a matrix
\begin{equation}\label{eq:VaJ-M}
    \hat{V}_A{}^J := \begin{pmatrix}
        v_a^j & 0 & 0 & 0 \\ 
        0 & 2!\,v_{[j_1}^{[a_1}\,v_{j_2]}^{a_2]} & 0 & 0 \\ 
        0 & 0 & 5!\,v_{[j_1}^{[a_1}\dots v_{j_5]}^{a_5]} & 0 \\
        0 & 0 & 0 & 7!\,v_{[j_1}^{[a_1}\dots v_{j_7]}^{a_7]}\,v_{j'}^{a'}
    \end{pmatrix}\,.
\end{equation}
By comparing $\hat V_A{}^I$ with the ansatz \eqref{eq:ansatzoddframe}, we see that it is a natural generalisation of the matrix in the middle. To complete the analogous ansatz in exceptional generalised geometry, we still need an upper-triangular matrix that corresponds to the $B$-field shift on the right of \eqref{eq:ansatzoddframe}. One can easily check that it is generated by the matrix exponent $\exp\left( \frac1{2!} B_{i_1 i_2} R^{i_1 i_2} \right)$ where $R^{i_1 i_2}$ is given in \eqref{eq:gen-ODD}. Thus, we introduce its obvious M-theory counterpart 
\begin{equation}
    N_I{}^J = \left[ \exp \left(\tfrac{1}{3!}\,C_{i_1\dots i_3}\,R^{i_1\dots i_3}\right) \exp \left(\tfrac{1}{6!}\,C_{i_1\dots i_6}\,R^{i_1\dots i_6}\right) \right]{}_I{}^J\,,
\end{equation}
It employs all generators with exclusively raised indices in \eqref{eq:Enn-gen-M}, in perfect analogy with \eqref{eq:ODD-gen}. For example, when $n\leq 6$\,, the explicit matrix form of $N_I{}^J$ is
\begin{equation}\label{eq:NiJ-M}
    N_I{}^J = \begin{pmatrix}
        \delta_i^j & - C_{ij_1j_2} & - (C_{ij_1\dots j_5}+5\,C_{i[j_1j_2}\,C_{j_3j_4j_5]}) \\
        0 & 2!\,\delta^{i_1i_2}_{j_1j_2} & 20\,\delta^{i_1i_2}_{[j_1j_2}\,C_{j_3j_4j_5]} \\ 
        0 & 0 & 5!\,\delta^{i_1\dots i_5}_{j_1\dots j_5}
    \end{pmatrix}\,.
\end{equation}
The three-form $C_{i_1\dots i_3}$ and six-form $C_{i_1\dots i_6}$ form the gauge sector of M-theory and are sourced by M2- and M5-branes respectively. $N_I{}^J$ is chosen as the name for this matrix because its generated by the upper-triangular nilpotent elements of E$_{n(n)}$. After these considerations, we finally complete the ansatz for the generalised frame \eqref{eq:ansatzframe} with
\begin{equation}\label{eq:VaI}
    V_A{}^I = \hat V_A{}^J N_J{}^I\,. 
\end{equation}
This is all one needs to compute
\begin{equation}
    \begin{aligned}
        \hat{\Omega}_{aB}{}^C &= -D_a \hat{V}_B{}^J\,\hat{V}_J{}^C
            - \hat{V}_B{}^J\, \hat{V}_K{}^C\, D_a N_J{}^L \, (N^{-1})_L{}^K\\
        &= - \bigl(D_{(a} v_{b)}^j\,v_j^c + \tfrac{1}{2}\,[v_{a},\, v_{b}]^j\,v_j^c\bigr)\,(\widetilde{K}^b{}_c)_B{}^C
        - \hat{V}_B{}^J\, \hat{V}_K{}^C\, D_a N_J{}^L \, (N^{-1})_L{}^K
    \end{aligned}
 \end{equation}
 from the definition in \eqref{eq:Omega-hat}, where $\widetilde{K}^a{}_b:= K^a{}_b + \beta\,\delta^a_b\,t_0$\,. We also defined $D_a:= v_a^i\, \partial_i$\,, and $[ v_a, v_b ]$ denotes the Lie derivative of the vector fields $v_a$ and $v_b$\,. We then get rid of all the derivatives by taking into account the two identities
 \begin{align}\label{eq:geomflux-M}
    v^c\bigl([v_a,\,v_b]\bigr) &= f_{ab}{}^c - \tfrac{2}{2!}\, f_{[a}{}^{cd_1d_2}\, A_{b]d_1d_2} 
 - \tfrac{2}{5!}\, f_{[a}{}^{cd_1\dots d_5}\, A_{b]d_1\dots d_5} \,, \quad \text{and}\\
    \dd N \, N^{-1} &= \tfrac{1}{3!}\,\dd C_{i_1\dots i_3}\, R^{i_1 \dots i_3}
    + \tfrac{1}{6!}\,\bigl(\dd C_{i_1\dots i_6}-10\,\dd C_{i_1i_2i_3}\,C_{i_4i_5i_6}\bigr)\, R^{i_1\dots i_6}\,.\nn
\end{align}
Note that the first of them arises due to \eqref{eq:modified-MC}. There is no contribution from $w^A$, because $w^{\grave a} = 0$ and therefore in particular $w^a = 0$ holds. Not all terms of $\hat \Omega_{aB}{}^C$ will eventually contribute to $\bm X_{AB}{}^C$. More precisely, terms of the form
\begin{itemize}
    \item $S_{ad}{}^e\,(\widetilde{K}^d{}_e)_B{}^C \in \hat{\Omega}_{aB}{}^C$ with $S_{ab}{}^c=S_{(ab)}{}^c$ or
    \item $S_{ad_1\dots d_n} \,(R^{d_1\dots d_n})_B{}^C \in \hat{\Omega}_{aB}{}^C$ with $S_{[ad_1\dots d_n]}=0$\quad $(n=3,6)$
\end{itemize}
are projected out by \eqref{eq:bmXabC}. We account for them by the similarity relation $\hat \Omega_{aB}{}^C \sim \hat \Omega'_{aB}{}^C$ between two $\Omega$'s that only differ by those terms. They eventually give rise to the same $\bm X_{AB}{}^C$ and we treat them as equal. Moreover, it makes equations more readable if one suppresses the last two indices of $\Omega_{aB}{}^C$. Taking all this into account, we obtain
\begin{equation}\label{eq:Omega-hat-M}
    \begin{aligned}
        \Omega_a \sim \bigl(&- \tfrac{1}{2}\,f_{ab}{}^c + \tfrac{1}{2!}\, f_{[a}{}^{cd_1d_2}\, A_{b]d_1d_2} + \tfrac{1}{5!}\, f_{[a}{}^{cd_1\dots d_5}\, A_{b]d_1\dots d_5}\bigr) \,\widetilde{K}^b{}_c \\
        & - \tfrac{1}{4!}\,F_{abcd}\,R^{bcd} - \tfrac{1}{7!}\, F_{ab_1\dots b_6} \,R^{b_1\dots b_6}
    \end{aligned}
\end{equation}
with the two field strengths
\begin{equation}\label{eq:defF4&F7}
    F_{ijkl} := 4\,\partial_{[i} C_{jkl]} \qquad \text{and} \qquad
    F_{i_1\dots i_7} := 7\,\partial_{[i_1} C_{i_2\dots i_7]} -\tfrac{35}{2}\,F_{[i_1\dots i_4}\,C_{i_5i_6i_7]}
\end{equation}
appearing in the second line. Our remaining task is to check that they can be fixed such that $\bm X_{AB}{}^C$ becomes $X_{AB}{}^C$. To this end, we evaluate the term in the brackets of \eqref{eq:Omega-hat},
\begin{align}
    & \hat{\Omega}_{a} + A_a{}^B\,X_B \sim  X_a - \tfrac12 \, f_{ab}{}^c \widetilde{K}^b{}_c \nn\\
    &\quad - \tfrac{1}{4!}\,\bigl(F_{abcd} - 12 \, f_{[ab}{}^{e} \,A_{cd]e} + 36\, Z_{[a}\,A_{bcd]} + 2\,f_{[a}{}^{efg}\,A_{bcd]efg} + \tfrac{1}{30}\,f_d{}^{e_1\dots e_6}\,A_{abe_1\dots e_6,c} \bigr)\,R^{bcd} \nn\\
    &\quad - \tfrac{1}{7!}\, \bigl(F_{ab_1\dots b_6} + 105\,f_{[ab_1b_2b_3}\,A_{b_4b_5b_6]} + 105\, f_{[ab_1}{}^{e} \,A_{b_2\dots b_6]e} \nn\\
    &\quad\qquad\quad + 252\,Z_{[a}\,A_{b_1\dots b_6]} + 21\,f_{[a}{}^{c_1c_2d}\,A_{b_1\dots b_6] c_1c_2,d} \bigr)\,R^{b_1\dots b_6} \,,
\end{align}
where we have used
\begin{equation}
    A_a{}^B\,X_B =  X_a + \tfrac{1}{2!}\,A_{ab_1b_2}\,X^{b_1b_2} + \tfrac{1}{5!}\,A_{ab_1\dots b_5}\,X^{b_1\dots b_5} + \tfrac{1}{7!}\,A_{ab_1\dots b_7,b'}\,X^{b_1\dots b_7,b'}
\end{equation}
with
\begin{equation}
    \begin{aligned}
        X_a &= - \tfrac{1}{6!}\,f_{ab_1\dots b_6}\,R^{b_1\dots b_6} - \tfrac{1}{3!}\,f_{ab_1b_2b_3}\,R^{b_1b_2b_3}\\
        &\quad + f_{ab}{}^c\,\widetilde{K}^b{}_c - Z_a\,\bigl(\widetilde{K}^b{}_b+t_0\bigr) - \tfrac{1}{3!}\,f_{a}{}^{b_1b_2b_3}\,R_{b_1b_2b_3} - \tfrac{1}{6!}\,f_{a}{}^{b_1\dots b_6}\,R_{b_1\dots b_6}\,\\
        X^{a_1a_2} &= f_b{}^{ca_1a_2}\,\widetilde{K}^b{}_c  - f_{c_1c_2}{}^{[a_1}\,R^{a_2]c_1c_2} + 3\,Z_b\,R^{ba_1a_2} - \tfrac{1}{4!}\,f_{b_1\dots b_4}\,R^{a_1a_2b_1\dots b_4}\,,\\
        X^{a_1\dots a_5} &= f_b{}^{ca_1\dots a_5}\,\widetilde{K}^b{}_c + \tfrac{5!}{3!\,2!}\,f_b{}^{[a_1a_2a_3}\,R^{a_4a_5]b} - \tfrac{5!}{2!\,4!}\,f_{bc}{}^{[a_1}\,R^{a_2\dots a_5]bc} + 6\,Z_b\,R^{ba_1\dots a_5}\,,\\
        X^{a_1\dots a_7,a'} &= -7\,f_b{}^{[a_1\dots a_6}\,R^{a_7]a'b} - 21\,f_b{}^{a'[a_1a_2}\,R^{a_3\dots a_7]b} \,.
    \end{aligned}
\end{equation}
If it falls in the same equivalence class as the constants,  
\begin{align}
 W_a = W_a{}^{\dot{\bm{\alpha}}}\,t_{\dot{\bm{\alpha}}} = X_a - \tfrac{1}{2}\,f_{ab}{}^c \,\widetilde{K}^b{}_c + \tfrac{3}{4!}\,f_{ab_1b_2b_3}\,R^{b_1b_2b_3} + \tfrac{6}{7!}\,f_{ab_1\dots b_6}\,R^{b_1\dots b_6} \,,
\end{align}
we defined in \eqref{eq:WA-M}, we have successfully realised our objective. Because $W_a$ is invariant under $G$-action mediated by $M_A{}^B$ and its inverse, in this case $\Omega_a = W_a = \hat\Omega_a + A_a{}^B X_B$ holds. It implies the desired result $\bm X_{AB}{}^C = X_{AB}{}^C$ after comparing \eqref{eq:bmXabC} and \eqref{eq:X-general}. Indeed, $W_a \sim \hat{\Omega}_{a} + A_a{}^B\,X_B$ can be satisfied if, and only if,
\begin{align}\tikzmarkin{F4F7-M}(1.8,-0.4)(-0.35,0.5)
    F_{abcd} &= -3\,f_{abcd} + 12\,f_{[ab}{}^e\,A_{cd]e} - 36\,Z_{[a}\,A_{bcd]} \nn\\
        &\quad - 2\,f_{[a}{}^{efg}\,A_{bcd]efg} - \tfrac{1}{30}\,f_d{}^{e_1\dots e_6}\,A_{abe_1\dots e_6,c}
    \quad \text{and}\\
    F_{a_1\dots b_7} &= - 6\,f_{a_1\dots a_7} - 105\,f_{[a_1\dots a_4}\,A_{a_5a_6a_7]} - 105\, f_{[a_1a_2}{}^{b} \,A_{a_3\dots a_7]b} \nn\\
        &\quad - 252\,Z_{[a_1}\,A_{a_2\dots a_7]} - 21\,f_{[a_1}{}^{b_1b_2c}\,A_{a_2\dots a_7] b_1b_2,c}\tikzmarkend{F4F7-M}
\end{align}
hold. That fixes the field strengths $F_4$ and $F_7$ introduced in \eqref{eq:defF4&F7} to
\begin{align}
    F_4 &= -\tfrac{3}{4!}\,f_{abcd}\,v^a\wedge\dots\wedge v^d \label{eq:F4-desired} \\
        &\quad + \tfrac{1}{2}\,\bigl(f_{ab}{}^d\,A_{cd} + 3\,Z_{a}\,A_{bc} + \tfrac{1}{3!}\, f_{a}{}^{e_1e_2e_3}\,A_{bce_1e_2e_3} - \tfrac{2}{6!}\,f_a{}^{e_1\dots e_6}\,A_{be_1\dots e_6,c}\bigr)\wedge v^a\wedge v^b \wedge v^c \quad \text{and} \nn\\
    F_7 &= -\tfrac{6}{7!}\,f_{a_1\dots a_7} \,v^{a_1}\wedge\dots\wedge v^{a_7} \nn\\
        &\quad - \tfrac{1}{2}\,\bigl(\tfrac{1}{4!}\, f_{a_1\dots a_4}\,A_{a_5a_6} + \tfrac{1}{4!}\, f_{a_1a_2}{}^{b} \,A_{a_3\dots a_6 b} \nn\\
        &\quad\qquad  - \tfrac{1}{10}\,Z_{a_1}\,A_{a_2\dots a_6} - \tfrac{1}{5!}\,f_{a_1}{}^{b_1b_2c}\,A_{a_2\dots a_6 b_1b_2,c} \bigr)\wedge v^{a_1}\wedge\dots\wedge v^{a_6}\,. \label{eq:F7-desired}
\end{align}

We are not yet completely done because $F_4$ and $F_7$ have to satisfy the Bianchi identities
\begin{equation}\label{eq:BI-M}
    \dd F_4 = 0 \qquad \text{and} \qquad
    \dd F_7 + \tfrac{1}{2}\,F_4 \wedge F_4 = 0\,.
\end{equation}
Recalling the definition of the field strengths \eqref{eq:defF4&F7}, they represent the integrability conditions for the potentials $C_3$ and $C_6$\,, which we need to construct $V_A{}^I$ and with it $E_A{}^I$. The idea is the same as in section~\ref{sec:ODDmotiv}, if and only if both Bianchi identities \eqref{eq:BI-M} are satisfied, it is possible to construct $C_3$ and $C_6$ at least locally. Since we are considering the case $n\leq 7$\,, the second Bianchi identity is trivially fulfilled. Hence, we just have to show that the right-hand side of \eqref{eq:F4-desired} is always closed for geometric gaugings. To this end, we first note that the corresponding field strength $F_4$ can be written in the more compact form\footnote{If we use a notation $v_A{}^B = v^C\,(T_C)_A{}^B$, $F_{p+2}$ ($p=2,5$) can be expressed also as
\begin{equation*}
        F_{p+2} = \tfrac{1}{(p+2)!}\,f_{a_1\cdots a_{p+2}}\,v^{a_1}\wedge\dots\wedge v^{a_{p+2}} + \tfrac{1}{(p+1)!}\,v_{a,b_1\cdots b_p}\wedge v^a\wedge v^{b_1}\wedge \cdots \wedge v^{b_p}\,.
\end{equation*}}
\begin{equation}\label{eq:F4compact}
    F_4 = \tfrac{1}{4!}\,f_{abcd}\,v^{a}\wedge\dots\wedge v^{d} - \tfrac{1}{3!}\,X_{A,b,c_1c_2}\,v^A\wedge v^{b}\wedge v^{c_1}\wedge v^{c_2}
\end{equation}
and similarly
\begin{equation}\label{eq:F7compact}
    F_7 = \tfrac{1}{7!}\,f_{a_1\dots a_7}\,v^{a_1}\wedge\dots\wedge v^{a_7} - \tfrac{1}{6!}\,X_{A,b,c_1\dots c_5}\,v^A\wedge v^{b}\wedge v^{c_1}\wedge\dots \wedge v^{c_5}\,.
\end{equation}
Taking the exterior derivative of the right-hand side, $\dd v^A$ contains the two-form field $w^A$ through the modified Maurer-Cartan equation \eqref{eq:modified-MC}. However, the contraction $\dd v^A\,X_{A,\dots}$, $w^A\, X_{AB}{}^C$ ensures that $w^A$ totally disappears from the expression and we do not need to care about anything else than $v^A$. After remembering the decomposition \eqref{eq:decomp-T}, $T_A= \begin{pmatrix} T_a & T_{\check \alpha}\end{pmatrix}$ where $T_{\check\alpha}=\begin{pmatrix} T^{a_1a_2} & T^{a_1\dots a_5} & T^{a_1\dots a_7,a'}\end{pmatrix}$, and the properties
\begin{equation}
    X_{AB}{}^c = -X_{BA}{}^c\,,\quad 
    X_{\check\alpha\check\beta}{}^c = 0\,,\quad\text{and}\quad
    X_{AB}{}^C\,X_C = -X_{BA}{}^C\,X_C\,,
\end{equation}
we eventually find
\begin{equation}\label{eq:dF4}
    \begin{aligned}
        \dd F_4 &= -\tfrac{1}{12}\,\bigl(X_{ab}{}^G\,X_{G,c,de}+2\,f_{ab}{}^g\,f_{gcde}\bigr)\,v^{a}\wedge v^{b}\wedge v^{c}\wedge v^{d}\wedge v^{e}\\
        &\quad -\tfrac{1}{12}\,\bigl(4\,f_{gbcd}\,X_{e\check\alpha}{}^g + 2\,X_{e\check\alpha}{}^G\,X_{G,b,cd} - 3\,f_{bc}{}^g\,X_{\check\alpha,g,de} \bigr)\,v^{\check\alpha}\wedge v^{b}\wedge v^{c}\wedge v^{d}\wedge v^{e}\\
        &\quad -\tfrac{1}{2\cdot 3!}\,\bigl(X_{\check\alpha\check\beta}{}^G\,X_{G,c,de} +    6\,X_{\check\alpha c}{}^f\,X_{\check\beta,f,de}\bigr)\,v^{\check\alpha}\wedge v^{\check\beta}\wedge v^{c}\wedge v^{d}\wedge v^{e}\,.
    \end{aligned}
\end{equation}
Our insights from section~\ref{sec:ODDmotiv} suggest that it should be possible to rewrite the terms quadratic in the structure constants in terms of the Leibniz identity \eqref{eq:LeibnizID}. In section~\ref{sec:ODDmotiv}, this is already worked out in \eqref{eq:dH=0}. In the same vein, we introduce
\begin{equation}\label{eq:LI}
    L_{ABC}{}^D := X_{AC}{}^E\,X_{BE}{}^D - X_{BC}{}^E\,X_{AE}{}^D + X_{AB}{}^E\,X_{EC}{}^D = 0\,.
\end{equation}
By looking carefully at each component of \eqref{eq:dF4}, we indeed find that it can be rewritten as
\begin{equation}
    \begin{aligned}
        \dd F_4 &= -\tfrac{1}{20}\,L_{a,b,c,de}\,v^{a}\wedge v^{b}\wedge v^{c}\wedge v^{d}\wedge v^{e}\\
        &\quad +\tfrac{1}{24}\,\bigl(L_{\check\alpha,b,c,de}+4\,L_{b,\check\alpha,c,de} \bigr)\,v^{\check\alpha}\wedge v^{b}\wedge v^{c}\wedge v^{d}\wedge v^{e}\\
        &\quad -\tfrac{1}{2\cdot 3!}\,L_{\check\alpha,\check\beta,c,de}\,v^{\check\alpha}\wedge v^{\check\beta}\wedge v^{c}\wedge v^{d}\wedge v^{e} = 0\,.
    \end{aligned}
\end{equation}
We therefore conclude that due to the Leibniz identity, the field strength $F_4$ (and also $F_7$) satisfy their respective Bianchi identities and one can always find the associated potentials $C_3$ and $C_6$, at least in a local patch. With them, $E_A{}^I$ is totally fixed and the construction of the M-theory generalised frame field satisfying \eqref{eq:framealgebra} for all geometric gaugings is completed.

\subsection{Type IIB section}\label{sec:frame-IIB}
For type IIB sections, the constructions follow mostly along the same line. We keep the discussion here brief and refer the reader to section~\ref{sec:frame-M-th} for an extended version. We shall begin by splitting the components of the one-form fields $v_{\sfm}^A$ and $A_{\sfa}{}^B$ from the $R_1$ representation of E$_{n(n)}$ to the manifest symmetry group GL($n-1$)$\times$SL(2) of type IIB sections. Taking into account the notation introduced in \eqref{eq:gen-IIB}, one finds
\begin{equation}
    \begin{aligned}
        v_{\sfm}^A &= \begin{pmatrix}
            v_{\sfm}^{\sfa} & A_{\sfm\sfa_1}^\alpha & A_{\sfm\sfa_1\sfa_2\sfa_3} & A_{\sfm\sfa_1\dots\sfa_5}^\alpha & A_{\sfm\sfa_1\dots \sfa_6,\sfa'}
        \end{pmatrix} \qquad \text{and} \\
        A_{\sfa}{}^B &= \begin{pmatrix}
            \delta_{\sfa}^{\sfb} & A^\beta_{\sfa\sfb} & A_{\sfa\sfb_1\sfb_2\sfb_3} & A^\beta_{\sfa\sfb_1\dots\sfb_5} & A_{\sfa\sfb_1\dots \sfb_6,\sfb'}
        \end{pmatrix}\,.
    \end{aligned}
\end{equation}
We keep the ansatz for $V_A{}^I$ from \eqref{eq:VaI}, but this time with the adapted
\begin{equation}\label{eq:VaJ-IIB}
    \hat{V}_A{}^J = \begin{pmatrix}
        v_{\sfa}^{\sfn} & 0 & 0 & 0 & 0 \\ 0 & \delta_\alpha^\nu\,v^{\sfa}_{\sfn} & 0 & 0 & 0\\
        0 & 0 & 3!\,v_{[\sfn_1}^{[\sfa_1}\,v_{\sfn_2}^{\sfa_2}\,v_{\sfn_3]}^{\sfa_3]} & 0 & 0\\
        0 & 0 & 0 & 5!\,\delta_\alpha^\nu\,v_{[\sfn_1}^{[\sfa_1}\dots v_{\sfn_5]}^{\sfa_5]} & 0\\
        0 & 0 & 0 & 0 & 6!\,v_{[\sfn_1}^{[\sfa_1}\dots v_{\sfn_6]}^{\sfa_6]}\,v_{\sfn'}^{\sfa'}
    \end{pmatrix} ,
\end{equation}
and
\begin{equation}\label{eq:NiJ-IIB}
    N_I{}^J=\left[ \exp \left(\tfrac{1}{2!}\,B^\mu_{\sfm\sfn}\,R^{\sfm\sfn}_\mu\right) \exp \left( \tfrac{1}{4!}\,D_{\sfm_1\dots \sfm_4}\,R^{\sfm_1\dots \sfm_4}\right) \exp\left(\tfrac{1}{6!}\,B^\mu_{\sfm_1\dots \sfm_6}\,R_\mu^{\sfm_1\dots \sfm_6}\right) \right]{}_I{}^J\,.
\end{equation}
Based on the components of the E$_{n(n)}$ generators in \eqref{eq:Enn-gen-IIB}, we have introduced two two-forms $B^\mu_{\sfm_1\sfm_2}$ and two six-forms $B^\mu_{\sfm_1\dots \sfm_6}$, both transforming in the fundamental of the S-duality group SL(2) labeled by $\mu=1, 2$, and the four-form $D_{\sfm_1\dots \sfm_4}$ that is an SL(2) singlet. They respectively couple to D1 branes, which are S-dual to the fundamental strings F1, ($B^\mu_{\sfm_1\sfm_2}$), D3-branes ($D_{\sfm_1\dots \sfm_4}$), and D5-branes or the S-dual NS5-branes ($B^\mu_{\sfm_1 \dots \sfm_6}$). This shows nicely that $N_I{}^J$ implements all expected degrees of freedom. In $n\leq 6$\,, it has the explicit form
{\small\begin{equation}\label{eq:NiJ-mat-IIB}
    N_I{}^J = \begin{pmatrix}
        \delta_{\sfm}^{\sfn} & -B^\nu_{\sfm\sfn} & -D_{\sfm\sfn_1\sfn_2\sfn_3}-\frac{3}{2} \epsilon_{\rho\sigma} B^\rho_{\sfm[\sfn_1} B^\sigma_{\sfn_2\sfn_3]} & - 5 \epsilon_{\rho\sigma} B^\rho_{\sfm[\sfn_1} B^\sigma_{\sfn_2\sfn_3} B_{\sfn_4\sfn_5]}^\nu \\
        0 & \delta_\mu^\nu \delta^{\sfm}_{\sfn} & 3 \epsilon_{\mu\rho} \delta^{\sfm}_{[\sfn_1} B^\rho_{\sfn_2\sfn_3]} & -5 \delta_\mu^\nu \delta^{\sfm}_{[\sfn_1} D_{\sfn_2\dots \sfn_5]}+15 \epsilon_{\mu\rho} \delta^{\sfm}_{[\sfn_1} B^\rho_{\sfn_2\sfn_3} B^\nu_{\sfn_4\sfn_5]} \\ 
        0 & 0 & 3! \delta^{\sfm_1\sfm_2\sfm_3}_{\sfn_1\sfn_2\sfn_3} & 60 \delta^{\sfm_1\sfm_2\sfm_3}_{[\sfn_1\sfn_2\sfn_3} B^\nu_{\sfn_4\sfn_5]} \\
        0 & 0 & 0 & 5! \delta_\mu^\nu \delta^{\sfm_1\dots \sfm_5}_{\sfn_1\dots \sfn_5}
    \end{pmatrix}\,.
\end{equation}}

Similar to the M-theory case, we next compute $\hat{\Omega}_{\sfa B}{}^C$ by using
\begin{equation}\label{eq:geomflux-IIB}
    v^{\sfc}\bigl([v_{\sfa},\,v_{\sfb}]\bigr) = f_{\sfa\sfb}{}^{\sfc} - 2\,f_{[\sfa|}{}^{\sfc\sfd}_\alpha\, A^\alpha_{|\sfb]\sfd} - \tfrac{2}{3!}\,f_{[\sfa}{}^{\sfc\sfd_1\sfd_2\sfd_3}\, A_{\sfb]\sfd_1\sfd_2\sfd_3} - \tfrac{2}{5!}\,f_{[\sfa}{}^{\sfc\sfd_1\dots\sfd_5}\,A_{\sfb]\sfd_1\dots\sfd_5}\,,\\
\end{equation}
in analogy with \eqref{eq:geomflux-M}, and
\begin{equation}\label{eq:dNN-1-IIB}
    \begin{aligned}
        \dd N \, N^{-1} &= \tfrac{1}{2!}\,\dd B^\mu_{\sfm\sfn}\,R^{\sfm\sfn}_\mu\\
        &\quad + \tfrac{1}{4!}\,\bigl(\dd D_{\sfm_1\dots \sfm_4} - 3\,\epsilon_{\mu\nu}\,\dd B^\mu_{\sfm_1\sfm_2}\,B^\nu_{\sfm_3\sfm_4}\bigr)\, R^{\sfm_1\dots \sfm_4}\\
        &\quad + \tfrac{1}{6!}\,\bigl(\dots\bigr)^\mu_{\sfm_1\dots \sfm_6}\, R_\nu^{\sfm_1\dots \sfm_6}\,.
    \end{aligned}
\end{equation}
Here, we do not specify terms contracted with $R_\mu^{\sfm_1\dots \sfm_6}$ explicitly, because they are removed from $\Omega_{\sfa B}{}^C$ under the equivalence relation $\sim$ resulting from dropping terms that satisfy
\begin{itemize}
    \item $S_{\sfa\sfd}{}^{\sfe}\,(\widetilde{K}^{\sfd}{}_{\sfe})_B{}^C \in \hat{\Omega}_{\sfa B}{}^C$ with $S_{\sfa\sfb}{}^{\sfc}=S_{(\sfa\sfb)}{}^{\sfc}$ (this is the same as in M-theory sections just with the indices adapted) or
    \item $S_{\sfa\sfd_1\dots \sfd_n} \,(R_{\bullet}^{\sfd_1\dots \sfd_n})_B{}^C \in \hat{\Omega}_{\sfa B}{}^C$ with $S_{[\sfa \sfd_1\dots \sfd_n]}=0$ for $n=2,4,6$ where $\bullet$ denotes the the additional SL(2) index $\alpha$ that has to be added for $n=2,6$\,.
\end{itemize}
As one might expect, $\sim$ follows the same intent as the ones we defined in section~\ref{sec:frame-M-th}. Applying it in combination with \eqref{eq:geomflux-IIB} and \eqref{eq:dNN-1-IIB}, we obtain
\begin{align}
    \hat{\Omega}_{\sfa B}{}^C &= -D_{\sfa} \hat{V}_B{}^J\,\hat{V}_J{}^C - \hat{V}_B{}^J\, \hat{V}_K{}^C\, D_a N_J{}^L \, (N^{-1})_L{}^K \nn\\
    &\sim \bigl[\bigl(- \tfrac{1}{2}\,f_{\sfa\sfb}{}^{\sfc} +  f_{[\sfa|}{}^{\sfc\sfd}_\alpha\, A^\alpha_{|\sfb]\sfd} + \tfrac{1}{3!}\,f_{[\sfa}{}^{\sfc\sfd_1\sfd_2\sfd_3}\, A_{\sfb]\sfd_1\sfd_2\sfd_3} + \tfrac{1}{5!}\,f_{[\sfa}{}^{\sfc\sfd_1\dots\sfd_5}\,A_{\sfb]\sfd_1\dots\sfd_5} \bigr) \,\widetilde{K}^{\sfb}{}_{\sfc} \bigr]_B{}^C \nn\\
    &\quad - \bigl[\tfrac{1}{3!}\,F^{\alpha}_{\sfa\sfb\sfc}\,R^{\sfb\sfc}_\alpha + \tfrac{1}{5!}\,F_{\sfa\sfb_1\dots \sfb_4}\,R^{\sfb_1\dots \sfb_4}\bigr]_B{}^C \,.
\end{align}
It contains the field strengths\footnote{In the following, we relate curved indices corresponding to the S-duality group SL(2), like $\mu,\nu$, and their flat counterparts,  $\alpha,\beta$, by using $\delta_\mu^\alpha$ and $\delta^\mu_\alpha$\,.}
\begin{equation}\label{eq:defF3&F5}
    \begin{aligned}
        F^\alpha_{\sfa\sfb\sfc} &:= 3\,v_{\sfa}^{\sfm}\,v_{\sfb}^{\sfn}\,v_{\sfc}^{\sfp}\,\delta^\alpha_\mu\, \partial_{[\sfm} B^\mu_{\sfn\sfp]} \qquad \text{and} \\
        F_{\sfa_1\dots \sfa_5} &:= v_{\sfa_1}^{\sfm_1}\dots v_{\sfa_5}^{\sfm_5}\,\bigl(5\,\partial_{[\sfm_1} D_{\sfm_2\sfm_3\sfm_4\sfm_5]} - 5\,\epsilon_{\mu\nu}\,F^{\mu}_{[\sfm_1\sfm_2\sfm_3}\,B^{\nu}_{\sfm_4\sfm_5]}\bigr)
    \end{aligned}
\end{equation}
corresponding to the potentials introduced in \eqref{eq:NiJ-IIB}. $B^\mu_{\sfm_1\cdots \sfm_6}$ is the only one among them lacking a field strength. As a six-form potential, it would have a seven-form field strength. But we only go up to $n-1\le 6$ dimensions here and therefore, there is no seven-form. Next, we again combine
\begin{equation}
    \begin{aligned}
        &\hat{\Omega}_{\sfa} + A_{\sfa}{}^B\,X_{B}\\
        &\sim X_{\sfa} - \tfrac{1}{2}\,f_{\sfa\sfb}{}^{\sfc} \,\widetilde{K}^{\sfb}{}_{\sfc}\\
        &\quad - \tfrac{1}{3!}\,\bigl(F^{\alpha}_{\sfa\sfb\sfc} + 3\,f_{[\sfa\sfb}{}^{\sfd}\,A^\alpha_{\sfc]\sfd} - 6\,f_{[\sfa|\beta}{}^\alpha\,A^\beta_{|\sfb\sfc]} + 12\,Z_{[\sfa}\,A^\alpha_{\sfb\sfc]} + 3\,\epsilon^{\alpha\beta}\,f_{[\sfa|}{}_{\beta}^{\sfd_1\sfd_2}\,A_{|\sfb\sfc]\sfd_1\sfd_2}\\
        &\qquad\qquad
        - \tfrac{1}{4}\, f_{[\sfa}{}^{\sfd_1\dots \sfd_4} \,A^\alpha_{\sfb\sfc]\sfd_1\dots\sfd_4} + \tfrac{1}{5!}\, \epsilon^{\alpha\beta}\,f_{\sfb}{}_\beta^{\sfd_1\dots \sfd_6} \,A_{\sfa\sfd_1\dots\sfd_6,\sfc}\bigr)\,R^{\sfb\sfc}_\alpha\\
        &\quad - \tfrac{1}{5!}\,\bigl(F_{\sfa\sfb_1\dots \sfb_4} + 30\,f_{[\sfa\sfb_1}{}^{\sfc} \,A_{\sfb_2\sfb_3\sfb_4]\sfc} + 80\,Z_{[\sfa}\,A_{\sfb_1\dots \sfb_4]}  + 10\,f_{[\sfa|}{}_{\alpha}^{\sfc_1\sfc_2} \,A^\alpha_{|\sfb_1\dots\sfb_4]\sfc_1\sfc_2}\\
        &\quad\qquad\quad
        - 20\,\epsilon_{\alpha\beta}\,A_{\sfa[\sfb_1}^\alpha\,f^\beta_{\sfb_2\sfb_3\sfb_4]} + \tfrac{10}{3}\,f_{[\sfb_1}{}^{\sfc_1\sfc_2\sfc_3\sfd}\,A_{\sfb_2\sfb_3\sfb_4]\sfc_1\sfc_2\sfc_3,\sfd}\bigl)\,R^{\sfb_1\dots \sfb_4}
    \end{aligned}
\end{equation}
which requires the components
\begin{equation}
    \begin{aligned}
        X_{\sfa} &= - \tfrac{1}{4!}\,f_{\sfa\sfb_1\dots \sfb_4}\,R^{\sfb_1\dots \sfb_4} - \tfrac{1}{2!}\,f^\beta_{\sfa\sfb_1\sfb_2}\,R_\beta^{\sfb_1\sfb_2}\\
        &\quad + f_{\sfa\sfb}{}^{\sfc}\,\widetilde{K}^{\sfb}{}_{\sfc} - Z_{\sfa}\,\bigl(\widetilde{K}^{\sfb}{}_{\sfb}+t_0\bigr) -f_{\sfa\alpha}{}^\beta\,R^\alpha{}_\beta\\
        &\quad - \tfrac{1}{2!}\,f_{\sfa}{}_\beta^{\sfb_1\sfb_2}\,R^\beta_{b_1b_2} - \tfrac{1}{4!}\,f_{\sfa}{}^{\sfb_1\dots \sfb_4}\,R_{\sfb_1\dots \sfb_4} - \tfrac{1}{6!}\,f_{\sfa}{}_\beta^{\sfb_1\dots \sfb_6}\,R^\beta_{\sfb_1\dots \sfb_6} \,,\\
        X^{\sfa}_{\alpha} &= f_{\sfb}{}_{\alpha}^{\sfc\sfa}\,\widetilde{K}^{\sfb}{}_{\sfc} - \tfrac{1}{2!}\,\bigl(\delta_\alpha^\beta\,f_{\sfb_1\sfb_2}{}^{\sfa} - 2\,\delta^{\sfa}_{[\sfb_1}\,f_{\sfb_2]\alpha}{}^\beta\bigr)\,R_\beta^{\sfb_1\sfb_2} - 2\,Z_{\sfb}\,R_\alpha^{\sfa\sfb}\\
        &\quad +\tfrac{1}{3!}\,\epsilon_{\alpha\beta}\,f^\beta_{\sfb_1\sfb_2\sfb_3}\, R^{\sfa\sfb_1\sfb_2\sfb_3} -\tfrac{1}{5!}\,f_{\sfb_1\dots \sfb_5}\, R_\alpha^{\sfa\sfb_1\dots \sfb_5}\,,\\
        X^{\sfa_1\sfa_2\sfa_3} &= f_{\sfb}{}^{\sfc\sfa_1\sfa_2\sfa_3}\,\widetilde{K}^{\sfb}{}_{\sfc} +3\,\epsilon^{\beta\gamma}\,f_{\sfb}{}_{\beta}^{[\sfa_1\sfa_2}\, R_\gamma^{\sfa_3]\sfb} - \tfrac{3}{2}\,f_{\sfb_1\sfb_2}{}^{[\sfa_1}\, R^{\sfa_2\sfa_3]\sfb_1\sfb_2} - 4\,Z_{\sfb}\,R^{\sfa_1\sfa_2\sfa_3\sfb}\\
        &\quad + \tfrac{1}{3!}\,f^\alpha_{\sfb_1\sfb_2\sfb_3}\,R_{\alpha}^{\sfa_1\sfa_2\sfa_3\sfb_1\sfb_2\sfb_3}\,,\\
        X_\alpha^{\sfa_1\dots \sfa_5} &= f_{\sfb}{}_{\alpha}^{\sfc\sfa_1\dots \sfa_5}\,\widetilde{K}^{\sfb}{}_{\sfc} + 5\, f_{\sfb}{}^{[\sfa_1\dots \sfa_4}\, R_\alpha^{\sfa_5]\sfb} - 10\,f_{\sfb}{}_{\alpha}^{[\sfa_1\sfa_2}\, R^{\sfa_3\sfa_4\sfa_5]\sfb}\\
        &\quad -\tfrac{5}{2}\, f_{\sfb_1\sfb_2}{}^{[\sfa_1}\, R_\alpha^{\sfa_2\dots \sfa_5]\sfb_1\sfb_2} + f_{\sfb\alpha}{}^\beta\, R_\beta^{\sfa_1\dots\sfa_5\sfb} - 6\,Z_{\sfb}\, R_\alpha^{\sfa_1\dots\sfa_5\sfb}\,, \\
        X^{\sfa_1\dots \sfa_6,\sfa'} &= - \epsilon^{\beta\gamma}\, f_{\sfb}{}_{\beta}^{\sfa_1\dots\sfa_6}\,R_\gamma^{\sfa' \sfb} + 20\,f_{\sfb}{}^{\sfa'[\sfa_1\sfa_2\sfa_3}\,R^{\sfa_4\sfa_5\sfa_6]\sfb} - 6\,\epsilon^{\beta\gamma}\, f_{\sfb}{}_{\beta}^{\sfa'[\sfa_1}\,R^{\sfa_2\dots\sfa_6]\sfb}_\gamma
    \end{aligned}
\end{equation}
of the structure constants. In analogy with the M-theory case, the desired $\bm{X}_{AB}{}^C = X_{AB}{}^C$ holds if and only if $\hat{\Omega}_{\sfa} + A_{\sfa}{}^B\,X_{B}$ coincides with
\begin{equation}
    W_{\sfa} = X_{\sfa} - \tfrac{1}{2}\,f_{\sfa\sfb}{}^{\sfc} \,\widetilde{K}^{\sfb}{}_{\sfc} + \tfrac{2}{3!}\,f^\alpha_{\sfa\sfb_1\sfb_2}\,R_\alpha^{\sfb_1\sfb_2} + \tfrac{4}{5!}\,f_{\sfa\sfb_1\dots \sfb_4}\,R^{\sfb_1\dots \sfb_4} \,,
\end{equation}
under the equivalence relation $\sim$\,. As before, imposing this constraint completely fixes the field strengths introduced in \eqref{eq:defF3&F5} to
\begin{equation}
    \begin{aligned}\tikzmarkin{F3F5}(1.3,-0.3)(-0.2,0.5)
        F^{\mu}_3 &= -\tfrac{2}{3!}\,f^\mu_{\sfa\sfb\sfc}\,v^{\sfa}\wedge v^{\sfb}\wedge v^{\sfc} - \bigl(\tfrac{1}{2}\,f_{\sfa\sfb}{}^{\sfd}\,A^\mu_{\sfd} + f_{\sfa\beta}{}^\mu\,A^\beta_{\sfb} - 2\,Z_{\sfa}\,A^\mu_{\sfb}\bigr) \wedge v^{\sfa}\wedge v^{\sfb} \qquad \text{and}\\
        &\quad + \bigl(\tfrac{1}{2!}\,\epsilon^{\mu\beta}\,f_{\sfa}{}_{\beta}^{\sfc_1\sfc_2}\,A_{\sfb\sfc_1\sfc_2} - \tfrac{1}{4!}\, f_{\sfa}{}^{\sfc_1\dots \sfc_4} \,A^\mu_{\sfb \sfc_1\dots\sfc_4} - \tfrac{1}{6!}\, \epsilon^{\mu\beta}\,f_{\sfa}{}_\beta^{\sfc_1\dots \sfc_6} \,A_{\sfc_1\dots\sfc_6,\sfb}\bigr)\wedge v^{\sfa}\wedge v^{\sfb} \,,\\
        F_5 &= -\tfrac{4}{5!}\,f_{\sfa_1\dots \sfa_5}\,v^{\sfa_1}\wedge \dots \wedge v^{\sfa_5}\\
        &\quad - \bigl(\tfrac{1}{2!\,2!}\, f_{\sfa_1\sfa_2}{}^{\sfb} \,A_{\sfa_3\sfa_4 \sfb} - \tfrac{4}{3!}\,Z_{\sfa_1}\,A_{\sfa_2\dots \sfa_4} - \tfrac{1}{3!\,2!}\,f_{\sfa_1}{}_{\alpha}^{\sfb_1\sfb_2} \,A^\alpha_{\sfa_2\dots\sfa_4\sfb_1\sfb_2}\\
        &\quad \qquad - \tfrac{1}{3!}\,\epsilon_{\alpha\beta}\,f^\alpha_{\sfa_1\sfa_2\sfa_3}\,A_{\sfa_4}^\beta + \tfrac{1}{36}\,f_{\sfa_1}{}^{\sfb_1\dots\sfb_4}\,A_{\sfa_2\sfa_3\sfa_4\sfb_1\sfb_2\sfb_3,\sfb_4} \bigr)\wedge v^{\sfa_1}\wedge\dots\wedge v^{\sfa_4}\,.\tikzmarkend{F3F5}
    \end{aligned}
\end{equation}
It remains to be check that they satisfy the Bianchi identities
\begin{align}
    \dd F^{\mu}_3=0 \qquad \text{and} \qquad
    \dd F_5 - \tfrac{1}{2}\,\epsilon_{\mu\nu}\,F^{\mu}_3\wedge F^{\nu}_3 = 0
\end{align}
for any geometric gauging with type IIB section. Only if this is the case, we obtain the corresponding potentials $B^\mu_2$, $D_4$, $B^\mu_6$ that are indispensable in constructing the generalised frame fields $E_A{}^I$ for the frame algebra \eqref{eq:framealgebra}. 

Similar to \eqref{eq:F4compact} and \eqref{eq:F7compact}, it is easier to verify the Bianchi identities after rewriting $F_3^\mu$ and $F_5$ in the compact form
\begin{align}
		F^\mu_3 &= \tfrac{1}{3!}\,f^\mu_{\sfa\sfb\sfc}\,v^{\sfa}\wedge v^{\sfb}\wedge v^{\sfc} - \tfrac{1}{2!}\,X^{\phantom{\mu}}_{A \sfb}{}_{\sfc\vphantom{A}}^\mu\,v^A\wedge v^{\sfb}\wedge v^{\sfc} \quad \text{and} \label{eq:B-F3}\\
    F_5 &= \tfrac{1}{5!}\,f_{\sfa_1\dots\sfa_5}\,v^{\sfa_1}\wedge\dots\wedge v^{\sfa_5} - \tfrac{1}{4!}\,X_{A}{}_{\sfb}{}_{\sfc_1\dots\sfc_3}\,v^A\wedge v^{\sfb}\wedge v^{\sfc_1}\wedge\dots \wedge v^{\sfc_3}\,. \label{eq:B-F5}
\end{align}
From the same argument as for M-theory sections in the previous subsection, we see that the two-form $w^A$ does not contribute to either $\dd F^\alpha_3$ or $\dd F_5$\,. Indeed using \eqref{eq:modified-MC}, we get
\begin{equation}
    \begin{aligned}
	    \dd F^\mu_3 &= -\tfrac{1}{4}\,\bigl(f_{\sfa\sfb}{}^{\sfe}\,f^\mu_{\sfe\sfc\sfd}+X_{\sfa,\sfb}{}^E\,X^{\phantom{\mu}}_{E,\sfc,}{}^\mu_d\bigr)\,v^{\sfa}\wedge \dots \wedge v^{\sfd}\\
					&\quad - \tfrac{1}{2}\,\bigl(f^\mu_{\sfg\sfb\sfc}\,X_{\check\alpha\sfd}{}^{\sfg} + X_{\check\alpha\sfb}{}^G\,X^{\phantom{\mu}}_{G,\sfc,}{}^\mu_{\sfd} - f_{\sfb\sfc}{}^{\sfg}\,X^{\phantom{\mu}}_{\check\alpha,\sfg,}{}^\mu_{\sfd} \bigr)\,v^{\check\alpha}\wedge v^{\sfb}\wedge v^{\sfc}\wedge v^{\sfd}\\
					&\quad - \tfrac{1}{2\cdot 2!}\,\bigl(X_{\check\alpha\check\beta}^E\,X^{\phantom{\mu}}_{E,\sfc,}{}^\mu_{\sfd}  + 4\,X^{\phantom{\mu}}_{\check\alpha,\sfe,}{}^\mu_{\sfc}\,X_{\check\beta,\sfd}{}^{\sfe} \bigr)\,v^{\check\alpha}\wedge v^{\check\beta}\wedge v^{\sfc}\wedge v^{\sfd} \,,
    \end{aligned}
\end{equation}
and
\begin{align}
    &\dd F_5 -\tfrac{1}{2}\,\epsilon_{\mu\nu}\,F^\mu_3\wedge F^\nu_3 \nn\\
    &= - \bigl(\tfrac{1}{16}\,f_{\sfa_1\sfa_2}{}^{\sfb}\,f_{\sfb\sfa_3\dots \sfa_6} +\tfrac{1}{48}\,X_{\sfa_1\sfa_2}{}^B\,X_{B,\sfa_3,\sfa_4\sfa_5\sfa_6}-\tfrac{1}{18}\,\epsilon_{\mu\nu}\,f^\mu_{\sfa_1\sfa_2\sfa_3}\,f^\nu_{\sfa_4\sfa_5\sfa_6}\bigr)\,v^{\sfa_1}\wedge \dots\wedge v^{\sfa_6} \nn\\
    &\quad - \bigl(\tfrac{1}{8}\,X_{\check\alpha\sfb_1}{}^{\sfc}\,f_{\sfc\sfb_2\dots\sfb_5} + \tfrac{1}{24}\,X_{\check\alpha\sfb_1}{}^{C}\,X_{C,\sfb_2,\sfb_3\sfb_4\sfb_5} \\
    &\quad\quad
- \tfrac{1}{12}\,f_{\sfb_1\sfb_2}{}^{\sfc}\,X_{\check\alpha,\sfc,\sfb_3\sfb_4\sfb_5} + \tfrac{1}{6}\,\epsilon_{\mu\nu}\,f^\mu_{\sfb_1\sfb_2\sfb_3}\,X^{\phantom{\nu}}_{\check\alpha,\sfb_4,}{}^\nu_{\sfb_5}\bigr)\,v^{\check\alpha}\wedge v^{\sfb_1}\wedge \dots\wedge v^{\sfb_5}\nn\\
	&\quad - \tfrac{1}{2\cdot 4!}\,\bigl(X_{\check\alpha\check\beta}{}^{G}\,X_{G,\sfc,\sfd\sfe\sff} +  8\,X_{\check\alpha\sfc}{}^{\sfg}\,X_{\check\beta,\sfg,\sfd\sfe\sff} - 6\,\epsilon_{\mu\nu}\,X^{\phantom{\mu}}_{\vphantom{d}\check\alpha,\sfc,}{}^\mu_{\sfd}\,X_{\check\beta,\sfe,}{}^\nu_{\sff} \bigr)\,v^{\check\alpha}\wedge v^{\check\beta}\wedge v^{\sfc}\wedge \dots\wedge v^{\sff} \nn
\end{align}
after decomposing the $R_1$ indices $T_A =\begin{pmatrix} T_a & T_{\check \alpha}\end{pmatrix}$. After a long but straightforward computation, we again find that the components of the two Bianchi identities,
\begin{equation}
    \begin{aligned}
			\dd F^\mu_3 &= -\tfrac{1}{8}\,L^{\phantom{\mu}}_{\sfa,\sfb,\sfc,}{}^\mu_{\sfd}\,v^{\sfa}\wedge \dots \wedge v^{\sfd} - \tfrac{1}{3}\,L^{\phantom{\mu}}_{\check\alpha,\sfb,\sfc,}{}^\mu_{\sfd}\,v^{\check\alpha}\wedge v^{\sfb}\wedge v^{\sfc}\wedge v^{\sfd}\\
        &\quad 
        - \tfrac{1}{2\cdot 2!}\,L_{\check\alpha,\check\beta,\sfc,}{}^\mu_{\sfd}\,v^{\check\alpha}\wedge v^{\check\beta}\wedge v^{\sfc}\wedge v^{\sfd}
    \end{aligned}
\end{equation}
and
\begin{align}
    \dd F_5 - \tfrac{1}{2}\,\epsilon_{\mu\nu}\,F^\mu_3\wedge F^\nu_3 
    &=-\tfrac{1}{72}\,L_{\sfa_1,\sfa_2,\sfa_3,\sfa_4\sfa_5\sfa_6}\,v^{\sfa_1}\wedge \dots\wedge v^{\sfa_6} - \tfrac{1}{30}\,L_{\check\alpha,\sfb_1,\sfb_2,\sfb_3\sfb_4\sfb_5}\,v^{\check\alpha}\wedge v^{\sfb_1}\wedge \dots\wedge v^{\sfb_5} \nn\\
    &\quad - \tfrac{1}{2\cdot4!}\,L_{\check\alpha,\check\beta,\sfc,\sfd\sfe\sff}\,v^{\check\alpha}\wedge v^{\check\beta}\wedge v^{\sfc}\wedge \dots\wedge v^{\sff}\,,
\end{align}
can be expressed in terms of components of the Leibniz identity \eqref{eq:LI}. We conclude that also for all geometric gaugings with IIB section, we constructed an explicit frame field $E_A$ satisfying the frame algebra \eqref{eq:framealgebra}.

\subsection{Generalised geometry revisited}
At this point, we close the circle and come back to our motivating example from section~\ref{sec:ODDmotiv}. We show that the tools we developed for the U-duality groups E$_{n(n)}$, also apply to the T-duality groups O($D$,$D$) and reproduce the expected result \eqref{eq:ansatzoddframe}. As the first step, we obtain
\begin{equation}
    v_i^A = \begin{pmatrix}
        v_i^a & A_{ia}
    \end{pmatrix} \qquad \text{and} \qquad
    A_a{}^B = v_a^i\,v_i^B = \begin{pmatrix}
        \delta_a^b & A_{ab}
    \end{pmatrix}
\end{equation}
from the branching O($D$,$D$)$\,\rightarrow\,$GL($D$) given in \eqref{eq:ODD->GLD}. Like before, we use the ansatz \eqref{eq:VaI} for $V_A{}^I$, this time with
\begin{equation}
    \hat{V}_A{}^J = \begin{pmatrix}
        v_a^j & 0 \\
        0 & v^a_j
    \end{pmatrix}
\end{equation}
and
\begin{equation}
    N_I{}^J = \exp\left(\frac{1}{2!}\,B_{ij}\,R^{ij}\right){}_I{}^J = 
    \begin{pmatrix}
        \delta_i^j & - B_{ij} \\
        0 & \delta^i_j
    \end{pmatrix}\,.
\end{equation}
They come with the two additional differential identities
\begin{align}
    v^c\bigl([v_a,\,v_b]\bigr) = f_{ab}{}^c - 2\, f_{[a}{}^{cd}\, A_{b]d}
        \qquad\text{and}\qquad
    \dd N \, N^{-1} = \tfrac{1}{2!}\,\dd B_{ij}\,R^{ij}
\end{align}
that are required to evaluate \eqref{eq:Omega-hat}. Using them, we find
\begin{equation}
    \begin{aligned}
        \hat{\Omega}_{aB}{}^C &= D_a \hat{V}_B{}^J\,\hat{V}_J{}^C + \hat{V}_B{}^J\, \hat{V}_K{}^C\, D_a N_J{}^L \, (N^{-1})_L{}^K\\
        &= -\bigl(D_{(a} v_{b)}^j\,v_j^c + \tfrac{1}{2}\,[v_{a},\, v_{b}]^j\,v_j^c\bigr)\,(K^b{}_c)_B{}^C  + \hat{V}_B{}^J\, \hat{V}_K{}^C\, D_a N_J{}^L \, (N^{-1})_L{}^K\\
        &\sim \bigl[\bigl(- \tfrac{1}{2}\,f_{ab}{}^c + f_{[a}{}^{cd}\, A_{b]d} \bigr) \,K^b{}_c - \tfrac{1}{3!}\,F_{abc}\,R^{bc} \bigr]_B{}^C\,,
    \end{aligned}
\end{equation}
with the three-from field strength
\begin{equation}
    F_{ijk} := 3\,\partial_{[i} B_{jk]}\,.
\end{equation}
The equivalence relation $\sim$ for the T-duality group that appears here is defined by dropping all contributions to $\hat\Omega_{aB}{}^C$ which are of the form
\begin{itemize}
    \item $S_{ab}{}^c\,(K^b{}_c)_B{}^C \in \hat{\Omega}_{aB}{}^C$ with $S_{ab}{}^c=S_{(ab)}{}^c$ or
    \item $S_{ade} \,(R^{de})_B{}^C \in \hat{\Omega}_{aB}{}^C$ with $S_{[ade]}=0$\,. 
\end{itemize}
Next is to compute the combination
\begin{equation}
    \begin{aligned}
        \hat{\Omega}_{a} + A_a{}^B\,X_B 
        &\sim \bigl(- \tfrac{1}{2}\,f_{ab}{}^c + f_{[a}{}^{cd}\, A_{b]d} \bigr) \,K^b{}_c - \tfrac{1}{3!}\,F_{abc}\,R^{bc} + X_a + A_{ab}\,X^{b}\\
        &\sim X_a - \tfrac{1}{2}\,f_{ab}{}^c \,K^b{}_c - \tfrac{1}{3!}\,\bigl(F_{abc} + 3\, f_{[ab}{}^{d} \,A_{c]d} \bigr)\,R^{bc}
    \end{aligned}
\end{equation}
which has to match the constant $W_a$ in \eqref{eq:Wa-ODD} under the equivalence relation $\sim$ for the construction to go through. This is the case, if and only if 
\begin{align}
    F_{abc} = - 2\,f_{abc} - 3\, f_{[ab}{}^{d} \,A_{c]d}
\end{align}
which fixes the three-from to
\begin{align}
    F_3 = - \tfrac{2}{3!}\,f_{abc}\,v^a\wedge v^b\wedge v^c - \tfrac{1}{2}\, f_{ab}{}^c \,A_c\wedge v^a\wedge v^b = H\,.
\end{align}
But this is nothing else than $H$ defined in \eqref{eq:H-ODD}. We have already shown at the end of section~\ref{sec:ODDmotiv} that it satisfies the Bianchi identity $\dd H = 0$.

\section{Conclusions and outlook}\label{sec:conclusions}
We identified the most general gaugings in maximal gauged supergravities that admit an uplift to ten- or eleven-dimensional maximal supergravities. This class of gaugings is called geometric because the corresponding generalised frame (or the twist matrix) satisfies the section conditions and therefore does not depend on the extended coordinates in exceptional field theory. For each higher-dimensional origin (either eleven-dimensional M-theory or type IIB supergravity in ten dimensions), we found all components of the embedding tensor $X_{AB}{}^C$ that are allowed for geometric gaugings in four dimensions or more. Note that two embedding tensors that are related to each other by a constant E$_{n(n)}$ transformation describe the same physics. This equivalence relation defines what is called duality orbits. Our constraints for geometric orbits are not duality covariant. Therefore, they results in distinguished representatives of the corresponding orbits which trivialise the section condition. Therefore to decide if a generic embedding tensor corresponds to a geometric gauging, one has to check if there exists an appropriate E$_{n(n)}$ transformation which results in of to the geometric representatives we found. This is another quadratic problem which is in general hard to solve. Therefore, it is beneficial to use the special representative we identified whenever possible. In addition to the structure constants of EDA, they contain new constants with totally antisymmetric indices. They are $f_{abcd}$ and $f_{a_1\cdots a_7}$ in M-theory and $f^\alpha_{\sfa\sfb\sfc}$, $f_{\sfa_1\cdots \sfa_5}$, $f^\alpha_{\sfa_1\cdots \sfa_7}$ in type IIB. 

The second major result is a systematic construction of the generalised frame fields for any geometric gaugings. They are of the form
\begin{equation}
    E_A{}^I = M_A{}^B \hat V_B{}^J N_J{}^I\,,
\end{equation}
where $\hat V_B{}^I$ and $N_J{}^I$ depend on the section. They are given for M-theory in \eqref{eq:VaJ-M} and \eqref{eq:NiJ-M}, while the type IIB expressions are given in \eqref{eq:VaJ-IIB} and \eqref{eq:NiJ-mat-IIB}. Starting from the Leibniz algebra, we obtained the coset representative $M_A{}^B$, a one-form $v^A$ and a two-form $w^A$\,. Using the first two, we fixed the generalised frame up to its gauge potentials. Next, we computed their field strengths, such as $F_4$ and $F_7$, from the Leibniz algebra and proved that they satisfy the respective Bianchi identities. Accordingly, we can use these field strengths to get at least locally the associated gauge potentials, like $C_3$ and $C_6$, and with them fix $N_I{}^J$. They are not exactly the potentials that appear in supergravity, but using them, we completed the construction of the generalised frame $E_A$ that satisfies the frame algebra \eqref{eq:framealgebra} from the introduction. Although our analysis is limited to the U-duality groups $E_{n(n)}$ with $n\leq 7$, its extension to $n=8$ (or three-dimensional gauged supergravity) is straightforward. For example, the expression of the field strengths, such as \eqref{eq:F4compact}, \eqref{eq:F7compact}, \eqref{eq:B-F3}, and \eqref{eq:B-F5}, will not change, and the only challenge is to verify the Bianchi identities. More details can be found in appendix~\ref{app:algebra}.

There are several new research opportunities that our results open up:
\begin{itemize}
    \item In general, the embedding tensor $X_{AB}{}^C$ in maximal gauged supergravities contains many components, and accordingly, its full classification has been worked out only for $n\leq 3$ \cite{Dibitetto:2012rk}. However, as is summarised in table~\ref{tab:dualitygroups}, if one restricts the discussion to the geometric gaugings, the number of the structure constants is reduced, and the classification becomes easier. Especially, because one can now resort to the analysis of the Lie algebra Lie($G$), together with the distinguished Leibniz representation, and its subalgebra Lie($H$). These insights will help in scanning for geometric gaugings that can be used to construct realistic phenomenological models. 
    \item The classification of embedding tensors is the classification of Leibniz algebras $E$. For each $E$, we can uniquely identify the corresponding gauge group $G$. However, any geometric gauging also has a subgroup $H$ whose choice in general is not unique. If one changes $H$ to another subgroup $H'$, a different generalised frame $E'_A{}^I$ on a different space $M'=G/H'$ arises. As already mentioned, this is the essence of generalised U-duality. For a given Lie group $G$, finding all inequivalent Lie subgroups $H$ is an important task for the future because it classifies all generalised $U$-dualities.
    \item One of the main motivation of our work here is to extend the definition of dressing cosets \cite{Klimcik:1996np} known in the context of generalised geometry to exceptional generalised geometry. Work by one of the authors \cite{Sakatani:2022auu} has made some progress in this direction, but the algebraic structure still stayed elusive. More recently, a systematic approach to study dressing cosets has been presented in \cite{Butter:2022iza} by the second author based on the idea of generalised cosets \cite{Demulder:2019vvh}. It starts from the generalised frame on some extended space. Thus, combining this approach and the result of the present article, it should be possible to formulate exceptional dressing cosets and thereby lift the concept of generalised cosets to exceptional generalised geometry. We hope to make progress in this direction in near future.
\end{itemize}

\subsection*{Acknowledgements}
We would like to thank Chris Blair and Fridrich Valach for helpful discussions. The work by YS is supported by JSPS Grant-in-Aids for Scientific Research (C) 18K13540. FH is supported by the SONATA BIS grant 2021/42/E/ST2/00304 from the National Science Centre (NCN), Poland.

\appendix

\section{Algebras of geometric gaugings for \texorpdfstring{$n\le8$}{n<=8}}\label{app:algebra}
In the M-theory section, the algebra for $n\le7$ can be expressed as
\begin{align}
    T_a \circ T_b &= f_{ab}{}^c\,T_c + f_{abc_2}\,T^{c_2} + f_{abc_5}\,T^{c_5}\,,\nn\\
    T_a \circ T^{b_2} &= -f_a{}^{b_2c}\,T_c + \delta^{b_2}_{de}\,f_{ac}{}^{d}\,T^{ec} + 3\,Z_a\,T^{b_2} - f_{a c_3}\,T^{b_2 c_3} + \delta^{b_2}_{ad}\,f_{c_7}\,T^{c_7,d} \,,\nn\\
    T_a \circ T^{b_5} &= f_a{}^{b_5c}\,T_c + \delta^{b_5}_{c_3d_2}\,f_{a}{}^{c_3}\,T^{d_2} - \delta^{b_5}_{de_4}\,f_{ac}{}^{d}\,T^{e_4c} + 6\,Z_a\,T^{b_5} + \delta^{b_5}_{d_4e}\,f_{ac_3}\,T^{c_3d_4,e}\,,\nn\\
    T_a \circ T^{b_7,b'} &= -\delta^{b_7}_{c_6 d}\,f_a{}^{c_6}\,T^{d b'} + \delta^{b_7}_{c_2d_5}\, f_a{}^{b' c_2}\,T^{d_5}\nn\\
    &\quad - \delta^{b_7}_{de_6}\,f_{ac}{}^{d}\,T^{e_6 c,b'} - f_{ac}{}^{b'}\,T^{b_7,c} + 9\,Z_a\,T^{b_7,b'}\,,\nn\\
    T^{a_2} \circ T_b &= f_b{}^{a_2c}\,T_c + \delta^{a_2}_{de}\,\delta_{c_2b}^{f_2e}\,f_{f_2}{}^{d}\,T^{c_2} - 3\,Z_c\,\delta^{ca_2}_{be_2}\,T^{e_2} + \delta^{a_2c_4}_{bd_5}\,f_{c_4}\,T^{d_5}\,,\nn\\
    T^{a_2} \circ T^{b_2} &= \delta^{b_2}_{de}\, f_c{}^{a_2d}\, T^{ec} - \delta^{a_2}_{de}\,f_{c_2}{}^{d}\,T^{eb_2c_2} + 3\,Z_c\,T^{a_2b_2c} + \delta^{b_2}_{de}\,f_{c_4}\,T^{a_2c_4d,e}\,,\nn\\
    T^{a_2} \circ T^{b_5} &= -\delta^{b_5}_{de_4}\,f_c{}^{a_2 d}\, T^{e_4c} - \delta^{a_2}_{de}\,\delta^{c_2e}_{f_2g}\,f_{c_2}{}^d\,T^{b_5f_2,g} + 3\,\delta^{a_2c}_{d_2e}\,Z_c\,T^{b_5d_2,e}\,,\nn\\
    T^{a_2} \circ T^{b_7,b'} &= -\delta^{b_7}_{de_6}\,f_c{}^{a_2d}\, T^{e_6 c,b'} - f_c{}^{a_2b'}\, T^{b_7,c} \,, \nn\\
    T^{a_5} \circ T_b &= - f_b{}^{a_5c}\,T_c - \delta^{a_5}_{c_3d_2}\,f_b{}^{c_3}\,T^{d_2} - \delta^{a_5}_{d_3bf}\,f_c{}^{d_3}\,T^{fc}\\
    &\quad + \delta^{a_5}_{de_4}\,f_{bc}{}^{d}\,T^{e_4c} + \delta^{a_5}_{dbf_3}\,f_{c_2}{}^{d}\, T^{f_3 c_2} - 6\,\delta^{ca_5}_{bd_5}\,Z_c\, T^{d_5}\,,\nn\\
    T^{a_5} \circ T^{b_2} &= - \delta^{b_2}_{de}\,f_c{}^{a_5 d}\,T^{ec} + \delta^{a_5}_{d_3e_2}\,f_c{}^{d_3}\, T^{e_2 b_2c}\nn\\
    &\quad + \delta^{a_5}_{de_4}\,\delta^{b_2}_{fg}\,f_{c_2}{}^{d}\,T^{e_4 c_2f,g} - 6\,\delta^{b_2}_{de}\,Z_c\,T^{c a_5d,e}\,, \nn\\
    T^{a_5} \circ T^{b_5} &= \delta^{b_5}_{de_4}\,f_c{}^{a_5 d}\, T^{e_4 c} + \delta^{a_5}_{f_3g_2}\,\delta^{g_2c}_{d_2e}\, f_c{}^{f_3}\,T^{b_5 d_2,e}\,,\nn\\
    T^{a_5} \circ T^{b_7,b'} &= \delta^{b_7}_{de_6}\,f_c{}^{a_5 d}\,T^{e_6 c,b'} + f_c{}^{a_5 b'}\,T^{b_7,c}\,,\nn\\
    T^{a_7,a'} \circ T_b &=  \delta^{a_7}_{e_6f}\,\delta^{fa'c}_{b d_2} \,f_c{}^{e_6}\, T^{d_2} + \delta^{a_7}_{e_2f_5}\,f_c{}^{a' e_2}\, \delta_{b d_5}^{f_5 c}\,T^{d_5}\,,\nn\\
    T^{a_7,a'} \circ T^{b_2} &= - \delta^{a_7}_{d_6e}\, f_c{}^{d_6}\,T^{e a' c b_2} + \delta^{a_7}_{d_2e_5}\,\delta^{b_2}_{fg}\,f_c{}^{a' d_2}\,T^{e_5 c f,g}\,,\nn\\
    T^{a_7,a'} \circ T^{b_5} &= - \delta^{a_7}_{f_6g}\,\delta^{ga'c}_{d_2e}\,f_c{}^{f_6}\,T^{b_5 d_2,e}\,,\nn\\
    T^{a_7,a'} \circ T^{b_7,b'} &= 0 \,.\nn
\end{align}
For completeness, we also write down the geometric gauging for $n\le 8$ (see \cite{Sakatani:2020wah} for details on the treatment of $n=8$)\footnote{The convention of the generators $t_{\bm{\alpha}}$ is the same as the one used in \cite{Sakatani:2020wah} wile the structure constants (other than $f_{ab}{}^c$ and $Z_a$) have the opposite sign, both in M-theory and type IIB.}. 
Using the shorthand notation explained below \eqref{eq:geoalg-M}, we decompose the generators of the Leibniz algebra as
\begin{align}
    T_A = \begin{pmatrix}
        T_a & T^{a_2} & T^{a_5} & T^{a_7,a'} & T^{a_8,a'_3} & T^{a_8,a'_6} & T^{a_8,a'_8,a''}
    \end{pmatrix} ,
\end{align}
and the structure constants that enter the embedding tensor are
\begin{equation}
 f_{a_7},\,f_{a_4},\,f_{a_2}{}^b,\,f_a{}^{b_3},\,f_a{}^{b_6},\,f_a{}^{b_8,b'}\,, \quad \text{and} \quad Z_a\,.
\end{equation}
The explicit form of the algebra $T_A\circ T_B=X_{AB}{}^C\,T_C$ is long, and here we only show all components of the embedding tensor $X_A$\,,
\begin{equation}
\begin{aligned}
    X_a &= - f_{a b_6}\,R^{b_6} - f_{a b_3}\,R^{b_3} + f_{ab}{}^c \, \widetilde{K}^b{}_c - Z_a\, (\widetilde{K}^b{}_b + t_0)\\
    &\quad - f_a{}^{b_3}\, R_{b_3} - f_a{}^{b_6}\, R_{b_6} - f_a{}^{b_8,c}\, R_{b_8,c} \,, \\
    X^{a_2}  &= f_c{}^{d a_2}\, \widetilde{K}^c{}_d -\delta^{a_2}_{de}\,f_{c_2}{}^{d}\,R^{ec_2} + 3\,Z_d\,R^{d a_2} - f_{b_4}\,R^{a_2 b_4} + \delta^{a_2}_{cd}\, f_{b_7}\,R^{b_7c,d} \,,\\
    X^{a_5} &= f_c{}^{d a_5} \, \widetilde{K}^c{}_d + \delta^{a_5}_{c_3d_2}\, f_b{}^{c_3}\, R^{d_2 b} - \delta^{a_4}_{de_4}\, f_{c_2}{}^{d}\,R^{e_4 c_2} + 6\,Z_b\,R^{b a_5} + \delta^{b_4}_{c_3d}\,f_{b_4}\,R^{a_5c_3,d}\,,\\
    X^{a_7,a'} &= \bigl(f_c{}^{d a_7,a'} - \tfrac{1}{4}\,f_c{}^{a_7a',d}\bigr) \, \widetilde{K}^c{}_d - \delta^{a_7}_{c_6 d}\,f_b{}^{c_6}\,R^{d a'b} + \tfrac{1}{2}\,\delta^{a_7a'}_{c_6 d_2}\,f_b{}^{c_6}\,R^{d_2 b}\\
    &\quad - \delta^{a_7}_{c_2d_5}\,f_b{}^{a' c_2}\,R^{d_5 b} - \tfrac{1}{4}\,\delta^{a_7 a'}_{c_3 d_5}\,f_b{}^{c_3}\,R^{d_5 b} - \delta^{a_7}_{c d_6}\,f_{b_2}{}^{c}\,R^{d_6 b_2,a'} + f_{bc}{}^{a'}\,R^{a_7b,c}\\
    &\quad + Z_b\,\bigl(9\,R^{b a_7,a'} + \tfrac{3}{4}\, R^{a_7a',b} \bigr)\,,\\
    X^{a_8, a'_3}  &= - \delta^{a'_3}_{cd_2}\,f_b{}^{a_8, c}\,R^{d_2 b} + \delta^{a_8}_{c_3 d_5}\,f_b{}^{a'_3 c_3}\,R^{d_5 b} - \delta^{a'_3}_{c_2d}\,f_b{}^{c_2 b}\,R^{a_8,d} - f_b{}^{a'_3} \, R^{a_8,b} \,,\\
    X^{a_8, a'_6} &= \delta^{a'_6}_{cd_5}\,f_b{}^{a_8,c}\,R^{d_5 b} - \delta^{a'_6}_{c_5d}\,f_b{}^{c_5 b}\,R^{8,d} - f_b{}^{a'_6}\,R^{a_8,b} \,,\\
    X^{a_8, a'_8,a''} &= - 2\,f_d{}^{a_8,a''}\, R^{a'_8,d} - f_d{}^{a_8,d}\, R^{a'_8,a''} \,.
\end{aligned}
\end{equation}
By using the explicit matrix representations of the generators $t_{\bm{\alpha}}$\,, one can write down the structure constants $X_{AB}{}^C$ more explicitly and with them the Leibniz algebra. For $n=8$\,, the embedding tensor is in the representation $(\bm{3875}\oplus\bm{1})\oplus\bm{248}$, but only $1110$ of these are geometric and the other $3014$ components correspond to (locally) non-geometric fluxes.

In the type IIB section, the algebra for $n\le7$ reads
{\small
\begin{align}
    T_{\sfa}\circ T_{\sfb} &=f_{\sfa\sfb}{}^{\sfc}\,T_{\sfc} + f^\gamma_{\sfa\sfb\sfc}\,T_\gamma^{\sfc} + f_{\sfa\sfb\sfc_3}\,T^{\sfc_3}\,,\nn\\
    T_{\sfa}\circ T^{\sfb}_\beta &= f_{\sfa}{}_{\beta}^{\sfb\sfc}\,T_{\sfc} - f_{\sfa\beta}{}^{\gamma}\,T_\gamma^{\sfb} - f_{\sfa\sfc}{}^{\sfb}\,T_\beta^{\sfc} + 2\,Z_{\sfa}\,T^{\sfb}_\beta - \epsilon_{\beta\gamma}\,f^{\gamma}_{\sfa\sfc_2}\, T^{\sfb\sfc_2} + f_{\sfa\sfc_4}\,T_\beta^{\sfb\sfc_4}\,,\nn\\
    T_{\sfa}\circ T^{\sfb_3} &= f_{\sfa}{}^{\sfb_3\sfc}\, T_{\sfc} - \delta^{\sfb_3}_{\sfc_2\sfd}\,\epsilon^{\gamma\delta}\,f_{\sfa}{}_{\gamma}^{\sfc_2}\, T_{\delta}^{\sfd} - \delta^{\sfb_3}_{\sfd\sfe_2}\,f_{\sfa\sfc}{}^{\sfd}\, T^{\sfe_2 \sfc} + 4\,Z_{\sfa}\,T^{\sfb_3} - f^\gamma_{\sfa\sfc_2} T_\gamma^{\sfb_3\sfc_2} - \delta^{\sfb_3}_{\sfc_2\sfd}\,f_{\sfa\sfe_4}\,T^{\sfc_2\sfe_4,\sfd}\,,\nn\\
    T_{\sfa}\circ T_\beta^{\sfb_5} &= f_{\sfa}{}_{\beta}^{\sfb_5\sfc}\,T_{\sfc} - \delta^{\sfb_5}_{\sfc_4\sfd}\, f_{\sfa}{}^{\sfc_4}\,T_\beta^{\sfd} + \delta^{\sfb_5}_{\sfc_2\sfd_3}\, f_{\sfa}{}_{\beta}^{\sfc_2}\,T^{\sfd_3}\nn\\
    &\quad - f_{\sfa\beta}{}^{\gamma}\, T_\gamma^{\sfb_5} - \delta^{\sfb_5}_{\sfd\sfe_4}\,f_{\sfa\sfc}{}^{\sfd}\,T_\beta^{\sfe_4\sfc} + 6\,Z_{\sfa}\,T^{\sfb_5}_\beta + \epsilon_{\beta\gamma}\,f^{\gamma}_{\sfa\sfc\sfd} T^{\sfb_5\sfc,\sfd}\,,\nn\\
    T_{\sfa}\circ T^{\sfb_6,\sfb'} &= \epsilon^{\gamma\delta}\,f_{\sfa}{}_{\gamma}^{\sfb_6}\, T_\delta^{\sfb'} - \delta^{\sfb_6}_{\sfc_3\sfd_3}\, f_{\sfa}{}^{\sfb' \sfc_3}\,T^{\sfd_3} + \delta^{\sfb_6}_{\sfc\sfd_5}\,\epsilon^{\gamma\delta}\,f_{\sfa}{}_{\gamma}^{\sfb'\sfc}\,T_\delta^{\sfd_5}\nn\\
    &\quad - \delta^{\sfb_6}_{\sfd\sfe_5}\,f_{\sfa\sfc}{}^{\sfd}\,T^{\sfc \sfe_5,\sfb'} - f_{\sfa\sfc}{}^{\sfb'}\,T^{\sfb_6,\sfc} + 8\,Z_{\sfa}\,T^{\sfb_6,\sfb'}\,,\nn\\
    T^{\sfa}_\alpha\circ T_{\sfb} &= - f_{\sfb}{}_{\alpha}^{\sfa\sfc} \, T_{\sfc} - \delta^{\sfa\sfd}_{\sfb\sfc}\,f_{\sfd\alpha}{}^{\gamma}\, T_\gamma^{\sfc} + f_{\sfb\sfc}{}^{\sfa}\,T_\alpha^{\sfc} + 2\,\delta^{\sfa\sfc}_{\sfb\sfd}\,Z_{\sfc}\, T^{\sfd}_\alpha - \delta^{\sfa\sfd_3}_{\sfb\sfc_3}\,\epsilon_{\alpha\gamma}\,f^{\gamma}_{\sfd_3}\,T^{\sfc_3} + \delta^{\sfa\sfc_5}_{\sfb\sfd_5}\,f_{\sfc_5}\,T_{\alpha}^{\sfd_5}\,,\nn\\
    T^{\sfa}_\alpha\circ T^{\sfb}_\beta &= f_{\sfc}{}_{\alpha}^{\sfa\sfb}\,T_\beta^{\sfc} + f_{\sfc\alpha}{}^\gamma\,\epsilon_{\gamma\beta}\,T^{\sfc \sfa\sfb} + \epsilon_{\alpha\beta}\,f_{\sfc_2}{}^{\sfa}\,T^{\sfc_2\sfb} - 2\,\epsilon_{\alpha\beta}\,Z_{\sfc}\,T^{\sfa\sfb\sfc} + \epsilon_{\alpha\gamma}\,f^{\gamma}_{\sfc_3}\,T_\beta^{\sfa\sfb\sfc_3} - \epsilon_{\alpha\beta}\,f_{\sfc_5}\,T^{\sfa\sfc_5,\sfb}\,,\nn\\
    T^{\sfa}_\alpha\circ T^{\sfb_3} &= \delta^{\sfb_3}_{\sfd\sfe_2}\,f_{\sfc}{}_{\alpha}^{\sfa \sfd}\,T^{\sfe_2 \sfc} + f_{\sfc\alpha}{}^\gamma\,T_\gamma^{\sfa\sfc \sfb_3} - f_{\sfc_2}{}^{\sfa}\,T_\alpha^{\sfc_2 \sfb_3} + 2\,Z_{\sfc}\,T^{\sfa \sfb_3\sfc}_\alpha + \delta^{\sfa\sfd_3}_{\sfe\sfc_3}\,\epsilon_{\alpha\gamma}\,f^{\gamma}_{\sfd_3}\,T^{\sfb_3\sfc_3,\sfe}\,,\nn\\
    T^{\sfa}_\alpha\circ T_\beta^{\sfb_5} &= \delta^{\sfb_5}_{\sfd\sfe_4}\,f_{\sfc}{}_{\alpha}^{\sfa \sfd}\,T_\beta^{\sfe_4\sfc} + \delta^{\sfa\sfc}_{\sfd\sfe}\,f_{\sfc\alpha}{}^\gamma\,\epsilon_{\gamma\beta}\,T^{\sfb_5\sfd,\sfe} - \epsilon_{\alpha\beta}\,f_{\sfc\sfd}{}^{\sfa}\,T^{\sfb_5 \sfc,\sfd} + 4\,\epsilon_{\alpha\beta}\,Z_{\sfc}\,T^{\sfa \sfb_5,\sfc}\,,\nn\\
    T^{\sfa}_\alpha\circ T^{\sfb_6,\sfb'} &= \delta^{\sfb_6}_{\sfd\sfe_5}\, f_{\sfc}{}_{\alpha}^{\sfa \sfd}\,T^{\sfc \sfe_5,\sfb'} + f_{\sfc}{}_{\alpha}^{\sfa\sfb'}\,T^{\sfb_6,\sfc}\,,\nn\\
    T^{\sfa_3} \circ T_{\sfb} &= - f_{\sfb}{}^{\sfa_3\sfc}\,T_{\sfc} - \delta^{\sfa_3}_{\sfd_2\sfe}\,\delta_{\sfb\sfc}^{\sfe\sff}\,\epsilon^{\gamma\delta}\,f_{\sff}{}_{\gamma}^{\sfd_2}\,T_\delta^{\sfc} + \delta^{\sfa_3}_{\sfd\sfe_2}\,f_{\sfb\sfc}{}^{\sfd}\, T^{\sfe_2\sfc} - \delta^{\sfa_3}_{\sfb\sfd\sfe}\, f_{\sfc_2}{}^{\sfd}\, T^{\sfe\sfc_2}\nn\\
    &\quad +4\,\delta^{\sfa_3\sfc}_{\sfb\sfd_3}\,Z_{\sfc}\, T^{\sfd_3} - \delta^{\sfa_3\sfd_3}_{\sfb\sfc_5}\,f^\gamma_{\sfd_3}\,T_\gamma^{\sfc_5}\,,\nn\\
    T^{\sfa_3} \circ T_\beta^{\sfb} &= f_{\sfc}{}^{\sfa_3 \sfb}\,T_\beta^{\sfc} - \delta^{\sfa_3}_{\sfd_2\sfe}\, f_{\sfc}{}_{\beta}^{\sfd_2}\,T^{\sfe \sfb\sfc} + \delta^{\sfa_3}_{\sfd\sfe_2}\, f_{\sfc_2}{}^{\sfd}\,T_\beta^{\sfe_2 \sfb \sfc_2} - 4\,Z_{\sfc}\,T_\beta^{\sfa_3\sfb\sfc} - \delta^{\sfa_3\sfd_3}_{\sfc_6}\, \epsilon_{\beta\gamma}\,f^\gamma_{\sfd_3}\,T^{\sfc_6,\sfb}\,,\nn\\
    T^{\sfa_3} \circ T^{\sfb_3} &= \delta^{\sfb_3}_{\sfd\sfe_2}\, f_{\sfc}{}^{\sfa_3 \sfd}\, T^{\sfe_2 \sfc} - \delta^{\sfa_3}_{\sfd_2\sfe}\, \epsilon^{\gamma\delta}\,f_{\sfc}{}_{\gamma}^{\sfd_2}\,T_\delta^{\sfe\sfb_3\sfc}\nn\\
    &\quad + \delta^{\sfa_3}_{\sfd\sfe\sff}\,f_{\sfc_2}{}^{\sfd}\, T^{\sfb_3\sfc_2\sfe, \sff}   + \delta^{\sfa_3}_{\sfe\sff_2}\,f_{\sfc\sfd}{}^{\sfe}\,T^{\sff_2 \sfb_3 \sfc, \sfd} + 4 \,\delta^{\sfa_3\sfc}_{\sfd_3\sfe}\,Z_{\sfc}\,T^{\sfb_3 \sfd_3,\sfe} \,,\nn\\
    T^{\sfa_3} \circ T_\beta^{\sfb_5} &= \delta^{\sfb_5}_{\sfd\sfe_4}\,f_{\sfc}{}^{\sfa_3 \sfd}\,T_\beta^{\sfe_4 \sfc} - \delta^{\sfa_3}_{\sff_2\sfg}\,\delta^{\sfg\sfc}_{\sfd\sfe}\,f_{\sfc}{}_{\beta}^{\sff_2}\,T^{\sfb_5 \sfd,\sfe} \,,\nn\\
    T^{\sfa_3} \circ T^{\sfb_6,\sfb'} &= \delta^{\sfb_6}_{\sfd\sfe_5}\,f_{\sfc}{}^{\sfa_3 \sfd}\,T^{\sfc \sfe_5, \sfb'} + f_{\sfc}{}^{\sfa_3 \sfb'}\,T^{\sfb_6, \sfc} \,,\nn\\
    T_\alpha^{\sfa_5} \circ T_{\sfb} &= - f_{\sfb}{}_{\alpha}^{\sfa_5 \sfc}\,T_{\sfc} - \delta^{\sfa_5}_{\sfd_4\sfe}\,\delta^{\sfe\sff}_{\sfb\sfc}\,f_{\sff}{}^{\sfd_4}\,T_\alpha^{\sfc} + \delta^{\sfa_5}_{\sfb\sfd_2\sfe_2}\,f_{\sfc}{}_{\alpha}^{\sfd_2}\, T^{\sfe_2\sfc} - \delta^{\sfa_5}_{\sfc_2\sfd_3}\,f_{\sfb}{}_{\alpha}^{\sfc_2}\,T^{\sfd_3} - \delta^{\sfa_5}_{\sfb\sfd_4}\,f_{\sfc\alpha}{}^{\gamma}\, T_\gamma^{\sfd_4\sfc}\nn\\
    &\quad + f_{\sfb\alpha}{}^{\gamma}\,T_\gamma^{\sfa_5} + \delta^{\sfa_5}_{\sfd\sfe_4}\,f_{\sfb\sfc}{}^{\sfd}\, T_\alpha^{\sfe_4\sfc} + \delta^{\sfa_5}_{\sfd\sfb\sfe_3}\,f_{\sfc_2}{}^{\sfd}\, T_\alpha^{\sfe_3 \sfc_2} + 6\,\delta^{\sfa_5\sfc}_{\sfb\sfd_5}\,Z_{\sfc}\, T_\alpha^{\sfd_5}\,,\nn\\
    T_\alpha^{\sfa_5} \circ T_\beta^{\sfb} &= f_{\sfc}{}_{\alpha}^{\sfa_5\sfb}\,T_\beta^{\sfc} + \delta^{\sfa_5}_{\sfd_4\sfe}\,\epsilon_{\alpha\beta}\,f_{\sfc}{}^{\sfd_4}\,T^{\sfe \sfb\sfc} - \delta^{\sfa_5}_{\sfd_2\sfe_3}\,f_{\sfc}{}_{\alpha}^{\sfd_2}\,T_\beta^{\sfe_3 \sfb\sfc}\nn\\
    &\quad - f_{\sfc\alpha}{}^\gamma\,\epsilon_{\gamma\beta}\,T^{\sfc \sfa_6,\sfb} - \delta^{\sfa_5}_{\sfd\sfe_4}\,\epsilon_{\alpha\beta}\,f_{\sfc_2}{}^{\sfd}\,T^{\sfe_4\sfc_2,\sfb} - 6\,\epsilon_{\alpha\beta}\,Z_{\sfc}\,T^{\sfa_5\sfc,\sfb}\,,\nn\\
    T_\alpha^{\sfa_5} \circ T^{\sfb_3} &= \delta^{\sfb_3}_{\sfd\sfe_2}\,f_{\sfc}{}_{\alpha}^{\sfa_5 \sfd}\, T^{\sfe_2\sfc} - \delta^{\sfa_5}_{\sfd_4\sfe}\,f_{\sfc}{}^{\sfd_4}\, T_\alpha^{\sfe \sfb_3\sfc} + \delta^{\sfa_5}_{\sff_2\sfg_3}\,\delta^{\sfg_3\sfc}_{\sfd_3\sfe}\,f_{\sfc}{}_{\alpha}^{\sff_2}\,T^{\sfb_3 \sfd_3,\sfe}\,,\nn\\
    T_\alpha^{\sfa_5} \circ T_\beta^{\sfb_5} &= \delta^{\sfb_5}_{\sfd\sfe_4}\,f_{\sfc}{}_{\alpha}^{\sfa_5 \sfd}\, T_\beta^{\sfe_4\sfc} + \delta^{\sfa_5}_{\sff_4\sfg}\,\delta^{\sfg\sfc}_{\sfd\sfe}\,\epsilon_{\alpha\beta}\,f_{\sfc}{}^{\sff_4}\,T^{\sfb_5\sfd,\sfe} \,,\nn\\
    T^{\sfa_5} \circ T^{\sfb_6,\sfb'} &= \delta^{\sfb_6}_{\sfd\sfe_5}\,f_{\sfc}{}_{\alpha}^{\sfa_5 \sfd}\, T^{\sfc \sfe_5,\sfb'} + f_{\sfc}{}_{\alpha}^{\sfa_5 \sfb'}\, T^{\sfb_6,\sfc} \,,\nn\\
    T^{\sfa_6,\sfa'} \circ T_{\sfb} &= \delta^{\sfa'\sfc}_{\sfb\sfd}\,\epsilon^{\gamma\delta}\,f_{\sfc}{}_{\gamma}^{\sfa_6}\, T^{\sfd}_\delta + \delta^{\sfa_6}_{\sfc_3\sfd_3}\,f_{\sfb}{}^{\sfa' \sfc_3}\,T^{\sfd_3} + \delta^{\sfa_6}_{\sfb\sfd_3\sfe_2}\,f_{\sfc}{}^{\sfa' \sfd_3}\, T^{\sfe_2\sfc}\nn\\
    &\quad - \delta^{\sfa_6}_{\sfc\sfd_5}\,\epsilon^{\gamma\delta}\,f_{\sfb}{}_{\gamma}^{\sfa' \sfc}\,T^{\sfd_5}_\delta - \delta^{\sfa_6}_{\sfb\sfd\sfe_4}\,\epsilon^{\gamma\delta}\,f_{\sfc}{}_{\gamma}^{\sfa' \sfd}\, T^{\sfe_4\sfc}_\delta \,,\nn\\
    T^{\sfa_6,\sfa'} \circ T_\beta^{\sfb} &= f_{\sfc}{}_{\beta}^{\sfa_6}\,T^{\sfa'\sfb\sfc} + \delta^{\sfa_6}_{\sfd_3\sfe_3}\,f_{\sfc}{}^{\sfa' \sfd_3}\,T_\beta^{\sfe_3\sfb\sfc} + \delta^{\sfa_6}_{\sfd\sfe_5}\,f_{\sfc}{}_{\beta}^{\sfa' \sfd}\,T^{\sfe_5\sfc,\sfb} \,,\nn\\
    T^{\sfa_6,\sfa'} \circ T^{\sfb_3} &= \epsilon^{\gamma\delta}\,f_{\sfc}{}_{\gamma}^{\sfa_6}\,T_\delta^{\sfa'\sfb_3\sfc} - \delta^{\sfa_6}_{\sff_3\sfg_3}\,\delta^{\sfg_3\sfc}_{\sfd_3\sfe}\,f_{\sfc}{}^{\sfa' \sff_3}\,T^{\sfb_3\sfd_3,\sfe} \,,\nn\\
    T^{\sfa_6,\sfa'} \circ T_\beta^{\sfb_5} &= \delta^{\sfa'\sfc}_{\sfd\sfe}\,f_{\sfc}{}_{\beta}^{\sfa_6}\,T^{\sfb_5 \sfd,\sfe} \,,\nn\\
    T^{\sfa_6,\sfa'} \circ T^{\sfb_6,\sfb'} &= 0 \,.
\end{align}}
To obtain a more general expression for $n\leq 8$\,, we decompose the generators as
\begin{equation}
   T_A = \begin{pmatrix}
        T_{\sfa} & T_\alpha^{\sfa} & T^{\sfa_3} & T_\alpha^{\sfa_5} & T^{\sfa_6,\sfb} & T^{\sfa_7}_{(\alpha_1\alpha_2)} & T^{\sfa_7,\sfa'_2}_\alpha & T^{\sfa_7,\sfa'_4} & T^{\sfa_7,\sfa'_6}_\alpha & T^{\sfa_7,\sfa'_7,\sfa''} 
    \end{pmatrix}\,.
\end{equation}
Then the embedding tensors' components can be determined as
{\small
\begin{equation}
\begin{aligned}
    X_{\sfa} &= - f^\beta_{\sfa \sfb_6}\,R^{\sfb_6}_\beta - f_{\sfa \sfb_4}\,R^{\sfb_4} - f^\beta_{\sfa\sfb_2}\,R_\beta^{\sfb_2} + f_{\sfa\sfb}{}^{\sfc}\,\widetilde{K}^{\sfb}{}_{\sfc} - Z_{\sfa}\,(\widetilde{K}^{\sfb}{}_{\sfb}+t_0)\\
    &\quad - f_{\sfa\beta}{}^\gamma \,R^\beta{}_\gamma - f_{\sfa}{}_{\beta}^{\sfb_2}\,R^\beta_{\sfb_2} - f_{\sfa}{}^{\sfb_4}\,R_{\sfb_4} - f_{\sfa}{}_{\beta}^{\sfb_6}\,R^\beta_{\sfb_6} - f_{\sfa}{}^{\sfb_7,\sfb'}\,R_{\sfb_7,\sfb'}\,,\\
    X^{\sfa}_{\alpha} &= f_{\sfb}{}_{\alpha}^{\sfc\sfa}\,\widetilde{K}^{\sfb}{}_{\sfc} - \bigl(\delta_\alpha^\beta\,f_{\sfb_2}{}^{\sfa} - \delta^{\sfa\sfc}_{\sfb_2}\,f_{\sfc\alpha}{}^\beta\bigr)\,R_\beta^{\sfb_2} - 2\,Z_{\sfb}\,R_\alpha^{\sfa\sfb}\\
    &\quad + \epsilon_{\alpha\beta}\,f^\beta_{\sfb_3}\, R^{\sfa \sfb_3} - f_{\sfb_5}\, R_\alpha^{\sfa \sfb_5} + \epsilon_{\alpha\beta}\,f^\beta_{\sfb_7}\, R^{\sfb_7,\sfa} \,,\\
    X^{\sfa_{3}} &= f_{\sfb}{}^{\sfc \sfa_3}\,\widetilde{K}^{\sfb}{}_{\sfc} + \epsilon^{\beta\gamma}\,\delta^{\sfa_3}_{\sfc_2\sfd}\, f_{\sfb}{}_{\beta}^{\sfc_2}\, R_\gamma^{\sfd \sfb} - \delta^{\sfa_3}_{\sfc\sfd_2}\,f_{\sfb_2}{}^{\sfc}\, R^{\sfd_2 \sfb_2} - 4\,Z_{\sfb}\,R^{\sfa_{3}\sfb}\\
    &\quad + f^\alpha_{\sfb_3}\,R_{\alpha}^{\sfa_3\sfb_3} + \tfrac{1}{2}\,\delta^{\sfa_3}_{\sfc_2\sfd}\,f_{\sfb_5}\,R^{\sfb_5\sfc_2,\sfd}\,,\\
    X_\alpha^{\sfa_5} &= f_{\sfb}{}_{\alpha}^{\sfc \sfa_5}\,\widetilde{K}^{\sfb}{}_{\sfc} + \delta^{\sfa_5}_{\sfc_4\sfd}\, f_{\sfb}{}^{\sfc_4}\, R_\alpha^{\sfd\sfb} - \delta^{\sfa_5}_{\sfc_2\sfd_3}\,f_{\sfb}{}_{\alpha}^{\sfc_2}\, R^{\sfd_3 \sfb}\\
    &\quad - \delta^{\sfa_5}_{\sfc\sfd_4}\, f_{\sfb_2}{}^{\sfc}\, R_\alpha^{\sfd_4\sfb_2} + f_{\sfb\alpha}{}^\beta\, R_\beta^{\sfa_5\sfb} - 6\,Z_{\sfb}\, R_\alpha^{\sfa_5\sfb} + \epsilon_{\alpha\beta}\,\delta^{\sfb_3}_{\sfc_2\sfd}\,f^\beta_{\sfb_3}\,R^{\sfa_5\sfc_2,\sfd}\,,\\
    X^{\sfa_6,\sfa'} &= \bigl(f_{\sfb}{}^{\sfc \sfa_6,\sfa'} - c_{7,1}\,f_{\sfb}{}^{\sfa_6\sfa',\sfc}\bigr)\,\widetilde{K}^{\sfb}{}_{\sfc} - \epsilon^{\beta\gamma}\, \bigl(f_{\sfb}{}_{\beta}^{\sfa_6}\,R_\gamma^{\sfa' \sfb} - c_{6}\,\delta^{\sfa_6\sfa'}_{\sfc_6\sfd}\,f_{\sfb}{}^{\sfc_6}_\beta\,R^{\sfd\sfb}_\gamma \bigr)\\
    &\quad + \delta^{\sfa_6}_{\sfc_3\sfd_3}\,f_{\sfb}{}^{\sfa' \sfc_3}\,R^{\sfd_3\sfb} - c_4\,\delta^{\sfa_6\sfa'}_{\sfc_4\sfd_3}\,f_{\sfb}{}^{\sfc_4}\,R^{\sfd_3\sfb} - \epsilon^{\beta\gamma}\,\bigl(\delta^{\sfa_6}_{\sfc\sfd_5}\,f_{\sfb}{}_{\beta}^{\sfa'\sfc}\,R^{\sfd_5 \sfb}_\gamma - c_2\,\delta^{\sfa_6\sfa'}_{\sfc_2\sfd_5}\,f_{\sfb}{}^{\sfc_2}_\beta\,R_\gamma^{\sfd_5\sfb} \bigr)\\
    &\quad - \delta^{\sfa_6}_{\sfc\sfd_5}\,f_{\sfb_2}{}^{\sfc}\,R^{\sfd_5 \sfb_2,\sfa'} - f_{\sfb\sfc}{}^{\sfa'}\,R^{\sfa_6\sfb,\sfc} + 8\,Z_{\sfb}\,R^{\sfa_6\sfb,\sfa'}-(1+c_{7,1})\,Z_{\sfb}\,R^{\sfa_6\sfa',\sfb}\,,\\
    X^{\sfa_{7}}_{(\alpha_1\alpha_2)} &= \delta^{\sfa_7}_{\sfc_6\sfd}\,f_{\sfb}{}^{\sfc_6}_{(\alpha_1}\,R^{\sfd\sfb}_{\alpha_2)} - \delta^{\sfa_7}_{\sfc_2\sfd_5}\,f_{\sfb}{}^{\sfc_2}_{(\alpha_1}\,R^{\sfd_5\sfb}_{\alpha_2)} - f_{\sfb(\alpha_1}{}^\beta\,\epsilon_{\alpha_2)\beta}\,R^{\sfa_7,\sfb}\,,\\
    X^{\sfa_{7},\sfa'_{2}}_\alpha &= \delta^{\sfa'_2}_{\sfc\sfd}\,f_{\sfb}{}^{\sfa_7,\sfc}\,R^{\sfd\sfb}_\alpha + \delta^{\sfa_7}_{\sfc_6\sfd}\, f_{\sfb}{}^{\sfc_6}_{\alpha}\,R^{\sfd \sfa'_2\sfb} - \delta^{\sfa_7}_{\sfc_2\sfd_5}\,f_{\sfb}{}^{\sfa'_2\sfc_2}\,R^{\sfd_5\sfb}_\alpha - \delta^{\sfa'_2}_{\sfc\sfd}\,f_{\sfb}{}^{\sfc\sfb}_\alpha\,R^{\sfa_7, \sfd} - f_{\sfb}{}^{\sfa'_2}_\alpha\,R^{\sfa_7,\sfb} \,,\\
    X^{\sfa_{7},\sfa'_{4}} &= \delta^{\sfa'_4}_{\sfc\sfd_3}\,f_{\sfb}{}^{\sfa_7,\sfc}\,R^{\sfd_3\sfb} + \epsilon^{\beta\gamma}\,\delta^{\sfa_7}_{\sfc_6\sfd}\,f_{\sfb}{}^{\sfc_6}_{\beta}\,R^{\sfd \sfa'_4\sfb}_\gamma - \delta^{\sfa'_4}_{\sfc_3\sfd}\,f_{\sfb}{}^{\sfc_3\sfb}\,R^{\sfa_7, \sfd} - f_{\sfb}{}^{\sfa'_4}\,R^{\sfa_7,\sfb}\,,\\
    X^{\sfa_{7},\sfa'_{6}}_\alpha &= \delta^{\sfa'_6}_{\sfc\sfd_5}\,f_{\sfb}{}^{\sfa_{7},\sfc}\,R_\alpha^{\sfd_5\sfb} - \delta^{\sfa'_6}_{\sfc_5\sfd}\,f_{\sfb}{}^{\sfc_5\sfb}_{\alpha}\,R^{\sfa_7,\sfd} - f_{\sfb}{}^{\sfa'_6}_\alpha\,R^{\sfa_7,\sfb} \,,\\
    X^{\sfa_7,\sfa'_7,\sfa''} &= - f_{\sfb}{}^{\sfa_7,\sfb}\,R^{\sfa'_7,\sfa''} - 2\, f_{\sfb}{}^{\sfa_7,\sfa''}\,R^{\sfa'_7,\sfb}\,,
\end{aligned}\end{equation}}%
where we have defined $c_2 := \tfrac{2}{7}+\tfrac{1}{7\sqrt{2}}$\,, $c_4 := \tfrac{4}{7} + \tfrac{2}{7\sqrt{2}}$\,, $c_6 := \tfrac{1}{7}-\tfrac{3}{7\sqrt{2}}$\,, and $c_{7,1} := \tfrac{1}{7}+\tfrac{4}{7\sqrt{2}}$\,. For $n=8$\,, only $954$ of the embedding tensors are geometric and the other $3170$ components correspond to (locally) non-geometric fluxes. 
In this case, a seven-form field strength
\begin{align}
 F^\mu_7 := \rmd B^\mu_6 - B^\mu_2\wedge \rmd D_4 - \tfrac{1}{3!}\,\epsilon_{\nu\rho}\,B^\mu_2\wedge B^\nu_2\wedge F^\rho_3\,,
\end{align}
appears, and in addition to the expressions \eqref{eq:B-F3} and \eqref{eq:B-F5}, we find the additional condition
\begin{align}
    F^\alpha_7 = \tfrac{1}{7!}\,f^\alpha_{\sfa_1\dots\sfa_7}\,v^{\sfa_1}\wedge\dots\wedge v^{\sfa_7}
 - \tfrac{1}{6!}\,X_{A,\sfb,}{}_{\sfc_1\dots\sfc_5}^\gamma\,v^A\wedge v^{\sfb}\wedge v^{\sfc_1}\wedge \dots \wedge v^{\sfc_5}\,.
\end{align}
It will be a hard but straightforward task to check the Bianchi identities for these field strengths in $n=8$\,.

\section{Elgebroids}\label{app:elgebroids}
An alternative approach to the frame algebra \eqref{eq:framealgebra} are $G$-algebroids \cite{Bugden:2021wxg,Bugden:2021nwl}. They are defined in terms of
\begin{enumerate}
    \item a bracket $[\cdot,\cdot]$,
    \item an anchor map $\rho$, and 
    \item an operator $\cD$\,.
\end{enumerate}
The anchor $\rho$ simply projects our generalised vector $V^I$ to its vector component, namely $\rho(V)=\rho^i{}_J\,V^J\,\partial_i = v^i\,\partial_i$ in M-theory sections or $\rho(V)=\rho^m{}_J\,V^J\,\partial_m=v^m\,\partial_m$ in the type IIB sections, and the bracket $[V,\,W]$ corresponds to our generalised Lie derivative $\gLie_V W^I$ defined in \eqref{eq:gLie}. Indeed, if we take M-theory sections, we find \cite{Sakatani:2017xcn}
\begin{equation}
    \gLie_V W^I = \begin{pmatrix}
        \Lie_v w^i \\
        (\Lie_v w_2 - \iota_w \rmd v_2)_{i_1i_2} \\
        (\Lie_v w_5 + \rmd v_2\wedge w_2 - \iota_w \rmd v_5)_{i_1\dots i_5} \\
        [\Lie_v w_{7,\,i'}+ (\iota_{i'}\rmd v_2)\wedge w_5 + (\iota_{i'}w_2)\wedge \rmd v_5]_{i_1\dots i_7}
    \end{pmatrix}\,,
\end{equation}
where $\Lie_v$ denotes the standard Lie derivative. This reproduces the bracket of \cite{Bugden:2021wxg} after restricting to $n\leq 6$ (up to the sign convention on the five-form part). In the type IIB section, we find \cite{Sakatani:2017xcn}
\begin{equation}
    \gLie_V W^I = \begin{pmatrix}
        \Lie_v w^{\sfm} \\
        (\Lie_v w_1^\mu -\iota_w \rmd v_1^\mu)_{\sfm} \\
        (\Lie_v w_3 - \epsilon_{\mu\nu}\, \rmd v_1^\mu\wedge w^\nu_1 - \iota_w \rmd v_3)_{\sfm_1\sfm_2\sfm_3}\\
        (\Lie_v w^\mu_5 + \rmd v_1^\mu\wedge w_3 - \rmd v_3\wedge w^\mu_1 - \iota_w \rmd v_5^\mu)_{\sfm_1\dots \sfm_5}\\
        [\Lie_v w_{6,\,\sfm'} + \epsilon_{\mu\nu}\,(\iota_{\sfm'}\rmd v^\mu_1)\wedge w^\nu_5 - (\iota_{\sfm'}\rmd v_3)\wedge w_3 + \epsilon_{\mu\nu}\,\rmd v^\mu_5\, w^\nu_{\sfm'}]_{\sfm_1\dots \sfm_6}
    \end{pmatrix}\,,
\end{equation}
and by considering $n\leq 6$, this is exactly the bracket used in \cite{Bugden:2021nwl}. 
Finally, $\cD$ is identified from the relation
\begin{align}\label{eq:genLieSym}
    [V,\,W]^I + [W,\,V]^I = Y^{IJ}_{KL}\,\bigl(\partial_J V^K\,W^L + \partial_J W^K\,V^L\bigr)\,.
\end{align}
For $n\leq 6$, the $Y$-tensor can be written as $Y^{IJ}_{KL}=\eta^{IJ;\cI}\,\eta_{KL;\cI}$ where $\eta_{IJ;\cK}=\eta_{(IJ);\cK}$ is known as the $\eta$-symbol\footnote{It is essentially the same as the projection $\times_N:E\times E\to N$ of \cite{Coimbra:2011ky} and the wedge product $\wedge:R_1\otimes R_1\to R_2$ of \cite{Malek:2017njj}.} \cite{Linch:2016ipx,Sakatani:2017xcn}. It relates the $R_1$ representation and the $R_2$ representation which governs the section condition. The $\eta$-symbol $\eta^{IJ;\cK}$ with raised indices has the same components as $\eta_{IJ;\cK}$. It is explicitly given in \cite{Sakatani:2017xcn}. To recover $\cD$ from our setup, we first rewrite \eqref{eq:genLieSym} as
\begin{equation}
    [V,\,W]^I + [W,\,V]^I = \cD^{I;\cJ} (V\times_N W)_{\cJ}\,,
\end{equation}
with
\begin{equation}
    \cD^{I;\cJ} := 2\,\eta^{IK;\cJ}\,\partial_K\,,
\end{equation}
and
\begin{equation}
    (V\times_N W)_{\cI} = \eta_{JK;\cI}\, V^J\,W^K\,.
\end{equation}
It is not hard to check that the differential operator $\cD$ arising this way indeed satisfies the defining property of its elgebroids counterpart\cite{Bugden:2021wxg}
\begin{equation}
    \cD^{I;\cJ}(f\,N_{\cJ}) = f\,\cD^{I;\cJ} N_{\cJ} + 2\,\eta^{IK;\cJ}\,\partial_Kf\,N_{\cJ} = f\,\cD^{I;\cJ} N_{\cJ} + \partial_K f\,N^{KI} + \partial_K f\,N^{IK} \,. 
\end{equation}
In particular, we verify the property $\rho\circ\cD=0$ \cite{Bugden:2021wxg} by using $\eta^{ij;\cK}=0$ (M-theory) or $\eta^{\sfm\sfn;\cK}=0$ (type IIB). 

For $n=7$\,, the definition of the $G$-algebroid has to be altered because the $Y$-tensor is not symmetric in its raised indices. Instead it becomes
\begin{align}
 Y^{IJ}_{KL} = \eta^{IJ;\cI}\,\eta_{KL;\cI} -\tfrac{1}{2}\,\Omega^{IJ}\,\Omega_{KL}\,,
\end{align}
where $\Omega_{IJ}$ is a matrix satisfying $\Omega_{IJ}=-\Omega_{JI}$ and $\Omega^{IJ}=\Omega_{IJ}$\,. Again, its explicit form is given in \cite{Sakatani:2017xcn}. 
Due to this modification, we have
\begin{equation}
    \begin{aligned}
        [V,\,W]^I + [W,\,V]^I &= \mathfrak{D}^I(V\otimes W)\\
        :\!&= \cD^{I;\cJ} (V\times_N W)_{\cJ} -\tfrac{1}{2}\,\Omega_{JK}\,\bigl(\tilde{\cD}^I V^J\,W^K - V^J\,\tilde{\cD}^I W^K\bigr)\,,
    \end{aligned}
\end{equation}
where $\tilde{\cD}^I := \Omega^{IJ}\,\partial_J$ and $\Omega^{IK}\,\Omega_{JK}=\delta^I_J$\,. Taking into account $\Omega^{ij}=0$, we check the property $\rho\circ \tilde{\cD}^I = 0$\,, and therefore see that $\mathfrak{D}$ plays a similar role as $\cD^{I;\cJ}$ that appears in elgebroids with $n\le 6$. However, we have not found an abstract index-free definition of the operator $\mathfrak{D}$ and this may explain the difficulties to formulate elgebroids in $n=7$. 

In the language of $G$-algebroids, our main result can be summarised as
\begin{enumerate}
    \item identifying the most general form of the structure constants $X_{AB}{}^C$ that can arise in the algebra
    \begin{align}
        [E_A,\,E_B] = -X_{AB}{}^C\,E_C\,,
    \end{align}
    and
    \item the explicit construction of the generalised frame $E_A$\,.
\end{enumerate}
Our class of Leibniz algebras $T_A\circ T_B=X_{AB}{}^C\,T_C$, called geometric gaugings, precisely corresponds to the elgebra studied in \cite{Bugden:2021wxg,Bugden:2021nwl}.\footnote{In \cite{Bugden:2021wxg,Bugden:2021nwl}, a twisted bracket is used. It is constructed as follows in our setting: First, we decompose the generalised frame $E_A{}^I$ as $E_A{}^I=\hat{E}_A{}^J\,N_J{}^I$ and define the twisted bracket $[\cdot,\cdot]_{\text{twist}}$ as $[E_A,\,E_B]^I = [\hat{E}_A,\,\hat{E}_B]^J_{\text{twist}}\,N_J{}^I$\,. The resulting bracket $[\cdot,\cdot]_{\text{twist}}$ contains the field strengths such as $F_3$ and $F_6$ and reproduces the twisted bracket \cite{Bugden:2021wxg,Bugden:2021nwl}. By definition, the frame $\hat{E}_A{}^I$ obeys the same algebra $[\hat{E}_A,\,\hat{E}_B]_{\text{twist}} = -X_{AB}{}^C\,\hat{E}_C$.}
In exceptional field theory, one can furthermore consider generalised parallelisations that violate the section condition. In this case, the gaugings $X_{AB}{}^C$ are called non-geometric, and the corresponding generalised frames necessarily require a coordinate dependence in the extended geometry. Therefore, one is forced to go beyond the regime of the elgebroids.

\section{Enhanced Leibniz algebras}\label{app:ELA}
Here, we connect the structure revealed in section~\ref{sec:group+diff} to enhanced Leibniz algebras \cite{Strobl:2016aph,Strobl:2019hha} (see also \cite{Bugden:2021nwl} for comments on this relation). As shown in \cite{Strobl:2019hha}, each enhanced Leibniz algebra is associated with a (semi-strict) Lie 2-algebra \cite{Baez:2003fs}, or two-term $L_\infty$-algebra. 

An enhanced Leibniz algebra is defined by three ingredients:
\begin{enumerate}
    \item a map $t:\mathbb{W}\to \mathbb{V}$\,,
    \item a bracket $[\cdot,\,\cdot]$\,, and
    \item a (bilinear) product $\circ:\mathbb{V}\times \mathbb{V}\to \mathbb{W}$\,.
\end{enumerate}
Here, $(\mathbb{V},\,[\cdot,\,\cdot])$ is a Leibniz algebra, saying that the bracket satisfies the Leibniz identity
\begin{equation}
    [V_1,\,[V_2,\,V_3]] = [[V_1,\,V_2],\,V_3] + [V_2,\,[V_1,\,V_3]]
\end{equation}
for all $V_1,\,V_2,\,V_3\in \mathbb{V}$. Additionally, the relations
\begin{equation}
    [t(w),\,V] = 0\,,\qquad
    t(w)\circ t(w) = 0\,,\qquad
    U\overset{s}{\circ}[V,\,V] = V\overset{s}{\circ}[U,\,V]\,,
    \qquad \text{and} \qquad
    [V,\,V]=t(V\circ V)\,,
\end{equation}
have to hold for all $U,\,V \in \mathbb{V}$ and $w\in \mathbb{W}$. Here, it is convenient to define the symmetrised product $u\overset{s}{\circ} v := \frac{1}{2}\,(u\circ v+v\circ u)$\,. For the special case, where the product is symmetric ($\circ=\overset{s}{\circ}$), the enhanced Leibniz algebra is called symmetric. 

This definition fits our situation perfectly. As is discussed in \cite{Strobl:2016aph,Strobl:2019hha} (see Proposition 1.7 of \cite{Strobl:2019hha}), for any Leibniz algebra, we can construct a canonical symmetric enhanced Leibniz algebra. In our notation, the Leibniz algebra $(\mathbb{V},\,[\cdot,\,\cdot])$ is denoted as $(E,\,\circ)$ and $\mathbb{W}$ is taken as the ideal $\cI$ (which is denoted as Sq($\mathbb{V}$) in \cite{Strobl:2019hha}). Finally, the map $t$ captures its embedding into $E$,
\begin{equation}
    t(v^{\acute{a}}) = V^A = (0,\,v^{\acute{a}})\,,
\end{equation}
and the symmetric product $\overset{s}{\circ}$ is defined as
\begin{equation}
    U\overset{s}{\circ} V = U^A\,V^B\,Z_{AB}{}^{\acute{c}}\,T_{\acute{c}}\,.
\end{equation}
With this identification, it is easy to check that all of the defining properties above are indeed satisfied. 

An element $x\in \mathrm{Lie}(G) = E/\cI$ corresponds to a class of elements of $E$, denoted as $[x] = x+\cI$\,. We then define the action of $x\in \mathrm{Lie}(G)$ on $m\in \cI$ as
\begin{equation}
    x \cdot m := [x+\cI,\,m] = [x,\,m]\,.
\end{equation}
Since the representation property $x\cdot (y\cdot m)-y\cdot (x\cdot m) = [x,\,y]\cdot m$ is satisfied, we can construct the matrix representation of Lie($G$) on $\cI$\,. This is precisely our representation $R_{\cI}$\,,
\begin{equation}
    (T_{\grave a})_{\acute{\beta}}{}{\acute{\gamma}} = - X_{\grave{a}\acute\beta}{}^{\acute\gamma}\,. 
\end{equation}
Then, combining the adjoint representation of Lie($G$) and $R_{\cI}$\,, we express the Leibniz bracket for $(x,\,m),\,(y,\,n) \in E$ as
\begin{align}
    [(x,\,m),\,(y,\,n)] = ([x,\,y],\, x\cdot n + \alpha(x,y))\,.
\end{align}
Here, $\alpha\in \text{Lie($G$)}^*\otimes \text{Lie($G$)}^*\otimes \cI$, is in our notation denoted as
\begin{equation}
    \alpha(T_{\grave{a}},\,T_{\grave{b}}) = X_{\grave{a}\grave{b}}{}^{\acute{\gamma}}\,T_{\acute{\gamma}}\,,
\end{equation}
which is the quantity appearing in the off-diagonal block of the Leibniz representation. 
Using these notations, the Leibniz identity \eqref{eq:intertwine} can be expressed as
\begin{align}
    x\cdot\alpha(y,\,z) - y\cdot\alpha(x,\,z) = \alpha([x,y],\,z) + \alpha(y,\,[x,z])-\alpha(x,\,[y,z])\,.
\end{align}
It represents a cocycle condition $\dd_L\alpha = 0$ with the Loday coboundary operator $\dd_L$ (see \cite{Strobl:2019hha} for details). Each tuple of a Leibniz algebra $T_A\circ T_B = X_{AB}{}^C\,T_C$ and its ideal $\cI$ is in one-to-one correspondence with a Lie($G$)-module $\cI$ and a cohomology class $[\alpha]$\,. This is stated in Proposition 2.4 of \cite{Strobl:2019hha} and our correspondence \eqref{eq:one-to-one} is a rephrasing of this proposition. 

For Lie algebras, the left-invariant one-form $A$ satisfies the Maurer-Cartan equation, or in other words, has a vanishing field strength
\begin{align}
 F = \rmd A + \tfrac{1}{2}\,[A\overset{\wedge}{,}\,A] =0 \,.
\end{align}
Enhanced Leibniz algebras (or semi-strict Lie 2-algebras) possess the natural generalisations of the field strength \cite{Grutzmann:2014hkn}
\begin{equation}
    \begin{aligned}
        F &= \dd A + \tfrac{1}{2}\,[A\overset{\wedge}{,}\,A]_- - t(B)\,, \qquad \text{and}\\
        G &= \dd B + t(B) \circ A - \tfrac{1}{3!}\,\gamma(A\overset{\wedge}{,}\,A\overset{\wedge}{,}\,A) - A \circ F\,,
    \end{aligned}
\end{equation}
where the wedge product is omitted on the product $\circ$ and we have introduced the anomaly,
\begin{align}
    \gamma(V_1,\,V_2,\,V_3) := - X_{[AB}{}^E\,Z_{C]E}{}^D\,V_1^A\,V_2^B\,V_3^C \,.
\end{align}
The vanishing of $F$ and $G$ can be understood as generalised Maurer-Cartan equations. They precisely match our equations \eqref{eq:modified-MC1} and \eqref{eq:dw} under the identification, $A^A = v^A$ and $w^A=-B^A$\,. Therefore, $A$ is the one-form $v^A$, which plays a central role in the construction of the generalised frame $E_A$\,, and $B$ is the two-form $w^A$.

\bibliography{literature}
   
\bibliographystyle{JHEP}

\end{document}